\documentclass[12pt]{article}

\newlength{\abstractwidth}
\usepackage{amsmath, nccmath, amsthm}
\usepackage{amsfonts}
\usepackage{amssymb}
\usepackage{latexsym}
\usepackage{shuffle}

\usepackage{color}
\usepackage{graphicx}
\usepackage{tikz}
\usepackage{dsfont}

\usepackage{tikz}
\usetikzlibrary{calc} 
\usetikzlibrary{patterns,snakes} 
\usetikzlibrary{decorations.pathreplacing} 
\usetikzlibrary{decorations.markings} 
\usetikzlibrary{decorations.pathmorphing} 
\usetikzlibrary{positioning}
\usetikzlibrary{arrows.meta}

\usetikzlibrary{arrows,shapes,positioning}
\usetikzlibrary{decorations.markings}
\usepackage[rightcaption]{sidecap}
\tikzstyle arrowstyle=[scale=1]
\tikzstyle directed=[postaction={decorate,decoration={markings,
    mark=at position .65 with {\arrow[arrowstyle]{stealth}}}}]
\tikzstyle reverse directed=[postaction={decorate,decoration={markings,
    mark=at position .65 with {\arrowreversed[arrowstyle]{stealth};}}}]
\usetikzlibrary{positioning}

\usepackage{hyperref}
\definecolor{darkred}{rgb}{0.8,0.1,0.1}
\hypersetup{colorlinks=true, linkcolor=darkred, citecolor=blue, linktoc=page}

\numberwithin{equation}{section}

\renewcommand{\thefootnote}{\fnsymbol{footnote}}
\renewcommand{\thanks}[1]{\footnote{#1}}
\newcommand{\starttext}{
\renewcommand{\thefootnote}{\arabic{footnote}}}

\newcommand{\bea}{\begin{eqnarray}}
\newcommand{\eea}{\end{eqnarray}}
\newcommand{\be}{\begin{eqnarray}}
\newcommand{\ee}{\end{eqnarray}}
\newcommand{\bma}{\begin{matrix}}
\newcommand{\ema}{\cr\end{matrix}}

\newtheorem{thm}{Theorem}[section]
\newtheorem{lem}[thm]{Lemma}
\newtheorem{prop}[thm]{Proposition}
\newtheorem{cor}[thm]{Corollary}

\newtheorem{deff}[thm]{Definition}
\newtheorem{remark}[thm]{Remark}

\def\beq{\begin{equation}}
\def\eeq{\end{equation}}

\def\cG{{\cal G}}
\def\cH{{\cal H}}

\def\cJ{{\cal J}}
\def\cK{{\cal K}}

\def\cM{{\cal M}}
\def\cN{{\cal N}}
\def\cO{{\cal O}}

\def\cS{{\cal S}}
\def\cT{{\cal T}}
\def\cU{{\cal U}}
\def\cV{{\cal V}}
\def\cW{{\cal W}}
\def\cX{{\cal X}}

\def\bG{{\bf G}}

\def\bJ{{\bf J}}
\def\bK{{\bf K}}

\def\mA{\mathfrak{A}}
\def\mB{\mathfrak{B}}

\def\mJ{\mathfrak{J}}

\def\mT{\mathfrak{T}}

\def\mg{\mathfrak{g}}

\def\mt{\mathfrak{t}}

\def\ZZ{{\mathbb Z}}

\def\MM{{\mathbb M}}
\def\CC{{\mathbb C}}

\def\sfY{{\mathsf{Y}}}
\def\sfZ{{\mathsf{Z}}}
\def\sfR{{\mathsf{R}}}

\def\Im{{\rm Im \,}}

\def\half{{1\over 2}}

\def\p{\partial}

\def\a{\alpha}
\def\b{\beta}

\def\ep{\varepsilon}
\def\om{\omega}

\def\KK{{\bf K}}
\def\OO{{\bf W}}
\def\XX{{\bf X}}
\def\Ber{{\rm Ber}}
\def\newL{{\cal S}}
\def\anotherL{\Upsilon}

\def\vI{{\vec{I}}}

\def\vP{{\vec{P}}}
\def\vQ{{\vec{Q}}}

\def\vX{{\vec{X}}}
\def\vY{{\vec{Y}}}

\def\no{\nonumber}
\def\sm{\smallskip}

\def\pbx{\p_{\bar x}}
\def\pby{\p_{\bar y}}
\def\pbz{\p _{\bar z}}

\def\xx{{\boldsymbol x}}
\def\yy{{\boldsymbol y}}
\def\zz{{\boldsymbol z}}

\def\bPhi{\boldsymbol{\Phi}}

\begin{document}
\starttext
\setcounter{footnote}{0}

\begin{flushright}
2026 February 01  \\
revised November 2026\\
UUITP--37/25
\end{flushright}

\vskip 0.1in

\begin{center}

{\Large \bf Single-valued flat connections in several variables}

\vskip 0.1in

{\large \bf on arbitrary Riemann surfaces}

\vskip 0.2in

{\large Eric D'Hoker${}^{a}$, Oliver Schlotterer\footnote{Corresponding author}${}^{b}$} 

\vskip 0.1in

{ \sl ${}^{a}$Mani L. Bhaumik Institute for Theoretical Physics}\\
{\sl  Department of Physics and Astronomy}\\
{\sl University of California, Los Angeles, CA 90095, USA}

\vskip 0.1in

{\sl ${}^b$Department of Physics and Astronomy,} \\
  { \sl Department of Mathematics,} \\
  { \sl Centre for Geometry and Physics,} \\ 
  {\sl Uppsala University, 75120 Uppsala, Sweden}

\vskip 0.15in 

{\tt \small dhoker@physics.ucla.edu, oliver.schlotterer@physics.uu.se}

\vskip 0.2in

\begin{abstract}
\vskip 0.1in
Polylogarithms on Riemann surfaces may be constructed systematically in terms of flat connections  that can enjoy various algebraic and analytic properties. In this paper, we present a single-valued and modular invariant connection ${\cal J}_\text{DHS}$ on the configuration space $\text{Cf}_n(\Sigma)$ of an arbitrary number $n$ of points on an arbitrary compact Riemann surface $\Sigma$ with or without punctures. The connection ${\cal J}_\text{DHS}$ generalizes an earlier construction for a single variable and is built out of the same integration kernels. We show that ${\cal J}_\text{DHS}$  is flat on $\text{Cf}_n(\Sigma)$. For the case without punctures, we relate it to the meromorphic multiple-valued Enriquez connection ${\cal K}_\text{E}$ in $n$ variables  on the universal cover $\tilde \Sigma$ of $\Sigma$ by the composition of a gauge transformation and an automorphism of the Lie algebra in which ${\cal J}_\text{DHS}$ and ${\cal K}_\text{E}$ take values.  In a companion paper, we shall establish the equivalence between the flatness of these connections and the corresponding interchange and Fay identities, for arbitrary compact Riemann surfaces.
\end{abstract}
\end{center}

\newpage

\setcounter{tocdepth}{2} 
\tableofcontents

\baselineskip=15pt
\setcounter{equation}{0}
\setcounter{footnote}{0}

\newpage

\section{Introduction}
\setcounter{equation}{0}
\label{sec:1}

Progress in performing perturbative computations in quantum field theory and string theory relies on  increasingly sophisticated methods for carrying out integrals on a variety of different geometries. Gaining a systematic understanding of the spaces of functions to which such integrals belong has proven particularly fruitful. One of the earliest and simplest cases involves integration on the Riemann sphere where the completion of the space of 
rational functions by multiple polylogarithms leads to an algebra of functions that is closed under taking derivatives and primitives \cite{Lappo:1953, Goncharov:1995, Brown:2009qja} (see \cite{Panzer:2015ida, Abreu:2022mfk} for general overviews). Analogous function spaces of elliptic polylogarithms on the torus were constructed in joint efforts of mathematicians \cite{Levin:1997, Levin:2007, BrownLevin, Enriquez:2023} and physicists \cite{Broedel:2014vla, Broedel:2017kkb} (see \cite{Bourjaily:2022bwx, Berkovits:2022ivl} for an overview of applications to scattering amplitudes). 
Generalizations to the case of higher genus Riemann surfaces were recently established in \cite{Enriquez:2021, Enriquez:2022, DHoker:2023vax, Baune:2024biq}  through a confluence of ideas from algebraic geometry, number theory, string theory and particle physics, and further developed in \cite{DHoker:2024ozn, DHoker:2024desz, Enriquez:next}.

\sm

The different types of polylogarithms can be constructed from flat connections on Riemann surfaces, with or without punctures, and their universal covers. The flatness condition ensures that the connection gives rise to homotopy invariant iterated integrals, namely, definite integrals that depend on the homotopy class of the path between the endpoints, but are independent of the specific path within a given homotopy class. While the relevant connections on the sphere, such as the KZ connection \cite{Knizhnik:1984nr}, can simultaneously be meromorphic and single-valued, these two properties enter into conflict starting at genus one. As a result, a variety of flat connections that enjoy different algebraic and analytic properties have been advanced for genus $\geq 1$. They include the meromorphic, but multi-valued, connections of  \cite{Calaque:2009, Enriquez:2014} at genus one and of \cite{Enriquez:2011} at arbitrary higher genus, which were further elaborated in \cite{Baune:2024biq, DHoker:2025dhv, Ichikawa:2025kbi}, as well as the single-valued and modular invariant, but non-meromorphic, connection of \cite{BrownLevin} at genus one and of \cite{DHoker:2023vax} at arbitrary genus. Connections that are simultaneously meromorphic and single-valued may be achieved by including multiple punctures or a higher order pole at a single  puncture   \cite{Levin:2007, Enriquez:2021, Enriquez:2022, Enriquez:2023}.

\sm

Both the mathematical understanding and the physical applications of  polylogarithms greatly benefit from extending their flat connections for a single variable on a punctured Riemann surface $\Sigma$ to the configuration space $\text{Cf}_n(\Sigma)$ of multiple variables on a compact Riemann surface without punctures. First, functional identities 
and translations between different iterated integral representations (referred to as \textit{fibration bases}) of  polylogarithms are most conveniently derived from connections that are one-forms in multiple variables.
Second, flat connections on configuration spaces play a vital role in formulating certain algebraic structures on the corresponding spaces of polylogarithms, such as the motivic coaction \cite{Gon:mot1, Gon:mot2, Brown:mot1, Brown:mot2}  and the single-valued map \cite{Brown:svpoly, Schnetz:2013hqa, Brown:2013gia, Brown:2018omk,Brown:2019wna}.
For instance, the multi-variable KZ connection is a key ingredient in reformulating  the motivic coaction \cite{Frost:2023stm, Frost:2025lre}  and the single-valued map on the sphere. This reformulation facilitates their generalization to higher genus and, among other things,  guides the explicit construction of single-valued elliptic polylogarithms \cite{Schlotterer:2025qjv}.

\sm

As a first main result of this work, we present a single-valued and modular invariant connection  
$\cJ_\text{DHS}$ on the configuration space $\text{Cf}_n (\Sigma)$ of $n \geq 2$ points on a compact Riemann surface of genus $h \geq 1$ and prove its flatness. The  connection $\cJ_\text{DHS}$ generalizes the DHS connection on punctured Riemann surfaces  \cite{DHoker:2023vax} to multiple variables and is built out of the same single-valued integration kernels, which transform as tensors under the modular group $Sp(2h,\mathbb Z)$ of a Riemann surface of genus $h$. While the single-variable  DHS connection with one puncture takes values in a freely generated Lie algebra, its multi-variable generalization without punctures takes values in the Lie algebra $\hat \mt_{h,n}$ introduced in  \cite{Enriquez:2011}, which is not freely generated.   In fact, the structure relations of $\hat \mt_{h,n}$ play a crucial role in the properties of $\cJ_\text{DHS}$, including the proof of its  flatness. For completeness, we derive the single-valued and modular invariant DHS connection $\cJ_\text{DHS}$ for the general case of $n$ variables and $p$ punctures with values in the Lie algebra $\hat \mt_{h,n,p}$ on a Riemann surface of arbitrary genus $h \geq 1$ by reduction from the connection with $n+p$ variables and no punctures.

\sm

A meromorphic flat connection $\cK_\text{E}$ on the configuration space $\text{Cf}_n (\tilde \Sigma)$ of $n$ points on the universal cover $\tilde \Sigma$ of a compact Riemann surface $\Sigma$ without punctures, which also takes values in $\hat \mt_{h,n}$, was presented some time ago by Enriquez  in \cite{Enriquez:2011}.

\sm

As a second main result of this work, we relate the multi-variable connections $\cJ_\text{DHS}$ and $\cK_\text{E}$ on a Riemann surface without punctures  
to one another via the composition of a gauge transformation and an automorphism of the Lie algebra $\hat \mt_{h,n}$, both of which we determine explicitly. This relation between multi-variable flat connections without punctures  generalizes the relation between the DHS and Enriquez  connections in one variable and one puncture, established in \cite{DHoker:2024desz}. The gauge transformation and Lie algebra automorphism for the multi-variable case are closely related to those required for the single-variable case and are built out of the same integration kernels and polylogarithms. 

\sm

The relation between the multi-variable connections $\cJ_\text{DHS}$ and $\cK_\text{E}$ carries over to the relations between their associated polylogarithms, just as for the single-variable case. In particular, the higher-genus polylogarithms obtained from the DHS connection are expressible in terms of those of the Enriquez connection with iterated integrals of antiholomorphic Abelian differentials as coefficients, see Theorem 6.2 of \cite{DHoker:2024desz} for further details.

\sm

We shall provide an alternative new proof of the flatness of Enriquez's connection based on direct computations and the explicit form of the respective connections, thereby offering complementary perspectives on Enriquez's earlier  proof~\cite{Enriquez:2011}.

\sm

In a companion paper \cite{DS-Fay}, we shall establish the equivalence between the flatness of the 
connection $\cJ_\text{DHS}$ in $n\geq 3$ variables and the  interchange and Fay identities formulated in terms of the DHS integration kernels $f^{I_1 \cdots I_r}{}_J(x,y)$.  In parallel, we shall also deduce the interchange and Fay identities formulated in terms of the Enriquez integration kernels $g^{I_1 \cdots I_r}{}_J(x,y)$ from the flatness of the connection $\cK_\text{E}$ in $n\geq 3$ variables.

\subsection*{Future applications and open questions}

The single-valued flat connection $\cJ_\text{DHS}$ in $n$ variables presented in this work offers a
convenient setting for  producing functional relations among polylogarithms on higher-genus Riemann surfaces.  Flatness of $\cJ_\text{DHS}$ allows us to equate the path-ordered exponentials of $\cJ_\text{DHS}$ evaluated along different homotopic paths between given endpoints in ${\rm Cf}_n(\Sigma)$ and to relate the associated polylogarithms in different ``fibration bases" \cite{Brown:2009qja, BrownLevin}.  A key benefit of $\cJ_\text{DHS}$ is to produce change of fibration-basis identities for polylogarithms by way of systematic generating functions, thereby streamlining the iterative process using the Fay identities, described in \cite{DHoker:2024ozn}, that was used to prove closure of the space of polylogarithms under taking primitives. 

\sm

The connection $\cJ_\text{DHS}$ of this work and its meromorphic counterpart in Enriquez' work call for extensions to include variations of the complex-structure moduli of the surface. The quest for such flat connections on 
the Teichm\"uller space ${\cal T}_{h,n}$ of compact genus-$h$ Riemann surfaces with $n$ marked points was 
initiated in \cite{DHoker:2026mxk}: The reference provides a flatness-preserving extension of $\cJ_\text{DHS}$ by $(1,0)$-form components in holomorphic moduli directions (using adapted coordinates and working locally in ${\cal T}_{h,n}$)  while leaving the inclusion of the anti-holomorphic moduli components as well as the analogous extension of the Enriquez connection for future investigations.

\subsection*{Organization}

The remainder of this paper is organized as follows. 
The single-variable DHS and Enriquez connections on a Riemann surface of arbitrary genus $h$ with or without punctures and their interrelation by a composition of a gauge transformation and a Lie algebra automorphism are reviewed in section \ref{sec1var}. The structure of the multi-variable Enriquez connection and the Lie algebra in which it  takes values are summarized  in section \ref{secnvar}. The new single-valued, modular invariant and flat DHS connection in multiple variables is introduced in section \ref{sec:33}. In section \ref{sec:99}, we prove the relation between the multi-variable Enriquez and DHS connections by a composition of a gauge transformation and a Lie algebra automorphism.

\sm

The main theorem and their proofs can be found as follows:
\begin{itemize}
\item The properties of the multi-variable Enriquez connection are summarized in Theorem~\ref{7.thm:10}, in the form of  a useful variant of Enriquez's theorem \cite{Enriquez:2011}, which is proven by alternative methods that are more direct than those used in \cite{Enriquez:2011} (the lengthy proof of an auxiliary lemma is relegated to appendix~\ref{sec:55}).
\item Flatness of the multi-variable DHS connection on configuration space is proven in 
Theorem \ref{2.thm:1} (see appendices \ref{sec:A} and \ref{sec:B} for the proofs of auxiliary Lemmas \ref{2.lem:3} and \ref{2.lem:4}, respectively) and its modular invariance is proven in Theorem \ref{thm:mod}.
\item The link between the multi-variable Enriquez and DHS connections by a composition of a gauge transformation and a Lie algebra automorphism is established in Corollary~\ref{exuniq} on the basis of Theorems~\ref{5.thm:1} and \ref{5.thm:35}  (see appendices \ref{sec:C1} for the proof of the automorphism properties as well as appendices \ref{sec:C} and  \ref{sec:D} for the proofs of auxiliary Lemmas \ref{3.lem:9} and \ref{7.lem:666}, respectively).
\end{itemize}


\subsection*{Acknowledgments}

We are deeply indebted to Benjamin Enriquez and Federico Zerbini for providing the original impetus for this  work and for highly valuable suggestions on a first draft version of this paper. Moreover, we thank Yoann Sohnle for valuable discussions on flat connections and their various applications. Furthermore, we are grateful to two anonymous referees for numerous insightful comments and suggestions for improving the clarity of the original manuscript. ED is grateful to the Center for Geometry and Physics at Uppsala University for the warm hospitality extended to him during the early stages of this work. The research of ED was supported in part by NSF grant PHY-22-09700.  The research of OS is funded by the European Union under ERC Synergy Grant MaScAmp 101167287. 

\sm

Views and opinions expressed are however those of the author(s) only and do not necessarily reflect those of the European Union or the European Research Council. Neither the European Union nor the granting authority can be held responsible for them. 

\newpage

\section{Review of single-variable connections}
\label{sec1var}

We begin with a brief summary of the properties of the DHS kernels in terms of the Arakelov Green function and Abelian differentials. The DHS and Enriquez connections are reviewed for the single-variable case, along with the freely generated Lie algebra in which the connections are valued. We also review the relation between the single-variable DHS and Enriquez connections through a gauge transformation and a Lie-algebra automorphism \cite{DHoker:2024desz}. Throughout, any punctures, if present, and the complex structure moduli of the underlying compact Riemann surface will be kept fixed.

\subsection{Basics}
\label{sec:twoone}

Throughout this paper, we consider a compact Riemann surface $\Sigma$ of arbitrary genus $h \geq 1$ whose homology group $H_1(\Sigma, \ZZ)$ is equipped with a non-degenerate intersection pairing $\mJ$. A canonical homology basis is spanned by cycles $\mA^I$ and $\mB_I$ whose intersection matrix is given by $\mJ(\mA^I, \mA^J) = \mJ(\mB_I, \mB_J) =0$ and $\mJ(\mA^I, \mB_J) = \delta ^I_J$ for $I,J=1,\cdots, h$. A canonical dual basis of $h$ holomorphic Abelian differentials $\om_I$ for  the Dolbeault cohomology group $H^{(1,0)} (\Sigma, \ZZ)$ may be chosen by normalizing their $\mA$-periods while their $\mB$-periods produce the components $\Omega_{IJ}$ of the period matrix  $\Omega$, 
\bea
\label{2.a.1}
\oint _{\mA^I} \om_J = \delta ^I_J \hskip 1in \oint _{\mB_I} \om_J = \Omega _{IJ}
\eea 
The Riemann relations ensure that the period matrix is symmetric $\Omega ^t=\Omega$  and its imaginary part $Y = \Im \Omega$ is positive definite. Indices are lowered and raised with the help of $Y$ and its inverse $Y^{-1}$, whose components are denoted by  $Y_{IJ}$ and $Y^{IJ}$, respectively.  For example, we~have the following relations,\footnote{\label{convfn} Henceforth, we adopt the Einstein summation convention for pairs of repeated upper and lower indices $I,J= 1, \cdots, h $, unless explicitly stated otherwise. Throughout, we shall make use of a system of local complex coordinates $x, \bar x$ on $\Sigma$ when convenient, denote by $d^2 x = { i \over 2} dx \wedge d \bar x$ the coordinate volume form and normalize the Dirac $\delta$ function on $\Sigma$ as follows $\int _\Sigma d^2 x \, \delta (x,y)  = 1$.} 
\bea
\label{2.a.2}
\om^I = Y^{IJ} \om_J \hskip 1in \om_I = Y_{IJ} \, \om^J
\hskip 1in Y_{IJ} Y^{JK} = \delta_I^K
\eea
In local complex coordinates $x, \bar x$ on $\Sigma$, we shall often use component functions, such as $\om_I = \om_I(x) \, dx$ and $\bar \om ^I = \bar \om^I(x) \, d\bar x$. The normalized canonical volume form $\kappa$  is given by,
\bea
\label{2.a.3}
\kappa = {i \over 2h} \om_I \wedge \bar \om^I = \kappa(x) d^2 x \hskip 1in
\int _\Sigma \kappa=1
\eea
The Arakelov Green function $\cG(x,y)$ is single-valued in $x,y \in \Sigma$, symmetric in the points $\cG(x,y) = \cG(y,x)$, and satisfies the following equations \cite{Arakelov, Faltings, Alvarez-Gaume:1986nqf}, 
\bea
\label{2.a.4}
\pbx \p_x \cG(x,y)  & = &  - \pi \delta (x,y) + \pi \kappa (x) 
\hskip 1in
\int_\Sigma d^2 z \, \kappa (z) \, \cG(z,y)  =0
\no \\
\pby \p_x \cG(x,y)  & = &   \pi \delta (x,y) - \pi \bar \om^K(y) \om_K(x)
\eea
of which the two equations on the first line suffice to define $\cG(x,y)$ uniquely.  Moreover, $\cG(x,y)$ is invariant under conformal and modular transformations, and may be expressed in terms of the prime form (see for example  section~3.2 of~\cite{DHoker:2017pvk}). The dependence of $\omega_I(x)$ and $\cG(x,y)$ on the moduli of the Riemann surface $\Sigma$ will be suppressed throughout.

\subsection{Review of the single-variable DHS connection}
\label{sec:2.2}

Flat DHS connections in a single variable on $\Sigma$ in the presence of an arbitrary number of punctures were constructed in \cite{DHoker:2023vax}. We begin by considering the case of a Riemann surface~$\Sigma$ in the presence of a single puncture for which the DHS connection takes values in the degree completion $\hat \mg$ of the Lie algebra $\mg$ that is freely generated by $a^I$ and $b_I$ for $I=1,\cdots , h$, all of which generators are assigned  degree~1.\footnote{The degree completion $\hat{\mathfrak{l}}$ of an infinite-dimensional graded Lie algebra $\mathfrak{l}$ extends the algebra by including infinite Lie series, which are needed here to express the various connections. The Lie algebra $\mg$ is also referred to as $\mt _{h,1,1}$ following the notation of  \cite{Enriquez:2011}.
More generally, $\mt_{h,n,p}$ and $\hat \mt_{h,n,p}$ will denote the Lie algebra on a surface of genus $h$ in $n$ variables and $p$ punctures and its degree completion, respectively (see  section \ref{sec:2.3} for the complete definition of $\mt_{h,n}$).
 A general reference on free Lie algebras is \cite{Reutenauer}. \label{foot2}}

\subsubsection{DHS kernels}

We start by reviewing the single-valued but non-meromorphic DHS kernels which are defined by setting $f ^\emptyset {}_J(x,y) = \om_J(x)$ and defining the higher rank functions by,
\bea
\label{2.a.5}
f^I{}_J(x,y) & = & \int _\Sigma d^2 z \, \p_x \cG(x,z) \Big ( \bar \om ^I(z) \om_J(z) - \delta (z,y) \, \delta ^I_J \Big )
\no \\
f^{I_1 \cdots I_r}{}_J (x,y) & = & \int _\Sigma d^2 z \, \p_x \cG(x,z) \, \bar \om^{I_1} (z) \, f^{I_2 \cdots I_r} {}_J(z,y)
\hskip 0.7in r \geq 2
\eea
They satisfy the following differential equations for $r\geq 1$,
\bea
\label{2.a.6}
\pbx f^{I_1 \cdots I_r}{}_J(x,y) & = & - \pi \, \bar \om^{I_1} (x) \, f^{I_2 \cdots I_r}{}_J(x,y) 
+ \pi \delta_{r,1} \, \delta ^{I_1}_J \, \delta(x,y)
\no \\
\pby f^{I_1 \cdots I_r}{}_J(x,y)  & = & \pi  f^{I_1 \cdots I_{r-1}} {}_K(x,y) \, \bar \om^K(y) \, \delta ^{I_r}_J 
- \pi \delta_{r,1} \, \delta ^{I_1}_J \, \delta(x,y)
\eea
For $r \geq 1$, we may decompose the DHS kernels into their trace and traceless parts, 
\bea
\label{2.a.7}
f^{I_1 \cdots I_r}{}_J(x,y) = \p_x \Phi ^{I_1 \cdots I_r}{}_J(x) - \p_x \cG^{I_1 \cdots I_{r-1} }(x,y) \, \delta ^{I_r }_J 
\eea
where the components $\Phi^{I_1 \cdots I_r}{}_J(x)$ and $\cG^{I_1 \cdots I_r}(x,y)$ satisfy $\cG^\emptyset(x,y) = \cG(x,y)$ and,
\bea
\label{2.a.8}
\Phi ^{I_1 \cdots I_{r-1}J }{}_J (x)=0 \hskip 1in \cG^{I_1 \cdots I_r} (x,y) = (-)^r \cG^{I_r \cdots I_1}(y,x)
\eea
Integral recursion relations for $\Phi ^{I_1 \cdots I_r}{}_J(x)$ and $\cG^{I_1 \cdots I_{r-1} }(x,y)$ may be derived from (\ref{2.a.5}) and were given, for example, in \cite{DHoker:2023vax} and \cite{DHoker:2024ozn}. These functions also satisfy the following differential equations that may be derived from (\ref{2.a.6}) for $r \geq 1$ (defining $ \p_x \Phi ^{\emptyset}{}_J(x) = \omega_J(x)$), 
\bea
\label{2.a.9}
\pbx \p_x \Phi ^{I_1 \cdots I_r }{}_J (x) & = & - \pi \bar \om^{I_1} (x) \p_x \Phi ^{I_2 \cdots I_r}{}_J(x) 
+ \pi \delta _{r,1} \, \delta ^{I_1}_J \, \kappa(x)
\no \\
\pbx \p_x \cG^{I_1 \cdots I_r} (x,y) & = & - \pi \bar \om^{I_1} (x) \cG^{I_2 \cdots I_r} (x,y) 
\no \\
\pby \p_x \cG^{I_1 \cdots I_r} (x,y) & = & - \pi \bar \om^K(y) f^{I_1 \cdots I_r} {}_K (x,y)
\eea
As will be detailed in section \ref{sec:4.mod}, the functions $\Phi ^{I_1 \cdots I_r}{}_J(x)$, $\cG^{I_1 \cdots I_r}(x,y)$ and the one-forms $f^{I_1 \cdots I_r}{}_J(x,y)$ transform as tensors under various non-linear representations of the modular group $Sp(2h,\ZZ)$ and under associated linear representations of $GL(h,\ZZ) \subset GL(h, \CC)$ \cite{DHoker:2023vax}.

\subsubsection{The DHS connections in terms of generating functions}

To formulate the DHS connections it will be convenient to use generating functions that succinctly encode all DHS kernels at once, and are defined by,\footnote{The generating series $\bG(x,y;B)$ and $\bJ_K(x,y;B)$ match the generating series $\cH(x,y;B)$ and $\Psi_K(x,y;B)$ introduced earlier  in section 3.4 of \cite{DHoker:2023vax}, respectively.}
\bea
\label{2.gen.1}
\bPhi _K(x;B) & = & \om_K(x) + \sum_{r=1}^ \infty \p_x \Phi ^{I_1 \cdots I_r} {}_K(x) B_{I_1} \cdots B_{I_r}
\no \\
\bG(x,y;B) & = &  \sum _{r=0}^ \infty \p_x \cG^{I_1 \cdots I_r} (x,y) B_{I_1} \cdots B_{I_r} 
\no \\
\bJ_K(x,y;B) & = &  \sum _{r=0} ^\infty f^{I_1 \cdots I_r}{}_K(x,y) B_{I_1} \cdots B_{I_r}
\eea
where $B_I X = [b_I, X]$ for any element $X \in \hat \mg$, while  $f^\emptyset {}_K(x,y) = \om_K(x)$ and $\cG^\emptyset (x,y) = \cG(x,y)$. The interrelation of the DHS kernels (\ref{2.a.7}) translates into the following relation between the corresponding generating functions,
\bea
\label{2.gen.2}
\bJ_K(x,y;B) = \bPhi_K(x;B) -  \bG(x,y;B) B_K
\eea
while the differential relations of (\ref{2.a.6}) and (\ref{2.a.9}) in the variable $x$ translate into, 
\bea
\label{2.gen.3}
\pbx \bPhi _K(x;B) & = & \pi \kappa(x) B_K - \pi \bar \om^I(x) B_I \, \bPhi_K (x;B)
\no \\
\pbx  \bG (x,y;B) & = & \pi \kappa (x) - \pi \delta(x,y) - \pi \bar \om ^I(x) B_I \, \bG(x,y;B)
\no \\
\pbx \bJ_K(x,y;B) & = & \pi \delta (x,y) B_K - \pi \bar \om^I(x) B_I \, \bJ_K (x,y;B) 
\eea
and those in the variable $y$ become,
\bea
\label{2.gen.4}
\pby  \bG(x,y;B) & = & \pi \delta (x,y) - \pi \bar \om^K(y) \, \bJ_K(x,y;B)  
\no \\
\pby \bJ_K (x,y;B) & = & - \pi \delta (x,y) B_K + \pi \bar \om^I(y) \, \bJ_I (x,y; B) B_K
\eea
In terms of these generating functions, we have the following result, proven in \cite{DHoker:2023vax}.

\sm

The single-valued and modular invariant connection $\cJ_\text{DHS}$ on $\Sigma$ with a single puncture $y$ taking values in $\hat \mg= \hat \mt_{h,1,1}$ and given by,
\bea
\cJ_\text{DHS} (x,y;a,b) = - \pi \bar \om^I (x) b_I d\bar x + \bJ_K(x,y;B) a^K dx
\label{defjdhs}
\eea
satisfies the following differential relation for all $x \in \Sigma$,
\bea
d _x \cJ_\text{DHS} (x,y;a,b) - \cJ_\text{DHS} (x,y;a,b) \wedge \cJ_\text{DHS} (x,y;a,b) 
= \pi d \bar x \wedge dx \, \delta(x,y) [b_I, a^I]
\eea
and is flat away from $x=y$. Restricting $\cJ_\text{DHS}$ to take values in the algebra $\hat \mt_{h,1}$ which is the quotient of $\hat \mt_{h,1,1}$ by the relation $[b_I, a^I]=0$, all dependence on the puncture $y$ cancels out and the connection satisfies $d _x \cJ_\text{DHS}  - \cJ_\text{DHS}  \wedge \cJ_\text{DHS} =0$ for all $x \in \Sigma$.

\subsubsection{Adding punctures}

Additional punctures $y_1, \cdots ,y_p$ on $\Sigma$ may be included by extending the Lie algebra to the one generated freely by $a^I, b_I$ and $p$ additional generators~$c_\a$, namely one for each puncture~$y_\a$ besides $y$, with $\a =1,\cdots ,p$.  The corresponding connection was obtained in \cite{DHoker:2023vax} and is given~by,
\bea
\label{2.gen.5}
\cJ_\text{DHS}(x,\yy;a,b,c) & = & 
- \pi \bar \om^I (x) b_I d\bar x + \bJ_K(x,y;B) a^K dx
\no \\ &&
- \sum _{\a=1}^p \Big ( \bG(x,y_\a;B) dx  - \bG(x,y;B) dx \Big ) c_\a
\eea
The connection $\cJ_\text{DHS} = \cJ_\text{DHS} (x,\yy;a,b,c)$ satisfies the differential relation,
\bea
\label{2.gen.6}
d _x \cJ_\text{DHS}  - \cJ_\text{DHS}  \wedge \cJ_\text{DHS}  
= \pi d \bar x \wedge dx \bigg \{  \delta(x,y) \Big ( [b_I, a^I] - \sum_{\a =1}^p c_\a \Big ) 
+ \sum _{\a=1}^p \delta (x,y_\a) c_\a \bigg \}
\eea
and involves $p{+}1$ punctures, $\yy=(y, y_1, \cdots, y_p)$. The puncture $y$ in the first entry of $\yy$ appears in both lines of (\ref{2.gen.5}) and therefore enters on special footing -- only the additional punctures $y_\alpha$ in the
second line of (\ref{2.gen.5}) are associated with separate generators $c_\alpha$ with $\a =1,\cdots ,p$.
The connection may be restricted to taking values in the  Lie algebra freely generated by $a^I, b_I, c_\a$ quotiented by the relation,
\bea
\label{2.gen.7}
[b_I, a^I] = \sum_{\a =1}^p c_\a
\eea
All dependence on the puncture $y$ then cancels out of $\cJ_\text{DHS} (x,\yy;a,b,c)$, and the connection involves only $p$ punctures, $y_1, \cdots, y_p$  that enter on an equal footing. This Lie algebra will be referred to as the degree completion $\hat \mt_{h,1,p}$ of the Lie algebra generated by $a^I, b_I, c_\a$ subject to the relation (\ref{2.gen.7}).

\subsection{The Enriquez connection in terms of generating functions}
\label{sec:9.1}
 
The Enriquez connection $\cK_\text{E}$ \cite{Enriquez:2011} in the single-variable case with one puncture $y$ takes values in the Lie algebra $\hat \mg= \hat \mt_{h,1,1}$ which is freely generated by the elements $a^I, b_I$ with $I=1, \cdots , h$. The connection $\cK_\text{E}$  is given~by,
\bea
\cK_\text{E} (x,y;a,b) & = & \bK_J (x,y;B) a^J
\label{keviaja}
\eea 
with $B_I X = [b_I, X]$ for all $X$ in the Lie algebra and points $x,y$ in the {\it universal cover} $\tilde\Sigma$ of a compact Riemann surface $\Sigma$ of arbitrary genus $h$. The generating function $\bK_J(x,y;B)$ is given in terms of the Enriquez kernels $g^{I_1 \cdots I_r}{}_J(x,y)$ by its expansion in powers of $B$,\footnote{Choosing the homology cycles $\mA^I$ and $\mB_I$ of $H_1(\Sigma, \ZZ)$  so as to share a common base point $p$ in $\Sigma$, we promote $\mA^I$ and $\mB_I$ to generators of the fundamental group $\pi_1(\Sigma, p)$, whose action on the variable $x$ will be denoted by $\mA^I \cdot x$ and $\mB_I \cdot x$, respectively. In terms of the canonical projection $ \pi : \tilde \Sigma \to \Sigma$, the Riemann surface $\Sigma$ may be presented in $\tilde \Sigma$ in terms of a simply connected fundamental domain $D_p$, (which may be obtained by cutting the surface $\Sigma$ along the homotopy generators $\mA^I$ and $\mB_I$ defined above) by identifying the boundary components  under the action of $\pi$. There exists a \textit{preferred fundamental domain}, which we shall continue to denote by $D_p$,  in which $g^{I_1 \cdots I_r}{}_J(x,y)$ with $r\geq 2$ are holomorphic in $x,y \in D_p$.  \label{foot1}}  
\bea
\bK_J(x,y;B) & = & \sum _{r=0}^\infty g^{I_1 \cdots I_r}{}_J(x,y) B_{I_1} \cdots B_{I_r}
\label{gfsKJ}
\eea   
One may uniquely define the Enriquez connection (\ref{keviaja}) and its kernels $g^{I_1 \cdots I_r}{}_J(x,y)$ through their meromorphicity, poles and monodromies \cite{Enriquez:2011, Enriquez:2021}, see (\ref{3.Kder}) to (\ref{3.Kmon2}) below. These defining conditions are met by the explicit expressions for 
$g^{I_1 \cdots I_r}{}_J(x,y)$ in \cite{DHoker:2025dhv} as iterated $\mA$-cycle convolutions with measure $\oint_{\mA^{I_j}} dz$ which closely resemble the DHS kernels (\ref{2.a.5}) given as iterated convolutions
with measure $\int_{\Sigma} d^2 z\, \bar \omega^{I_j}(z)$.\footnote{Alternatively, Enriquez kernels can be expressed in terms of DHS kernels and their polylogarithms \cite{DHoker:2024desz} or, when $\Sigma$ is a hyperelliptic surface, in terms of Poincar\'e series \cite{Baune:2024biq}.}

\sm

Similar to the DHS kernels $f^{I_1 \cdots I_r}{}_J(x,y) $, Enriquez kernels $g^{I_1 \cdots I_r}{}_J(x,y)$ may be decomposed into traces $ \chi ^{I_1 \cdots I_{r-1}}  (x,y) $ and traceless parts $\varpi^{I_1 \cdots I_r}{}_J(x) $ with respect to the last two indices, 
\bea
g^{I_1 \cdots I_r}{}_J(x,y) = \varpi^{I_1 \cdots I_r}{}_J(x) -  \chi ^{I_1 \cdots I_{r-1}}  (x,y) \, \delta_J^{I_r}
\label{partsg}
\eea
where we set $g^{\emptyset}{}_J(x,y) = \varpi ^\emptyset {}_J(x) = \om_J(x)$ and $\chi^\emptyset(x,y) = \chi(x,y)$. Note that $\varpi^{I_1 \cdots I_r}{}_J(x) $ is independent on $y$ and obeys $\varpi^{I_1 \cdots I_{r-1}J }{}_J(x) =0$
\cite{Enriquez:2011, DHoker:2024ozn}. 

\sm

Generating functions that capture each component $ \chi ^{I_1 \cdots I_{r-1}}  (x,y)$ and $\varpi ^{I_1 \cdots I_r} {}_J (x)$ of the Enriquez kernels in (\ref{partsg}) separately are given by,
\begin{align}
\label{gfsKW} 
\XX (x,y;B) & =  \sum _{r=0}^\infty \chi ^{I_1 \cdots I_r}  (x,y) B_{I_1} \cdots B_{I_r}
\no \\ 
\OO _J(x;B) & =  \sum _{r=0}^\infty \varpi ^{I_1 \cdots I_r} {}_J (x) B_{I_1} \cdots B_{I_r}
\end{align}
in terms of which $\KK_J(x,y;B)$ decomposes as follows,
\bea
\label{3.KXW}
\KK_J(x,y;B) & =  \OO_J(x;B) - \XX(x,y;B) B_J
\eea
In the \textit{preferred fundamental domain} $D_p$ for $\Sigma$ the kernels $g^{I_1 \cdots I_r}{}_J(x,y)$ for $r=0$ and $r \geq 2$ with $x,y \in D_p$  are holomorphic (see footnote \ref{foot1}), while $g^I{}_J(x,y)$ is meromorphic, having  a simple pole in $x \in D_p$ at $y \in D_p $ with residue $ \delta ^I_J$. Correspondingly, the generating function $\OO(x;B)$ is holomorphic in $x \in D_p$ while $\XX(x,y;B)$ and $\KK_J(x,y;B)$ are meromorphic in $x \in D_p$ with residues $-1$ and $B_J$, respectively. It will be useful to express these singularities in $x, y \in D_p$ and their residues equivalently via the following distributional differential equations, which parallel the corresponding relation (\ref{2.gen.3}) for DHS kernels, 
\begin{align}
\label{3.Kder}
\pbx \KK_J(x,y;B) & =  \pi \delta (x,y) B_J & \pbx \XX(x,y;B) & =  - \pi \delta (x,y) & \pbx \OO_J(x;B) & = 0
\qquad \no \\
\pby \KK_J(x,y;B) & = - \pi \delta (x,y) B_J & \pby \XX(x,y;B) & =   \pi \delta (x,y) 
\end{align}
Beyond the preferred fundamental domain $D_p$, the triviality of the $\mA$ monodromy along with the non-trivial $\mB$ monodromy relations (see footnote \ref{foot1} for the notations used below),\footnote{Throughout, the functions of $B_L$ should be understood as formal Taylor series in $B_L$ so that we have, in the present case,  $(e^{ 2 \pi i B_L} -1)/B_L= \sum _{k=0} ^\infty  ( 2 \pi i )^{k+1} { B_L^k \over (k+1)!}$.}
\bea
\label{3.Kmon1}
\KK_J(\mB_L \cdot x, y;B) & = & e^{- 2 \pi i B_L} \, \KK_J (x,y;B)
\no \\
\KK_J(x, \mB_L \cdot y;B) & = & \KK_J (x,y;B) + \KK_L (x,y;B) \, { e^{ 2 \pi i B_L} -1 \over B_L} \, B_J
\eea
may be used to extend the differential equations of (\ref{3.Kder}) to the universal covering space  $\tilde \Sigma$ for the canonical projection $ \pi : \tilde \Sigma \to \Sigma$. In particular, the $\mB$ monodromy of $\cK_\text{E}$ is given by,\footnote{Note that the repeated index $L$ on the right sides of (\ref{3.Kmon1}) and (\ref{3.Kmon2})
is not summed over. Here and in the remainder of this work, the convention of footnote \ref{convfn} 
to sum over repeated indices only applies to pairs of upper and lower indices and not to repeated lower indices. \label{foot7}}
\bea
\label{3.Kmon2}
\cK_\text{E} ( \mB_L \cdot x, y;a,b) = e^{-2 \pi i b_L} \, \cK_\text{E} ( x, y;a,b) \, e^{2 \pi i b_L}
\eea
Finally, the $\mA$-cycle periods of $\KK_J$ are given by, 
\bea
\label{3.Kper}
\oint _{\mA^L} dz \, \KK_J(z,y;B) =  \sfZ(B_L)  \, \delta ^L_J
\eea
where the function $\sfZ(B)$ is defined by,
\bea 
\label{3.Kper1}
\sfZ(B) = { - 2 \pi i B \over e^{-2 \pi i B} -1}  = \sum_{r=0}^\infty \frac{{\rm Ber}_r}{r!}  \, ({-}2 \pi i B)^k 
\eea
in terms of the Bernoulli numbers $\Ber_r$ (denoted so in order to avoid confusion with the Lie algebra generators $B$). The monodromy relations and $\mA$-cycle integrals for the generating functions $\XX(x,y;B)$ and $\OO_J(x;B)$ may be readily deduced from (\ref{3.Kmon1}) and (\ref{3.Kper}), respectively, using the relation (\ref{3.KXW}).

\subsection{Relating single-variable Enriquez and DHS connections}
\label{sec:1var}

In the single-variable case, the relation between the DHS connection $\cJ_\text{DHS}$ and the Enriquez connection $\cK_\text{E}$ was established in terms of the composition of a gauge transformation and a Lie algebra automorphism \cite{DHoker:2024desz}.
Two different, but equivalent, constructions of the gauge transformation were presented in the reference, one in terms of the DHS connection, and the other in terms of the Enriquez connection. Here, we shall review only the former.

\sm 

For both connections we shall denote the variable by $x$ and puncture by $y$. The equivalence relation between the two connections may then be expressed as follows, 
\begin{align}
 {\cal K}_{\rm E}(x,y;a,b) &= {\cal U}_{\rm DHS}^{-1}(x,y) \Big(
    {\cal J}_{\rm DHS}(x,y;\hat a, \hat b) \, {\cal U}_{\rm DHS}(x,y)- d_x\, {\cal U}_{\rm DHS}(x,y) \Big)
\label{mauto.04}
\end{align}
Here, the generators $\hat a= (\hat a^1, \cdots, \hat a^h)$ and $ \hat b = (\hat b_1 , \cdots, \hat b_h)$  of  $\hat \mt_{h,1,1}$ are related to the generators $a= (a^1, \cdots ,a^h)$ and $ b = (b_1 , \cdots, b_h)$  of $\hat \mt_{h,1,1}$ by an automorphism $\phi : \hat \mt_{h,1,1} \to \hat \mt_{h,1,1} $, which is independent of the variable $x$ but generally does depend on the puncture $y$ and the moduli of the  compact Riemann surface $\Sigma$. 

\sm

The gauge transformation ${\cal U}_{\rm DHS}(x,y)$ in (\ref{mauto.04}) is constructed from the path-ordered exponential of ${\cal J}_{\rm DHS}(z,y;\xi,\eta)$ with generators $\xi^J,\eta_I$ that are Lie series in (a non-faithful homomorphism of) $ \hat \mt_{h,1,1} $, 
  \begin{align}
{\cal U}_{\rm DHS}(x,y)
&= {\rm Pexp} \bigg( \int^x_y {\cal J}_{\rm DHS}(z,y;\xi, \eta)  \bigg)
 \label{mauto.05}
 \end{align}
 Producing the monodromies (\ref{3.Kmon2}) of $\cK_\text{E}$ from the single-valued connection $\cJ_\text{DHS}$ imposes the following monodromy conditions on the gauge transformation,
\begin{align}
{\cal U}_{\rm DHS}(\mA^K \cdot x,y) &= {\cal U}_{\rm DHS}( x,y) 
 \label{mauto.06} \\
 {\cal U}_{\rm DHS}(\mB_K \cdot x,y) &= {\cal U}_{\rm DHS}( x,y) \, e^{2\pi i b_K}
 \notag
\end{align}
As detailed in section 3.1 of \cite{DHoker:2024desz}, these monodromy relations imply trivial monodromy of ${\cal U}_{\rm DHS}( x,y)$ as $x$ is taken around the boundary of the preferred fundamental domain $D_p$. Moreover, by flatness of $\cJ_\text{DHS}$ away from $x=y$, this implies trivial monodromy as $x$ is taken around a small circle enclosing $y$. As a result, the generators $\xi^J$ and $\eta_I$ obey,
\begin{align}
[ \eta_J , \xi^J ] = 0
 \label{mauto.07} 
 \end{align}
As a consequence, the integrand ${\cal J}_{\rm DHS}(z,y;\xi, \eta)$ in (\ref{mauto.05}) is actually independent of the puncture $y$, as was shown in   section 3.1 of~\cite{DHoker:2024desz}. Furthermore, the absence of a $(0,1)$ component to the connection $\cK_\text{E}$ in (\ref{mauto.04}) imposes the relation, 
\begin{align}
\eta_I = \hat b_I
 \label{mauto.08} 
 \end{align}
while the monodromy conditions of (\ref{mauto.06}) are solved order by order in the generators by,
 \bea
\xi^J & = &  \pi \hat b ^J + \sum_{r=2}^{\infty} {\cal X}^{J I_1 \cdots I_r} \, \hat b_{I_1} \cdots  \hat b _{I_r} 
\no \\ 
\hat b_I & = & b_I - \sum_{r=2}^{\infty} {\cal M}_\shuffle^{J_1 \cdots J_r}{}_I(y) \, b_{J_1} \cdots  b_{J_r} 
 \label{mauto.09} 
 \eea
As a result, the gauge transformation $ {\cal U}_{\rm DHS}(x,y) $ depends only on $b_I$ and is independent of~$a^I$. While both sets of coefficients  ${\cal X}^{J I_1 \cdots I_r}$ and ${\cal M}_\shuffle^{J_1 \cdots J_r}{}_I(y)$  depend on the moduli of the compact surface $\Sigma$, the former are actually independent of $y$ (see section 3.3.1 of \cite{DHoker:2024desz} for a derivation and a discussion of their symmetry properties, and explicit formulas to low orders can be found in the appendices of the reference).  Finally, the $(1,0)$ component of the relation (\ref{mauto.04}) may be reduced to the following form,
\begin{align}
 {\cal K}_{\rm E}(x,y;a,b) &= {\cal U}_{\rm DHS}^{-1}(x,y)  
    {\cal J}^{(1,0)}_{\rm DHS}(x,y;\hat a{-}\xi, \hat b) \, {\cal U}_{\rm DHS}(x,y)
\label{mauto.12} 
\end{align}
where ${\cal J}^{(1,0)}_{\rm DHS}(x,y;\hat a{-}\xi, \hat b) $ refers to the $(1,0)$ component
$\bJ_K(x,y;\hat B) (\hat a^K{-}\xi^K) dx$ of the DHS connection (\ref{defjdhs}) with $\hat B_I X = [\hat b_I, X]$ for any $X\in \hat \mt_{h,1,1}$. Linearity of $\cK_\text{E}(x,y;a,b)$ in $a$ and linearity of $ {\cal J}^{(1,0)}_{\rm DHS}(x,y;\hat a{-}\xi, \hat b)$ in $\hat a {-} \xi$ requires the following relation between their Lie series, 
 \begin{align}
 \hat a^K{-}\xi^K &= a^K + \sum_{r=1}^{\infty} {\cal M}^{K I_1 \cdots I_r}{}_J (y)
B_{I_1} \cdots B_{I_r} a^J 
      \label{mauto.13} 
 \end{align}
 while matching the residues of the simple pole in $x$ at $y$ on both sides requires,
 \bea
 \label{resm.01}
{} \big [ \hat b_K, \hat a^K \big ] =  \big [ b_K, a^K \big ] 
\eea
Combining (\ref{mauto.13}) and (\ref{resm.01}) with the second equation in (\ref{mauto.09}), shows that 
the  coefficients $ {\cal M}^{K I_1  \cdots  I_r}{}_J $ may be expressed in terms of the 
${\cal M}_\shuffle^{J_1 \cdots J_r}{}_I $ in (\ref{mauto.09}). For example, we have 
\bea
{\cal M}^{J_1 J_2}{}_I= {\cal M}_\shuffle^{J_1  J_2}{}_I \, , \ \ \ \ \ \ 
{\cal M}^{J_1 J_2 J_3}{}_I= {\cal M}_\shuffle^{J_1  J_2 J_3}{}_I
+ {\cal M}^{J_1 J_2}{}_K {\cal M}^{K J_3}{}_I
\label{cmvsshuff}
\eea
at $r\leq 3$, and we refer to section 3.3.3 of \cite{DHoker:2024desz} for an all-rank formula.

\newpage

\section{Enriquez's multi-variable connection}
\label{secnvar}

In preparation for presenting the single-valued flat multi-variable connection $\cJ_\text{DHS}$ in section~\ref{sec:33}, we review the meromorphic multi-variable Enriquez connection and its Lie algebra $\hat \mt_{h,n}$ where both connections take values and present an alternative proof of its flatness. In order to give an overview of the notations used for the meromorphic and single-valued connections as well as their composing generating series of kernels, we shall start this section with a list of symbols and equation pointers: \begin{itemize}
\item meromorphic but multi-valued:
\begin{itemize}
\item[$(i)$] multi-variable Enriquez connection ${\cal K}_{\rm E}$ organized into $dx_i$-components $K_i$ in (\ref{7.1.a});
\item[$(ii)$] components $K_i$ in terms of generating functions ${\bf K}_J$ and ${\bf X}$ in (\ref{7.1.b});
\item[$(iii)$] generating functions ${\bf K}_J$ and ${\bf X}$, ${\bf W}_J$ of Enriquez kernels in (\ref{gfsKJ}) and (\ref{gfsKW}), respectively;
\end{itemize}
\item single-valued but non-meromorphic:
\begin{itemize}
\item[$(i)$] multi-variable DHS connection ${\cal J}_{\rm DHS}$ organized into $dx_i$- and $d\bar x_i$-components $J_i^{(1,0)}$ and $J_i^{(0,1)}$ in (\ref{defcJ.1});
\item[$(ii)$] components $J_i^{(1,0)}$ in terms of generating functions ${\bf J}_K$ and ${\bf G}$ in (\ref{defcJ.3});
\item[$(iii)$] generating functions ${\bf J}_K$ and ${\bf G}$, $\bPhi _K$ of DHS kernels in (\ref{2.gen.1});
\end{itemize}
\end{itemize}
Apart from  $J_i^{(0,1)}$, the quantities with the same item numbers in the first and second 
bullet point may be informally viewed
as meromorphic and single-valued counterparts of one another. The reader is reminded that, following 
footnote \ref{foot7}, repeated indices $I,J,K,\cdots$ are only summed over $\{1,\cdots,h\}$ if they form 
a pair of upper and lower indices but not for repeated lower indices.

\subsection{The Lie algebra for the multi-variable connections}
\label{sec:2.3}

The Lie algebra in which the Enriquez connection for $n$ variables on a compact Riemann surface of genus $h$ without punctures  takes values was constructed in \cite{Enriquez:2011} and denoted $\hat \mt_{h,n}$. We shall show in section \ref{sec:33} that a single-valued and modular invariant multi-variable DHS connection may be constructed taking values in the same Lie algebra $\hat \mt_{h,n}$. We shall now discuss the definition and basic properties of $\hat \mt_{h,n}$. 

\sm

In passing from the single-variable to the multiple-variable case, the Lie algebra $ \hat \mg$ is generalized by associating a pair of generators $a^I_i, b_{iI}$  to each one of the variables $x_i$ for $i=1,\cdots , n$. Contrarily to the single-variable case, the resulting Lie algebra is no longer freely generated. Instead, its generators include the above mentioned $a_i^I$ and $ b_{iI}$ generators, as well as additional generators $t_{ij} = t_{ji}$, for $i,j = 1, \cdots, n$ with $j \not= i$, subject to several non-trivial structure relations.

{\deff
\label{2.def:1}
The Lie algebra $\mt_{h,n}$ may be defined in terms of a triplet of generators $(a,b,t)$ with components $a_i^I, b_{jJ}$ and $t_{ij}$ for $i,j=1,\cdots, n$ and $I,J=1,\cdots, h$ that satisfy the following structure relations \cite{Enriquez:2011},\footnote{For $h \geq 2$, the last relation (\ref{11.3}) follows from (\ref{11.1}) and (\ref{11.1a}). Indeed,  for any value of $I$  we can choose a $K \not= I$ so that we have  
$[ a^I_i, t_{jk} ] \delta ^J_K =  [ b_{kK}, [ a^I_i, a^J_j    ]  ] +  [ a^J_j, [b_{kK}, a^I_i ]  ]$ with the help of the Jacobi identity. 
Both terms on the right vanish by (\ref{11.1}) and (\ref{11.1a})  since  we have $k \not = i$ and  $K\not= I$. The argument for the vanishing of $[b_{iI}, t_{jk}]$ is analogous. For $h=1$, the relations of (\ref{11.3}) are independent of (\ref{11.1}), (\ref{11.1a}) and (\ref{11.2})  and must be imposed as part of the structure relations of $\mt_{1,n}$.}
\begin{align}
& \big [ a^I_i, a^J_j \big ]  =  \big [ b_{iI} , b_{jJ} \big ] =  0 && i \not = j 
\label{11.1}
\\
&\big [b_{iI } , a^J _j \big ]  =  \delta ^J _I \, t_{ij} && i \not = j 
\label{11.1a}
 \\
& \big [ b_{iI} , a^I _i \big ] + \sum_{j \not = i} t_{ij}  =0
\label{11.2}
  \\
& \big [ a^I_i, t_{jk} \big ]  = \big [b_{iI}, t_{jk} \big] = 0 && i,j,k \hbox{ mutually distinct}
\label{11.3}
\end{align}
The algebra $\mt_{h,n}$ admits a positive bi-grading by assigning the degrees $|a_i^I|=(1,0)$, $|b_{iI}|=(0,1)$ and $| t_{ij}|=(1,1)$. The algebra $\hat \mt_{h,n}$ is the degree completion of $\mt_{h,n}$ (see footnote \ref{foot2}).}

\sm

Immediate consequences of Definition \ref{2.def:1}, that will be used throughout the sequel, are
provided by the following proposition, which was proven in \cite{Enriquez:2011}.
{\prop 
\label{2.lem:1}
The relations below follow directly  from Definition \ref{2.def:1} \cite{Enriquez:2011}, 
\begin{align}
& t_{ji} = t_{ij} && i \not= j 
\label{11.lem.0}
\\
& \big [ a^I_i + a^I_j, t_{ij} \big ] =   \big [ b_{iI} + b_{jI} , t_{ij} \big ] = 0 && i \not= j 
\label{11.lem.1}
\\
& B_{i I_1} \cdots B_{i I_r } \, t_{ij} = (-)^r B_{j I_r  } \cdots B_{j I_1} \, t_{ij} && i \not= j 
\label{11.lem.2}
\\
& \big [ t_{ij} + t_{ik} , t_{jk} \big ] = 0 &&  i,j,k \hbox{  mutually distinct }
\label{11.lem.3}
\\
& \big [ t_{ij}, t_{k\ell} \big ] =0 && i,j,k, \ell \hbox{  mutually distinct }
\label{11.lem.4}
\end{align}
where $B_{iI} X = [b_{iI}, X]$ for arbitrary $X \in \hat \mt_{h,n}$ in (\ref{11.lem.2}).}

\subsection{The multi-variable Enriquez connection}
\label{sec:mEcon}

In a seminal paper \cite{Enriquez:2011}, Enriquez proved the existence and uniqueness  of a meromorphic flat connection $\cK_\text{E}$ with values in the Lie algebra $\hat \mt_{h,n}$ on the configuration space $\text{Cf}_n(\tilde \Sigma)  = \tilde \Sigma ^n \setminus \{ \hbox{diagonals} \}$ of $n$ points $\xx= (x_1, \cdots, x_n) $ in the universal cover $\tilde \Sigma$ of the compact Riemann surface $\Sigma$ (without punctures), subject to certain monodromy and residue conditions.  The resulting Enriquez connection $\cK_\text{E} (\xx;a,b,t)$ is a sum of $(1,0)$ forms in each variable, 
\begin{align}
\label{7.1.a}
\cK_\text{E} (\xx;a,b,t) &= \sum_{i=1}^n K_i(\xx;a,b,t) \, dx_i 
 \end{align}
The components $K_i$  may be expressed in terms of the generating functions that were used to decompose the single-variable Enriquez connection in section \ref{sec:9.1} and a triplet  $(a,b,t)$ subject to the structure relations of $\mathfrak{t}_{h,n}$ in Definition \ref{2.def:1}. Two equivalent representations for $K_i$ in terms of generating functions are given as follows,
\begin{align}
\label{7.1.b}
K_i(\xx;a,b,t)  &=   {\bf W}_J(x_i; B_i) a_i^J + \sum_{j \neq i} {\bf X}(x_i,x_j;B_i)  t_{ij} 
\no \\
K_i(\xx;a,b,t)  &= 
 {\bf K}_J(x_i,z; B_i) a_i^J + \sum_{j \neq i} \big \{  {\bf X}(x_i,x_j;B_i) - {\bf X}(x_i,z;B_i)  \big \}  t_{ij}  
 \end{align}
 where $B_i X = [b_i,X]$ for arbitrary $X \in \hat \mt_{h,n}$. Their equivalence follows from expressing $\bK_J$ on the second line in terms of $\OO_J$ and $\XX$ with the help of (\ref{3.KXW}) and then using the Lie algebra relation (\ref{11.2}) to cancel the $z$-dependent terms against one another.

\sm 
 
 An alternative, but equivalent,  formulation of Enriquez's theorem of \cite{Enriquez:2011} is  as follows.  
 {\thm
 \label{7.thm:10}
 The connection $\cK_{\rm E}$ of (\ref{7.1.a}) has the following properties,
 \begin{itemize}
\itemsep=0in
\item[(1)] The $\mA$ monodromy of $\cK_{\rm E}$ is trivial, while its $\mB$ monodromy  is given by,
 \bea
 \label{3.Kmon3}
 \gamma ^{(i)} _K \, \cK_{\rm E}(\xx; a,b,t) & = & e^{- 2 \pi i b_{iK}} \cK_{\rm E}(\xx; a,b,t) \, e^{ 2 \pi i b_{iK}}
\eea
where the monodromy operator $\gamma^{(i)} _L$ around a $\mB_L$ cycle  is defined by its action on $\xx$,
\bea
\gamma^{(i)} _L \xx = (x_1, \cdots , x_{i-1}, \, \mB_L \cdot x_i, \, x_{i+1}, \cdots , x_n)
\label{defgmo}
\eea
\item[(2)] The only poles of $\cK_{\rm E}$ in the preferred fundamental domain $D_p$  are simple poles at $x_i=x_j$ with residues $(dx_j-dx_i)t_{ij}$ and  $\cK_{\rm E}$ satisfies the following differential equation,
\bea
\label{7.1.1}
d \cK_{\rm E} =  \pi \sum_{j \not= i} d\bar x_i  \wedge (dx_j - d x_i)  \, \delta(x_i,x_j) \, t_{ij}
\eea
where $d$ denotes the sum of the total differentials in all $n$ variables, $d= d_{x_1} + \cdots + d_{x_n}$.
\item[(3)] The connection $\cK_{\rm E}$ obeys the following algebraic equations,
\bea
\label{7.1.2}
 \cK_{\rm E} \wedge \cK_{\rm E} = \half \sum_{i \not= j}  dx_i \wedge dx_j \, [K_i, K_j] =0
 \qquad
\eea
and therefore  is a meromorphic  flat connection on the configuration space ${\rm Cf}_n(\tilde \Sigma)$ of $n$ points in~$\tilde \Sigma$ taking values in the Lie algebra $\hat \mt_{h,n}$. 
\end{itemize}
}
Note that the residue determined by (\ref{7.1.1}) is consistent with the symmetry $t_{ij}=t_{ji}$
of (\ref{11.lem.0}) and the antisymmetry of $dx_j-dx_i$ under $i\leftrightarrow j$.

\subsection{Two vanishing Lemmas}

The following closely related  lemmas will provide key ingredients in the proofs of a number of important results to be derived in the remainder of this paper. Lemma \ref{4.lem:100} concerns meromorphic expressions relevant to the Enriquez connection, while Lemma \ref{4.lem:300} concerns single-valued expressions relevant to the DHS connection. In both cases, we assume that $\b_1, \cdots , \b_h$  freely generate a non-Abelian Lie algebra, which is graded by the $\b$-degree and assigns the value 1 to each generator $\b_I$. 

{\lem
\label{4.lem:100}
Let $\mu = \mu(z) dz$ be a holomorphic $(1,0)$ form defined on the universal cover $\tilde \Sigma$ of a compact Riemann surface $\Sigma$ and which is expressible as a formal Taylor series in the generators $\b_{I}$ of a free Lie algebra with $I=1,\cdots, h$. If the $\mA$ monodromies of $\mu$ vanish; its $\mB$ monodromies are given by,
\bea
\label{4.lem.100}
\mu (\mB_L \cdot z) = e^{- 2 \pi i \b_L} \mu (z) \, e^{ 2 \pi i \b_L}
\eea
and its $\mA$ periods vanish for $L=1,\cdots,h$,
\bea
\label{4.lem.200}
\oint _{\mA^L} dz \, \mu (z) =0
\eea
then $\mu(z) =0$. }

\sm

We shall prove the lemma by contradiction, namely, assuming that $\mu$ satisfies the properties stated in the lemma and is non-vanishing leads to a contradiction.

\sm

Since $\mu$ is expressible as a multiple  Taylor series in $\b_I$ the $\b$-degree of each term in the series  must be positive or zero. If $\mu \not= 0$, then the Taylor series must contain a non-vanishing term $\mu_m$ of $\b$-degree $m\geq 0$, such that all terms in the Taylor series of $\mu$ with $\b$-degree strictly less than $m$ vanish. In other words, $\mu_m$ is the non-vanishing term of lowest $\b$-degree in the Taylor series of $\mu$.

\sm

The $\mB$ monodromy of $\mu_m$ may be read off from (\ref{4.lem.100}) and is found to be trivial, since any non-trivial contribution arises from the exponentials and  must have $\b$-degree strictly greater than $m$. Thus, $\mu_m$ has trivial $\mA$ and $\mB$ monodromies and therefore  is single-valued on $\Sigma$. Since $\mu$ is holomorphic by the assumptions of the lemma, $\mu_m$ is also holomorphic and  a linear combination of holomorphic differentials on~$\Sigma$, i.e.~$\mu_m(z) = \mu^L_m \omega_L(z)$ with $z$-independent  coefficients $\mu^L_m$.
In view of the $\mA$ periods (\ref{4.lem.200}), the coefficients 
$ \mu^L_m = \oint_{\mA^L} dz \, \mu_m(z)$ 
of the differential $\mu_m$ must all vanish,\footnote{We use the normalization (\ref{2.a.1}) of holomorphic differentials in expressing the expansion coefficients in $\mu_m(z) = \mu^L_m \, \omega_L(z)$ via $\mA$ periods $ \mu^L_m = \oint_{\mA^L} dz \, \mu_m(z)$.} which contradicts our initial assumption that there exists a non-vanishing term $\mu_m$ of lowest $\beta$-degree.
 Hence $\mu=0$, which proves the lemma.

\sm

{\lem
\label{4.lem:300}
Let $\nu = \nu(z) dz$ be a single-valued $(1,0)$ form on a compact Riemann surface~$\Sigma$ which is expressible as a formal Taylor series in the generators $\b_{I}$ with $I=1,\cdots, h$ of a free Lie algebra. If the anti-holomorphic derivative of $\nu$ is given by,
\bea
\label{4.lem.400}
\pbz \nu (z) = - \pi \bar \om^I (z) [ \b_I , \nu(z) ]
\eea
and its integrals over the surface against antiholomorphic differentials vanish for $I=1,\cdots,h$,
\bea
\label{4.lem.500}
\int _{\Sigma} d^2z \,\bar \om^I(z) \, \nu (z) =0
\eea
then $\nu(z) =0$. }

\sm

The proof of Lemma \ref{4.lem:300} parallels the proof of Lemma \ref{4.lem:100}, with monodromy conditions replaced by conditions on the anti-holomorphic derivative and $\mA$ periods replaced by surface integrals against $\bar \omega^I$. 
Proceeding again by contradiction, we assume that $\nu$ satisfies the conditions of Lemma \ref{4.lem:300} and that $\nu \not=0$, so that its Taylor series in the $\b_I$ must contain a non-vanishing term $\nu_n$ of lowest $\b$-degree $n\geq 0$.
 Since the right side of (\ref{4.lem.400}) has $\b$-degree at least equal to $n{+}1$, we conclude that $\nu_n$ is holomorphic and single-valued on $\Sigma$. As a consequence, there exist $z$-independent coefficients $\nu_n^J$ such that $\nu_n = \nu_n^L \omega_L(z)$, but (\ref{4.lem.500}) together with $\int_{\Sigma}\bar\om^J(z) \omega_L(z) = \delta^J_L$ enforces $ \nu_n^L=0$. We therefore find the  term $\nu_n$ of lowest $\b$-degree to vanish which contradicts our initial assumption. Hence $\nu=0$, which proves the lemma.

\subsection{Alternative proof of Theorem \ref{7.thm:10} by direct methods}
\label{sec:6}

In this subsection, we present a proof of Theorem \ref{7.thm:10} as an alternative formulation of Enriquez's theorem \cite{Enriquez:2011}. The difference between the formulations of the two theorems  is that in the present case, the form of the connection is adopted from the outset, while in \cite{Enriquez:2011} its existence  is demonstrated from its monodromy,  residue and $\mA$ period properties. The proof of Theorem \ref{7.thm:10} given below will proceed by explicit calculation while the proof of Enriquez's theorem in \cite{Enriquez:2011} relies on more formal arguments. The ``easy" parts in items \textit{(1)} and \textit{(2)} of Theorem \ref{7.thm:10} will be proven below, while the proof of item \textit{(3)} is quite involved and will be relegated to appendix \ref{sec:55}.

\subsubsection{Proof of item {\it (1)}}

Item \textit{(1)} holds for the $\mA$ monodromies of each generating function as their $\mA$ monodromy is trivial. The $\mB$ monodromy is proven by decomposing $K_i$ into generating functions (\ref{7.1.b}), using the $\mB$ monodromies of $\bK_J$ given in (\ref{3.Kmon1}) on the first term on the right side of (\ref{7.1.b})  and the $\mB$ monodromies of  $\XX$ on the second term on the right side of (\ref{7.1.b}),
\bea
\label{E.5}
\XX(\mB_L \cdot x, y;B) & = & e^{- 2 \pi i B_L} \, \XX(x,y;B) - { 1 \over h} { e^{-2 \pi i B_L} -1 \over B_L} \, \om_L(x)
\no \\
\XX(x, \mB_L \cdot  y;B) & = & \XX(x,y;B) - \bK_L(x,y;B) \, { e^{2 \pi i B_L} -1 \over B_L}
\eea
The latter may be deduced from (\ref{3.Kmon1}) by decomposing $\bK_J$ into Enriquez kernels using~(\ref{gfsKJ}), extracting the trace part of their $\mB$ monodromies using (\ref{partsg}) and noting that the term $\omega_J$ of $\bK_J$ is not incorporated into $\XX$.
Evaluating the $\mB_L$ monodromy of $K_i$  in the variable $x_i$ we use the first relation in (\ref{3.Kmon1}) for the first term and the first relation of (\ref{E.5}) for the second term to establish the statement (\ref{3.Kmon3}) of item \textit{(1)} for $j=i$.  Evaluating its $\mB_L$ monodromy in the variables $x_j$ with $j \not= i$ proceeds as follows. 
By the second line of (\ref{E.5}), we~have, 
\bea
\label{E.7}
\gamma _L ^{(j)} K_i(\xx;a,b,t)  = K_i(\xx;a,b,t) - \bK_L(x_i,x_j;B_i) \,  { e^{2 \pi i B_{iL}} -1 \over B_{iL}} \,  t_{ij}  
\eea
On the other hand, using the consequence of (\ref{11.1a}),
\bea
\label{E.9}
e^{ - 2 \pi i b_{jL}} \, a_i^J \, e^{2 \pi i b_{jL}} = a_i^J +  \delta ^J_L \, {e^{ - 2 \pi i B_{jL}}-1 \over B_{jL}} t_{ij}
\eea
and the fact that $b_{jL}$ commutes with $t_{ik}$ when $j \not = i,k$, we obtain, 
\bea
\label{E.10}
e^{- 2 \pi i B_{jL}} K_i
=
K_i + {e^{ - 2 \pi i B_{jL}}-1 \over B_{jL}} \bK_L(x_i,x_j;B_i) t_{ij}
\eea
which is seen to match the right side of (\ref{E.7}) upon using the relations $[B_{iI}, B_{jJ} ]=0$ and  $B_{iL} t_{ij} = - B_{jL} t_{ij}$. This proves item \textit{(1)} also for arbitrary $j \not= i$.

\subsubsection{Proof of item {\it (2)}}

The pole structure of item \textit{(2)} follows from the fact that $\chi(x,y)$ has a single pole in $x$ at $y$ with residue $-1$ and no other poles in the preferred fundamental domain $D_p$. The formula for $d \cK_\text{E}$ of (\ref{7.1.1}) is easily shown  by evaluating, 
\bea
\label{E.3.1}
d \cK_\text{E} =
\sum _{i,j} d\bar x_i \wedge dx_j \, \bar \p_i K_j + \half \sum _{i \not = j} dx_i \wedge dx_j \big ( \p_i K_j - \p_j K_i \big )
\eea
The second sum evaluates to, 
\bea
\label{E.3.2}
\p_i K_j - \p_j K_i = \p_i \XX(x_j,x_i;B_j) t_{ij} - \p_j \XX(x_i,x_j;B_i) t_{ij}
\eea
which vanishes in view of the expansion (\ref{gfsKW}) of $\XX$ in powers of $B$, use of formula (\ref{11.lem.2}) to convert $B_j$ into $B_i$, and of the following relation,
\bea
\label{E.3.3}
\p_i \chi^{J_1 \cdots J_r} (x_j,x_i) & = & (-)^r \p_j \chi^{J_r \cdots J_1} (x_i,x_j)
\eea
The first term on the right of (\ref{E.3.1}) is evaluated using the relations of (\ref{3.Kder}) and gives (\ref{7.1.1}),  thereby completing the proof of item \textit{(2)}.

\subsubsection{Proof of item {\it (3)}}
\label{sec:quote}

The heart of Theorem \ref{7.thm:10} lies in item \textit{(3)}, which will be proven with the help  of  Lemma \ref{4.lem:100} and the lemma below.

{\lem
\label{E.lem:1}
The commutator $\big [ K_i, K_j \big ]$ of the components $K_i$ in (\ref{7.1.b}) satisfies the following properties, 
\begin{itemize}
\itemsep=-0.05in
\item[(a)] it is a $(1,0)$ form in $x_i$ and $x_j$ and a scalar in all $x_k$ with $k \not= i,j$;
\item[(b)] its $\mA$ monodromies are trivial, while its  $\mB$ monodromies
are given as follows,
\bea
\label{E.7.p}
\gamma ^{(k)} _L \left ( \big [ K_i, K_j \big ] \right ) = 
e^{- 2 \pi i b_{kL}} \, \big [ K_i, K_j \big ] \, e^{2 \pi i b_{kL}}
\eea
for $k=1,\cdots,n$ and $L=1,\cdots,h$;
\item[(c)] it is holomorphic in all its arguments $x_k$ for $k=1,\cdots, n$;
\item[(d)] its $\mA^L$ periods, which we shall denote by $\cV_{ij}^L$ and are defined by, 
\bea
\label{E.7.q}
\cV_{ij}^L = \oint _{\mA^L} dx_i \, \big [ K_i, K_j \big ]
\eea
vanish identically for $L=1,\cdots,h$.
\end{itemize}}

\sm

The proof of Lemma \ref{E.lem:1} is relegated to appendix \ref{sec:55}. The proof of parts \textit{(a)} and \textit{(b)} of the lemma is straightforward; the proof of \textit{(c)} is more complicated and the proof of \textit{(d)} is quite involved and occupies more than 14 pages of appendix \ref{sec:55}. Interestingly, the proof of part \textit{(d)}  appeals to an identity between sums of bilinears in  Bernoulli numbers (see Proposition \ref{4.prop:5}),
\bea
\label{3.bernoulli}
\sum _{\sigma=0}^{\nu} { \nu ! \, \Ber _{\sigma} \Ber_{\nu+ \rho + 1  -\sigma} \over \sigma ! \, (\nu +1 - \sigma)! }
+ \sum _{\sigma=0}^{\rho} { \rho ! \, \Ber _{\sigma} \Ber_{\nu +\rho +1 -\sigma} \over \sigma ! \, (\rho +1  - \sigma)!}
+  \Ber _{\nu + \rho} =0
\eea
that does not seem to have appeared in print elsewhere (see, for example, \cite{Zagier} for an extensive 
collection of this type of bilinear identities for Bernoulli numbers and polynomials).

\sm

Assuming the validity of Lemma \ref{E.lem:1}, we prove item \textit{(3)} of Theorem \ref{7.thm:10} by demonstrating the vanishing of $\big [ K_i, K_j \big ]$ as follows. Substituting the expression for $K_i$ of (\ref{7.1.b}) into $\big [ K_i, K_j \big ]$ and making use of the expansions (\ref{gfsKJ}) and (\ref{gfsKW}) shows that $\big [ K_i, K_j \big ]$ admits a formal multiple  Taylor expansion  in powers of $b_{kI}$ for $k=1,\cdots, n$ in which $a_{iI}, a_{jI}$ and $t_{ik}, t_{jk}$ are viewed as being  part of the coefficients in the expansion. We make use of Lemma \ref{4.lem:100} for the generators $\beta_I = b_{kI}$ and $\mu(z) = [K_i, K_j]$ considered as a holomorphic $(1,0)$ form in $z=x_i$.  
The conditions (\ref{4.lem.100}) and (\ref{4.lem.200}) are satisfied in view of the assumptions \textit{(b)} and \textit{(d)} of Lemma \ref{E.lem:1}, respectively, as is holomorphicity. As a result of Lemma \ref{4.lem:100}, we conclude that $[K_i, K_j]=0$ for all $i, j =1,\cdots, n$ and by (\ref{7.1.a}) that ${\cal K}_{\rm E}\wedge {\cal K}_{\rm E} =0$, which concludes the proof of item \textit{(3)}  of Theorem \ref{7.thm:10}.

\newpage

\section{The multi-variable DHS connection}
\label{sec:33}

Using the infinite-dimensional Lie algebra $\hat \mt_{h,n}$ introduced by Enriquez in \cite{Enriquez:2011} and reviewed in section \ref{sec:2.3}, we generalize the single-variable DHS connection introduced in \cite{DHoker:2023vax} and reviewed in section \ref{sec:2.2} to a  connection $\cJ_\text{DHS}$ in an arbitrary number $n$ of variables on an arbitrary compact Riemann surface $\Sigma$ with and without punctures. 

\sm

The connection $\cJ_\text{DHS}$  is introduced in section \ref{sec:2.4} and proven to be flat on configuration space $\text{Cf}_n(\Sigma) = \Sigma ^n \setminus \{\hbox{diagonals}\}$ in sections~\ref{sec:2.4prf} and~\ref{sec:2.5}. The flat DHS connection in $n$ variables on  a Riemann surface with $p$ punctures, which takes values in the Lie algebra $\hat \mt_{h,n,p}$,  is deduced from the DHS connection without punctures in section~\ref{sec:4.punctures}. We prove that these DHS connections are modular invariant for suitable $Sp(2h,\mathbb Z)$ transformations of the corresponding Lie algebra generators in section~\ref{sec:4.mod}.

\subsection{Stating the main theorem} 
\label{sec:2.4}

In this subsection, we propose a single-valued connection $\cJ_\text{DHS}(\xx;a,b,t)$ dependent on  $n$ variables $\xx= (x_1, \cdots , x_n) \in \Sigma ^n$ with values in $\hat \mt_{h,n}$, on an arbitrary  Riemann surface of genus $h \geq 1$  without punctures, in terms of the generating functions that were introduced earlier in section~\ref{sec:2.2}. The multi-variable connection $\cJ_\text{DHS}$ is defined by,
\bea
\cJ_\text{DHS}(\xx;a,b,t)  = \sum _{i=1}^n  \Big ( dx_i \, J^{(1,0)}_i (\xx;a,b,t) + d \bar x_i \, J^{(0,1)} _i (\xx;a,b,t) \Big )
\label{defcJ.1}
\eea
where the component $J^{(0,1)} _i$ is given by,
\bea
J^{(0,1)} _i (\xx;a,b,t) = - \pi \bar \om^I(x_i) \, b_{iI}
\label{defcJ.2}
\eea
and depends on fewer variables and generators than what is displayed on the left side
for the sake of uniformity. The component $J^{(1,0)} _i$ admits
two different, but equivalent, expressions, 
\bea
J_i ^{(1,0)} (\xx;a,b,t) & = & \bPhi _K (x_i; B_i) \, a^K_i + \sum _{j \not = i} \bG (x_i,x_j;B_i) \, t_{ij}
\no \\
J^{(1,0)}_i (\xx;a,b,t) & = &
\bJ_K(x_i,z;B_i)  \, a^K_i + \sum _{j \not = i} \big[  \bG (x_i,x_j;B_i) -  \bG (x_i,z;B_i) \big] \, t_{ij}
\label{defcJ.3}
\eea
in terms of the generating functions $\bPhi _K,\bG, \bJ_K$ defined in (\ref{2.gen.1}).
Although some of the terms in the second expression of (\ref{defcJ.3}) individually depend on the point $z$, this dependence is actually spurious and cancels out in the sum thanks to the Lie algebra relation of item 2 in Definition \ref{2.def:1} and the decomposition of the generating functions in (\ref{2.gen.2}). 
All subsequent uses of the generating function identity refer to (\ref{2.gen.2}). The result of this cancellation is the expression on the first line in (\ref{defcJ.3}). While in the present paper we shall be concerned primarily by the first version of (\ref{defcJ.3}), the second version is used in the proof of Theorem \ref{5.thm:1} below and will be especially important in the companion paper \cite{DS-Fay}.

\sm

The following theorem is one of the main results of this paper.

{\thm 
\label{2.thm:1}
The curvature of the connection $\cJ_{\rm DHS}(\xx;a,b,t)$ defined in (\ref{defcJ.1}), with components given by  (\ref{defcJ.2}) and (\ref{defcJ.3}), evaluates to, 
\bea
\label{2.thm.1}
d \cJ_{\rm DHS}  - \cJ_{\rm DHS} \wedge \cJ_{\rm DHS}
=
\pi \sum_{i \not = j}  d \bar x_i \wedge \big ( d x_j - d x_i  \big )   \,  \delta (x_i, x_j) \, t_{ij} 
\eea
which reflects the simple poles at $x_i=x_j$ with residues $(dx_j {-} dx_i)t_{ij}$ for $i \not= j$, and vanishes on the configuration space ${\rm Cf}_n (\Sigma)  = \Sigma ^n \setminus \{ \hbox{diagonals} \}$ of $n$ points in~$\Sigma$. Therefore, $\cJ_{\rm DHS}(\xx;a,b,t)$ is a  flat connection on ${\rm Cf}_n (\Sigma)$. Here, $d$ denotes the sum of the total differentials in all $n$ variables, $d= d_{x_1} + \cdots + d_{x_n}$.}

\subsection{Proving Theorem \ref{2.thm:1}} 
\label{sec:2.4prf}

The proof of Theorem \ref{2.thm:1} draws crucially on the following lemma.

{\lem 
\label{2.lem:2}
The following commutators of $(1,0)$-form components (\ref{defcJ.3}) satisfy, 
\bea
\label{2.lem.2}
{} \left [ J^{(1,0)}_i (\xx;a,b,t), J^{(1,0)}_j (\xx;a,b,t) \right ] =0
\eea
identically on $\Sigma ^n$ for all $i \not = j$ and $i,j =  1, \cdots, n$. }
 
 \sm
 
 The proof of Lemma \ref{2.lem:2} constitutes the most difficult part of the proof of Theorem \ref{2.thm:1} and  is relegated to section \ref{sec:2.5} as well as appendices \ref{sec:A} and \ref{sec:B}.  

\sm

We now proceed to proving Theorem \ref{2.thm:1} with the help of Lemma \ref{2.lem:2}. To do so, we evaluate the curvature of the connection $\cJ= \cJ_\text{DHS}(\xx;a,b,t)$ in components, using (\ref{defcJ.1}), 
\bea
\label{2.e.1}
d \cJ- \cJ \wedge \cJ
& = &
\sum_i d \bar x_i \wedge d x_i \left (  \bar \p_i  J^{(1,0)}_i - \big [J^{(0,1)} _i ,  J^{(1,0)}_i \big ] \right )
\no \\ &&
+\sum_{i \not = j}  d \bar x_j \wedge d x_i \left (  \bar \p_j  J^{(1,0)}_i - \big [ J^{(0,1)} _j , J^{(1,0)}_i \big ] \right )
\no \\ &&
+ \half \sum _{i \not = j} dx_j \wedge dx_i \left (  \p_j J^{(1,0)}_i - \p_i J^{(1,0)}_j \right )
\no \\ &&
- \half \sum_{i \not = j} dx_i \wedge dx_j \big [ J^{(1,0)}_i, J^{(1,0)}_j \big ]
\eea 
where we have used the fact that the $(0,1)$ component (\ref{defcJ.2}) of $\cJ $ is closed.  The last line vanishes by Lemma \ref{2.lem:2} which, we recall, will be proven in section \ref{sec:2.5}. We shall now evaluate each one of the first three lines with the help of the expression for $J^{(1,0)}_i$ on the first line of (\ref{defcJ.3})  and prove that they add up to the right side of (\ref{2.thm.1}). 

\sm

\begin{itemize}
\item To evaluate the first line on the right side of (\ref{2.e.1}), we use the formulas for the derivatives of $\bPhi$ and $\bG$ given in (\ref{2.gen.3}), 
\bea
\label{2.d.4}
\bar \p_i  J^{(1,0)}_i 
& = &  \pi \kappa (x_i) [ b_{iJ},   \, a^J_i] - \pi \bar \om^I (x_i) B_{iI} \Phi _J (x_i;B_i) a^J_i 
\no \\ &&
+ \pi \sum_{j\not= i} \Big ( \kappa (x_i) - \delta (x_i,x_j) - \bar \om ^I (x_i) B_{iI} \bG(x_i, x_j;B_i) \Big ) t_{ij} 
\eea
The terms proportional to $\kappa(x_i)$ cancel one another  in view of (\ref{11.2}), while the remaining terms may be rearranged in terms of $J^{(1,0)}_i$ and $J^{(0,1)}_i$ as follows,
\bea
\label{2.derJi}
\bar \p_i  J^{(1,0)}_i - \left [ J^{(0,1)} _i, J^{(1,0)} _i \right ] & = & - \pi \sum _{j \not = i} \delta (x_i,x_j) t_{ij}
\eea
\item To evaluate the second line of (\ref{2.e.1}), we use the derivative of $\bG$  in (\ref{2.gen.4}) for $j \not= i$, 
\bea
\label{2.e.3}
\bar \p_j  J^{(1,0)}_i & = &  \pi  \delta (x_i, x_j) \, t_{ij}  - \pi  \bar \om^I(x_j) \bJ_I (x_i, x_j;B_i) t_{ij}
\no \\
- \big [ J^{(0,1)} _j , J^{(1,0)}_i \big ] & = & 
\pi \bar \om^I(x_j) B_{jI} \bigg \{ \bPhi _J(x_i;B_i) a^J_i + \sum _{k \not = i} \bG(x_i, x_k;B_i) t_{ik} \bigg \}
\eea
Since $[B_{jI}, B_{iK}]=0$ in view of the Jacobi identity and the second relation in (\ref{11.1}), we can move the operator $B_{jI}$ outside the braces on the second line past both functions $\bPhi$ and $\bG$. For the first term, we then use (\ref{11.1a}) to convert $B_{jI} a_i ^J = \delta ^J_I t_{ij}$.  For the second term, the only non-zero commutator is for $k=j$, and using $B_{jI} t_{ij} = - B_{iI} t_{ij}$ we obtain,
\bea
\label{2.d.5}
- \big [ J^{(0,1)} _j , J^{(1,0)}_i \big ] & = & 
\pi \bar \om^I(x_j) \Big \{ \bPhi _I(x_i;B_i) t_{ij}  - \bG(x_i, x_j;B_i) B_{iI} t_{ij} \Big \}
\eea
The sum inside the parentheses equals $\bJ_I(x_i,x_j;B_i) t_{ij}$ by (\ref{2.gen.2}) and cancels the corresponding term on the first line of (\ref{2.e.3}) in the combination on the second line of~(\ref{2.e.1}), 
\bea
\label{2.derJj}
\bar \p_j  J^{(1,0)}_i  - \big [ J^{(0,1)} _j , J^{(1,0)}_i \big ] & = & \pi  \delta (x_i, x_j) \, t_{ij} 
\eea
\item To evaluate the third line of (\ref{2.e.1}), we use the fact that the mixed derivative  $\p_j J^{(1,0)}_i $ receives a non-zero contribution only from $j$ in the first line of (\ref{defcJ.3}) so that we have, 
\bea
\label{2.d.6}
\p_j J^{(1,0)}_i - \p_i J^{(1,0)}_j
= \p_j \bG (x_i, x_j;B_i) t_{ij} - \p_i \bG(x_j,x_i;B_j) t_{ij}
\eea
Inserting the expansion of $\bG$ of (\ref{2.gen.1}), we obtain, 
\bea
\label{2.d.7}
\p_j J^{(1,0)}_i - \p_i J^{(1,0)}_j 
& = & 
\sum_{r=1}^\infty \Big ( \p_j \p_i \cG^{I_1 \cdots I_r} (x_i,x_j) B_{i I_1} \cdots B_{i I_r} - (i \leftrightarrow j) 
 \Big ) t_{ij}
\eea
Using the second identity in (\ref{2.a.8}), we recast the second term in the summand as follows,
\bea
\label{2.d.8}
\p_j J^{(1,0)}_i - \p_i J^{(1,0)}_j 
=
\sum_{r=1}^\infty  \p_j \p_i \cG^{I_1 \cdots I_r} (x_i,x_j) \Big ( B_{i I_1} \cdots B_{i I_r} 
- (-)^r B_{j I_r} \cdots B_{j I_1} \Big ) t_{ij}
\qquad
\eea
In view of (\ref{11.lem.2}) of Proposition \ref{2.lem:1}, the sum inside the parentheses vanishes for all values of $r$ so that we find,
\bea
\label{2.d.9}
\p_j J^{(1,0)}_i - \p_i J^{(1,0)}_j  =0
\eea 
\end{itemize}
Assembling in (\ref{2.e.1}) the contributions from (\ref{2.derJi}), (\ref{2.derJj}), (\ref{2.d.9}) and the result (\ref{2.lem.2}) of Lemma \ref{2.lem:2}, we obtain (\ref{2.thm.1}) and thereby conclude the proof of Theorem \ref{2.thm:1}.

\subsection{Vanishing of the commutator $[J_i^{(1,0)} , J_j^{(1,0)} ]$}
\label{sec:2.5}

This subsection is entirely devoted to the proof of Lemma \ref{2.lem:2}, namely to the vanishing of the commutator $[J_i^{(1,0)} , J_j^{(1,0)} ]$. We shall need the following lemmas.

{\lem
\label{2.lem:3}
The commutator satisfies the following differential equation for all $\xx \in \Sigma ^n$,
\bea
\label{3.2}
\bar \p_i \big [ J^{(1,0)}_i, J^{(1,0)}_j \big ]  = \Big [ J^{(0,1)} _i, \big [ J^{(1,0)}_i, J^{(1,0)}_j \big ]  \Big ]
\eea}

\noindent
Lemma \ref{2.lem:3} will be proven in appendix \ref{sec:A}. Note that the statement of this lemma applies to all $\xx \in \Sigma^n$ and not just to all $\xx \in {\rm Cf}_n(\Sigma)$.

{\lem
\label{2.lem:4}
The following integrals vanish identically for all values of $K=1,\cdots,h$,
\bea
\label{3.3}
\int _\Sigma d^2 x_i \, \bar \om^K(x_i) \big [ J^{(1,0)}_i, J^{(1,0)}_j \big ] =0
\eea}

\noindent
Lemma \ref{2.lem:4} will be proven in appendix \ref{sec:B}. 

\sm

We now proceed to proving Lemma \ref{2.lem:2}  assuming the validity of Lemmas~\ref{2.lem:3} and \ref{2.lem:4}. We note that the commutator $[ J^{(1,0)}_i, J^{(1,0)}_j  ]$ has $a$-degree 2 and consists of a Lie series in $b_{kI}$ with $k=1,\cdots,n$,  all terms in which have positive or vanishing $b$-degree and are single-valued on $\Sigma ^n$. Since we have $J_i^{(0,1)} = - \pi b_{iI} \bar \om^I(x_i)$, as well as the relations (\ref{3.2}) and (\ref{3.3}), the conditions of Lemma~\ref{4.lem:300} are met with $\beta_I = b_{kI}$ and $\nu (z) = [ J^{(1,0)}_i, J^{(1,0)}_j  ]$ as a single-valued $(1,0)$ form in $z=x_i$. As a result, the commutator must vanish for arbitrary $i,j=1,\cdots, n$. This completes the proof of Lemma~\ref{2.lem:2} using the results of Lemmas~\ref{2.lem:3} and \ref{2.lem:4}.

\subsection{The multi-variable DHS connection with punctures}
\label{sec:4.punctures}

The DHS connection $\cJ_\text{DHS} ^{(p)} (\xx, \yy;a,b,c,t)$ in $n$ variables $\xx=(x_1, \cdots ,x_n)$ on a Riemann surface $\Sigma_p$ with $p$ punctures $\yy=(y_1, \cdots , y_p)$ may be obtained from the DHS connection with $n{+}p$ variables $\zz=(z_1, \cdots, z_n, z_{n+1}, \cdots , z_{n+p})$ such that  $z_i=x_i$ for $i=1,\cdots, n$ and $z_{n+\a} = y_\a$ for $\a=1,\cdots, p$ in the absence of punctures by setting the differentials $dy_\a= d \bar y_\a =0$.\footnote{We are grateful to Benjamin Enriquez for suggesting this approach.} This procedure corresponds to reducing the \textit{mobile variables} $z_{n+1},\cdots,z_{n+p}$ in the connection on a Riemann surface without punctures to \textit{fixed punctures} in the connection on a surface with punctures $y_1,\cdots,y_p$. The resulting connection takes the following form,
\bea
\label{10.1}
\cJ_\text{DHS} ^{(p)} (\xx, \yy;a,b,c,t) = \sum_{i=1}^n \Big ( dx_i J_i ^{(1,0)} (\xx,\yy;a,b,c,t) + d \bar x_i J_i^{(0,1)} (\xx,\yy;a,b,c,t) \Big )
\eea
where the components are given by (displaying the same potential arguments of $J_i^{(1,0)} $ and $J_i^{(0,1)} $ for the sake of uniformity), 
\bea
\label{10.2}
J_i^{(1,0)} (\xx,\yy;a,b,c,t) & = & \bPhi _J(x_i;B_i) a^J_i + \sum_{j \not= i} \bG(x_i,x_j;B_i) t_{ij} + \sum_{\a =1}^p \bG(x_i,y_\a;B_i) c_{i\a}
\no \\
J_i^{(0,1)} (\xx,\yy;a,b,c,t) & = & - \pi \, \bar \om^I(x_i) \, b_{iI}
\eea
In this reduction process, the components $J_{n+\a} ^{(1,0)}$ and $J_{n+\a} ^{(0,1)} $ are removed from the connection and, along with them, the Lie algebra generators $a^J_{n+\a}, b_{n+\a, I}$ and $t_{n+\a, n+\b}$ for all $\alpha, \beta=1,\cdots,p$. The remaining generators are those of the Lie algebra $\mt_{h,n,p}$ (see footnote \ref{foot2}), namely, $a_i^I, \, b_{iI}, \, t_{ij}, \, c_{i \a} = t_{n+\a,i} = t_{i, n+\a} $ for $i, j =1,\cdots,n$ with $i \not=j$ and $\a=1,\cdots, p$. The structure relations are induced from those of $ \mt_{h,n+p}$ by omitting any structure relation in which one or both of the arguments of the commutator  is one of  the generators $a^J_{n+\a}, b_{n+\a, I}$ and $t_{n+\a, n+\b}$. This operator is well-defined since the ideal generated by the omitted generators is a two-sided ideal of $\hat \mt_{h,n+p}$ so that the quotient Lie algebra structure on the remaining generators is unambiguous.\footnote{We thank one of the referees for articulating this clarification.}

The resulting structure relations are given by, 
\begin{align}
\label{10.6}
[ a^I_i, a^J_j]  &= 0  
& [ b_{iI}, b_{jJ} ]  &= 0         
&  [ c_{i \a}, c_{j \b} ]  &= 0 
 \no \\
 [ b_{iI} , a^J_j]  &= \delta^J_I t_{ij}  & [ a^I_i , c_{j\a} ]  &= 0         
&  [ b_{iI} , c_{j \a} ]  &= 0 
 \no \\
[ a^I_i, t_{jk} ] &= 0  
&  [ b_{iI} , t_{jk} ] &= 0  
&  [ c _{ i  \a} , t_{jk}] &=0
\no  \\
 [a_i^I + a_j^I, t_{ij} ]  &= 0 
& [b_{iI}  + b_{jI} , t_{ij} ]  &= 0 
& [ c_{i \a} + c_{j \a} , t_{ij} ]  &= 0 
\no \\
 [ c_{ i\a }, c_{j \a} + t_{ij} ]  &= 0  
&  [ t_{ik} + t_{jk}, t_{ij} ] &=0  
&  [ t_{ij}, t_{k \ell} ]  &= 0 
\end{align}
where the indices $i,j,k,\ell$, or any subset thereof, are mutually distinct and $\b \not= \a$, as well as,  
\bea
\label{10.5}
{} [b_{iI}, a_i^I] = -  \sum_{j \not = i} t_{ij} - \sum _{\a=1}^p c_{i\a} =0
\eea
where the index $I$ is summed over but the index $i$ is not. The Lie algebra $\mt_{h,n,p}$ inherits the bi-grading of $\mt_{h,n+p}$ so that $a_i^I$, $b_{iI}$, $t_{ij}$ and $c_{i\a}$ have degree $(1,0), (0,1), (1,1)$ and $(1,1)$, respectively. The DHS connection (\ref{10.1}) takes values in the degree completion $\hat \mt_{h,n,p}$ of $\mt_{h,n,p}$  (see footnote \ref{foot2}). Flatness on ${\rm Cf}_n (\Sigma_p)$ is ensured by  flatness of the connection in~$n+p$ variables without punctures provided we set $dy_\a= d \bar y_\a =0$ in the differential $d_\xx= d_{x_1} + \cdots + d_{x_n}$, 
\bea
\label{10.3}
d_\xx  \cJ_\text{DHS} ^{(p)} - \cJ_\text{DHS} ^{(p)} \wedge \cJ_\text{DHS} ^{(p)}=0
\eea
Flatness may also be proven directly along the same lines that flatness was proven for the multivariable DHS connection in the absence of punctures using the structure relations of $\mt_{h,n,p}$ given in (\ref{10.6}) and (\ref{10.5}).  Finally, the connection $\cJ_\text{DHS} ^{(p)} (\xx, y_1,\cdots,y_p ;a,b,c,t) $ for $n=1$ reduces to the connection $\cJ_\text{DHS}(x,y,y_1,\cdots,y_p;a,b,c)$ of (\ref{2.gen.5}) in a single variable $\xx=x$ and $p{+}1$ punctures $y,y_1,\cdots,y_p$ by setting $t_{ij}=0$, $a_i^I=a^I, b_{iI}=b_I$ and $c_{i \a} = - c_\a$ and supplementing by the condition (\ref{2.gen.7}) that eliminates the dependence on $y$.

\subsection{Modular invariance of the multi-variable DHS connections}
\label{sec:4.mod}

In this subsection, we establish modular invariance of the multi-variable DHS connection  $\cJ_\text{DHS}(\xx;a,b,t)$ defined by (\ref{defcJ.1}), (\ref{defcJ.2}) and (\ref{defcJ.3}), as well as $\cJ_\text{DHS}^{(p)}(\xx,\yy;a,b,c,t)$ defined in (\ref{10.1}) and (\ref{10.2}). On a compact Riemann surface~$\Sigma$ of genus~$h$ with intersection pairing $\mJ$, a modular transformation maps a canonical basis of homology cycles,  assembled here into column matrices $\mA, \mB$ with components $\mA^I$ and $\mB_I$, respectively,  into another  canonical basis $\tilde \mA, \tilde \mB$ by an element $M \in Sp(2h,\ZZ)$,
\bea
\label{modsec.01}
\left ( \bma \mB  \cr \mA \ema \right ) 
\, \to \,
\left ( \bma \tilde \mB  \cr \tilde \mA \ema \right ) 
= M
\left ( \bma \mB  \cr \mA \ema \right )  
\hskip 0.7in 
M = \left ( \bma A & B \cr C & D \ema \right ) 
\hskip 0.7in 
M^t \mJ M = \mJ
\eea 
Under $Sp(2h,\ZZ)$, the row matrix  of holomorphic Abelian differentials $\om$, the period matrix $\Omega$, and its imaginary part $Y= \Im (\Omega)$ transform as follows, 
\bea
\label{modsec.05}
\om & \to & \tilde \om = \om (C \Omega {+}D)^{-1}
\no \\
\Omega & \to &  \tilde \Omega = (A\Omega{+}B)(C\Omega{+}D)^{-1}
\no \\
Y & \to & \tilde Y = (\bar \Omega C^t {+} D^t)^{-1} Y (C \Omega {+}D)^{-1} 
\eea
More generally, we define a modular tensor $\mT$ of rank $(r,s)$ with components $\mT^{I_1 \cdots I_r} {} _{J_1 \cdots J_s}$ to transform as follows under $M \in Sp(2h,\ZZ)$,
\bea
\label{modsec.02}
\mT^{I_1 \cdots I_r}{}_{J_1\cdots J_s}(\Omega) 
\, \to \, 
\tilde{\mT}^{I_1 \cdots I_r}{}_{J_1\cdots J_s}(\tilde \Omega) = Q^{I_1}{}_{\! K_1}  \cdots Q^{I_r}{}_{\! K_r}  \, 
\mT^{K_1 \cdots K_r}{}_{L_1\cdots L_s}(\Omega) \, 
R^{L_1}{}_{\! J_1}  \cdots R^{L_s}{}_{\! J_s} 
\qquad
\eea
where we use shorthands for the automorphy tensor $Q$ and its inverse $R$,
\bea
Q = C \Omega {+} D \hskip 0.6in R = Q^{-1} = (C \Omega {+} D)^{-1}
\label{modsec.03}
\eea
and suppress the dependence on the period matrix here and below.
One may extend the definition of modular tensors given in (\ref{modsec.02}) to tensors that transform by factors of $Q$ and $R$ as well as by factors of $\bar Q$ and $\bar R$, such as is the case for $Y$ in the last equation of (\ref{modsec.05}). However, by contracting with suitable factors of $Y$ or its inverse, all such tensors are equivalent to a linear combination of the modular tensors defined in (\ref{modsec.02}). We illustrate this procedure on the complex conjugates of the holomorphic Abelian differentials of (\ref{2.a.2}), 
\bea
\bar \om_I \to \tilde{\bar \om} _I = \bar \om_J \, \bar R^J{}_I
\hskip 1in 
\bar \om^I \to \tilde{\bar \om}^I = Q^I{}_J \, \bar \om ^J
\label{modomega}
\eea
As defined above, modular tensors are functions on Torelli space, the moduli space of compact Riemann surfaces equipped with a canonical basis of $\mA$ and $\mB$ cycles,  and are akin to vector valued modular forms \cite{vdG3}.

\sm

Building upon mathematical work in \cite{Kawazumi,  Kawazumi:lecture}, a variety of modular tensors were constructed
from convolutions of the Arakelov Green function defined by (\ref{2.a.4}) and its derivatives \cite{DHoker:2020uid}. In particular, the DHS kernels (\ref{2.a.5})  transform as modular tensors of rank $(r,1)$~\cite{DHoker:2023vax},
\bea
f^{I_1 \cdots I_r}{}_{J}(x,y) \, \rightarrow \, 
\tilde{f}^{I_1 \cdots I_r}{}_{J}(x,y)= Q^{I_1}{}_{K_1} \cdots Q^{I_r}{}_{K_r} f^{K_1 \cdots K_r}{}_{L}(x,y) R^{L}{}_{J} 
\label{modsec.04}
\eea
The above transformation laws lead to a modular invariant multi-variable DHS connection.

{\thm 
\label{thm:mod}
The DHS connections in $n$ variables, defined in (\ref{defcJ.1}) in the absence of punctures,  and defined in (\ref{10.1}) in the presence of $p$ punctures,  are modular invariant,
\bea
\cJ_{\rm DHS}(\xx;a,b,t)  & \rightarrow & \tilde{\cJ}_{\rm DHS}(\xx;\tilde a,\tilde b,\tilde t) = \cJ_{\rm DHS}(\xx;a,b,t)  
\no \\
\cJ_{\rm DHS}^{(p)} (\xx,\yy;a,b,c,t)  & \rightarrow & \tilde{\cJ}_{\rm DHS}^{(p)} (\xx,\yy;\tilde a,\tilde b, \tilde c,\tilde t) = \cJ_{\rm DHS}^{(p)} (\xx,\yy;a,b,c,t)  
\label{modsec.06}
\eea
provided that the transformation of the generators under $Sp(2h,\mathbb Z)$ is engineered as follows,
\bea
\label{modsec.00}
a_i^K \, & \rightarrow  &\, \tilde a_i^K = Q^K{}_L \, a_i^L 
\notag \\
b_{iK} \, & \rightarrow & \, \tilde b_{iK} = b_{iL} \, R^L{}_K  
\no \\
 t_{ij} \, & \rightarrow & \, \tilde t_{ij} ~ = t_{ij}
 \no \\
c_{i\a} \, & \rightarrow & \,  \tilde c_{i\a}  \! ~ = c_{i\a}
\eea
with $i,j=1,\cdots, n$ with $j \not= i$, $\a=1,\cdots, p$, $K=1,\cdots,h$ and automorphy factors 
$Q^K{}_L $,  $R^L{}_K$ given by (\ref{modsec.03}).
}

\sm

The proof of the theorem relies on the modular properties of the generating
functions ${\bf J}_K(x,y;B) $ and ${\bf G}(x,y;B) $  in (\ref{2.gen.1}) \cite{DHoker:2023vax},
\begin{align}
{\bf J}_K(x,y;B)   &\rightarrow \tilde{{\bf J}}_K(x,y;\tilde B) = {\bf J}_L(x,y;B) R^L{}_K \notag\\
{\bf G}(x,y;B)  &\rightarrow \tilde{{\bf G}}(x,y;\tilde B) = {\bf G}(x,y;B)
\label{modsec.07}
\end{align}
which in turn follow from the $Sp(2h,\mathbb Z)$ transformations of the DHS kernels and the Abelian differentials in (\ref{modsec.04}) and (\ref{modsec.05}). To prove the first relation in (\ref{modsec.06}), we observe that the $(0,1)$-form components $J^{(0,1)} _i (\xx;a,b,t)$ of  $\cJ_\text{DHS}(\xx;a,b,t)$ in (\ref{defcJ.2}) and the $(1,0)$-form components $J^{(1,0)}_i (\xx;a,b,t)$ in (\ref{defcJ.3}) are separately modular invariant by combining the transformations (\ref{modsec.00}) of the $\mt_{h,n}$ generators with (\ref{modomega}) for $\bar \omega^I(x_i)$ and (\ref{modsec.07}) for the generating functions, respectively. To prove the second relation in (\ref{modsec.06}), we use the fact that modular transformations leave the geometrical points on $\Sigma$ invariant, whether they are variables or punctures. Therefore, imposing the  relations $dy_\a=d \bar y_\a=0$ on the modular invariant connection in $n{+}p$ variables without punctures, to metamorphose these variables into punctures,  preserves the modular properties of the connection.

{\remark
\label{5.remark.1}
Note that the same steps were used in \cite{DHoker:2023vax} to demonstrate modular invariance
of the single-variable DHS connection (\ref{defjdhs}) under $Sp(2h,\mathbb Z)$
transformations $a^K \rightarrow  Q^K{}_L a^L$ and $b_{K} \rightarrow  b_{L} R^L{}_K $
of its generators. As argued in the reference, these $Sp(2h,\mathbb Z)$
properties of the flat connections imply the higher-genus polylogarithms generated by
their path-ordered exponentials to transform as modular tensors (\ref{modsec.02}).}

 \newpage

\section{Relating multi-variable flat connections}
\label{sec:99}

In this section, we  prove that, in the absence of punctures, the single-valued, modular invariant, but non-meromorphic, connection $\cJ_\text{DHS}$ in $n$ variables on a compact Riemann surface and the meromorphic, but multiple-valued, Enriquez connection $\cK_\text{E}$ in $n$ variables are related by the composition of a gauge transformation and an automorphism of their common Lie algebra $\hat \mt_{h,n}$. This result generalizes the one reviewed in section \ref{sec:1var} for the case of one variable and one puncture, which was originally obtained in \cite{DHoker:2024desz}.

\sm

Specifically, in sections \ref{sec:7.1}, \ref{secmon} and \ref{sec:5.poles} we derive, with the help of an auxiliary result of section \ref{sec:aux}, a number of necessary conditions for the existence of a combined gauge transformation and Lie algebra automorphism that relates the connections $\cJ_\text{DHS}$ and~$\cK_\text{E}$. 
In particular,  we prove that these necessary conditions completely determine the gauge transformation and the automorphism of $\hat \mt_{h,n}$ in terms of the same constituents that entered the relation between the single-variable versions of $\cJ_\text{DHS}$ and $\cK_\text{E}$ in section \ref{sec:1var}. In section \ref{sec:equiv} we prove that the collection of these necessary conditions is actually sufficient for the existence of the combined gauge transformation and Lie algebra automorphism. The $\hat \mt_{h,n}$ automorphism properties of the maps of this section between different types of generators are proven in sections \ref{strsec} and \ref{nwsc.57}.

\sm

An overview of notations and pointers to key equations can be found in the preamble of
section \ref{secnvar}, and we reiterate that Einstein summation convention for indices $I,J,\cdots = 1,\cdots,h$ 
only applies to pairs of upper and lower indices but not to repeated lower indices.

\subsection{The gauge transformation}
\label{sec:7.1}

In this subsection, we begin spelling out some of the necessary conditions for the existence of a gauge transformation and Lie algebra automorphism relating the multi-variable Enriquez connection $\cK_\text{E}$, reviewed in section \ref{sec:mEcon}, to the  multi-variable connection $ \cJ_\text{DHS}$ of section~\ref{sec:33}.

\sm

Applying an arbitrary gauge transformation $\cU$ to the flat connection $\cK_\text{E}$, which satisfies 
$d \cK_\text{E} - \cK_\text{E}  \wedge \cK_\text{E}=0$ in the configuration space $\text{Cf}_n(\tilde \Sigma)$ of $n$ points on the universal cover $\tilde \Sigma$ of the Riemann surface $\Sigma$, will  produce a connection $\cJ$ that is related to $\cK_\text{E}$ by,
\bea
\label{7.0}
d- \cK_\text{E} = \cU^{-1} ( d - \cJ) \, \cU
\eea 
and is automatically flat so that it satisfies $d \cJ- \cJ \wedge \cJ=0$. In the case at hand, however, the two connections must satisfy different monodromy relations. Specifically, we seek a gauge transformation from $\cK_\text{E}$, whose non-trivial monodromies are stated in Theorem~\ref{7.thm:10}, to $\cJ_\text{DHS}$ which is single-valued on $\Sigma$. Such a gauge transformation must be accompanied by a non-trivial automorphism $\phi: ( a, b, t) \to (\hat a, \hat b, \hat t)$ of $\hat \mt_{h,n}$ as is familiar  from the single-variable case in \cite{DHoker:2024desz} reviewed in section \ref{sec:1var} (see section~5 of the reference for a general argument), and takes the following~form, 
\bea
\label{7.1}
d- \cK_\text{E}  (\xx;a,b,t) = \cU(\xx, \yy) ^{-1} \Big ( d - \cJ_\text{DHS} (\xx; \hat a, \hat b, \hat t) \Big ) \cU(\xx,\yy)
\eea
The gauge transformation $\cU(\xx, \yy)$ may be constructed from the connection $\cJ_\text{DHS} $ for some other triplet $(\xi, \eta, \tau)$ of Lie algebra generators that will be determined in terms of $b_{iI}$ and turns out to obey the structure relations of  $\hat \mt _{h,n}$ (see Lemma \ref{5.lem:88a} below).\footnote{The gauge transformation $\cU$ may alternatively be constructed in terms of $\cK_\text{E}$ similar to the procedure in  section 4 of \cite{DHoker:2024desz} for the single-variable case.}  We shall carry out this construction  shortly.  Flatness of  $\cJ_\text{DHS} (\zz; \xi,\eta,\tau)$ yields $\cU(\xx,\yy)$ as a homotopy invariant path-ordered~exponential,
\bea
 \label{7.1.c} 
\cU (\xx, \yy) = {\rm Pexp}\bigg( \int^{\xx}_{\yy} \cJ_\text{DHS}  (\zz; \xi,\eta,\tau)  \bigg) 
\eea
The non-trivial monodromy of the gauge transformation $\cU(\xx,\yy)$ constructed above will allow us to relate the connections $\cK_\text{E}$ and $\cJ_\text{DHS}$, despite the fact that their monodromies are different from one another.  We note that the dependence of the right side of (\ref{7.1}) on the endpoint $\yy = (y_1,\cdots,y_n) \in \tilde \Sigma^n$  will be compensated by the $\yy$-dependence of the relation between the generators $(a,b,t)$ and $(\hat a, \hat b, \hat t)$, as will be made explicit in section \ref{sec:aux} below.

\sm

To derive the necessary conditions for the correspondence (\ref{7.1}), we assume that each one of the  triplets $(a,b,t)$, $(\hat a, \hat b, \hat t)$ and $(\xi, \eta, \tau)$ satisfies the structure relations of $\mt_{h,n}$. From this assumption, along with (\ref{7.1}), we shall show that $(\hat a, \hat b, \hat t)$ and $(\xi, \eta, \tau)$ are uniquely and explicitly determined by $(a,b,t)$. To begin, we prove the following lemma.

{\lem
\label{5.lem:1}
The relations of (\ref{7.1.d}) below are necessary conditions for the existence of a gauge transformation $\cU(\xx,\yy)$ of (\ref{7.1.c}) and a Lie algebra automorphism $(a,b,t) \to (\hat a, \hat b, \hat t)$ relating the flat connections $\cJ_{\rm DHS} $ and $\cK_{\rm E}$ through equation (\ref{7.1}), 
\bea
 \label{7.1.d} 
 \eta_{iJ}   = \hat b_{iJ}  
 \hskip 0.8in 
\tau_{ij}  = 0 
\hskip 0.8in
[ \eta_{iI}, \xi_i^I ]  = 0
\hskip 0.8in
[ \eta_{iI}, \xi_j^J ]  = 0
\eea
for all $i,j =1, \cdots, n$ with $i \not= j$.  The gauge transformation $\cU(\xx,\yy)$ factorizes,
\bea
\label{7.2.a}
\cU (\xx, \yy)  =  \cU_1 (x_1, y_1) \, \cdots \, \cU_n (x_n, y_n) 
\eea
where $\cU_i(x_i,y_i)$ depends only on the points $x_i,y_i \in \tilde \Sigma$ and on the generators $\eta_i$ and~$\xi_i$ (but not on $x_j, y_j, \xi_j$ and $ \eta_j$ for $j \not= i$). Each factor is given by the following path-ordered exponential,
\bea
\label{7.2.c}
\cU_i  (x, y)  =
{\rm Pexp} \int _{y} ^{x}  \Big( dz\, \bPhi_J(z; H_i) \, \xi_i^J  - \pi \, d \bar z  \, \bar \omega^I(z) \, \eta_{iI} \Big) 
\eea
where $H_{iJ} X = [\eta_{iJ},X]$ for $X \in \hat \mt_{h,n}$. The factors  satisfy $
 [ \, \cU_i (x,y) , \cU_j (x',y') \, ] = 0$ for $ i \not= j$ for arbitrary $x,y,x',y' \in \tilde \Sigma$.}

\sm

To prove the first relation in (\ref{7.1.d}) we insert the expression (\ref{7.1.c}) for $\cU(\xx,\yy)$ into
(\ref{7.1}) and use the fact that $\cK_\text{E}$ has vanishing $(0,1)$ component, resulting in,
\bea
\label{7.2.d}
J_i^{(0,1)} (\xx; \hat a, \hat b, \hat t) = \bar \p_i \, \cU(\xx,\yy) \, \cU(\xx,\yy)^{-1} = J_i^{(0,1)} (\xx; \xi, \eta, \tau) 
\eea
Making use of the explicit expression for $J_i^{(0,1)} $ in (\ref{defcJ.2}) then  proves the first relation of (\ref{7.1.d}). The second relation, $\tau_{ij}  = 0 $, follows from the required absence of monodromy as the point $x_i$ is moved in a small circle around another point $x_j$, by the same argument that was used in the single-variable case when the variable was moved around the puncture (see section 3.1.2 of \cite{DHoker:2024desz}). The third relation follows from the structure relation (\ref{11.2}) of Definition~\ref{2.def:1} satisfied by $\xi$ and $\eta$, while the fourth relation follows from  $[\eta_{iI}, \xi^J_j]= \delta ^J_I \tau_{ij}=0$. Finally, since $\tau_{ij}=0$ for all $i \not= j$, the $(1,0)$ components of the connection $\cJ_\text{DHS}$ in (\ref{defcJ.3})  reduce to,
\bea
J_i ^{(1,0)} (\zz;\xi, \eta, \tau) = \bPhi _K(z_i;H_i) \xi_i^K
\eea
These components commute with one another for different values of $i \not = j$ in view of the last relation of (\ref{7.1.d}) and the structure relations $[ \eta_{iI}, \eta_{jJ} ]  =[ \xi_{i}^I, \xi_j^J ]  = 0$ of $\hat \mt_{h,n}$ in (\ref{11.1}). Hence, 
their path-ordered exponentials factorize as in (\ref{7.2.a}) and the factors mutually commute,\footnote{Note, however, that  $\cU_i (x,y)$ and $ \cU_i (x',y') $ generally do not commute with one another.} which completes the proof of the lemma.

\subsection{Relating $b_i$ to $\hat b_i$ by the monodromy transformations}
\label{secmon}

The connections $\cJ_\text{DHS} (\xx;\hat a, \hat b, \hat t)$ and $\cK_\text{E}(\xx; a,b,t)$ have trivial monodromies  around $\mA$ cycles, while their monodromy (\ref{defgmo}) as a variable $x_i$ is moved around a $\mB_K$ cycle is trivial for $\cJ_\text{DHS} (\xx;\hat a, \hat b, \hat t)$ but non-trivial for  $\cK_\text{E}(\xx; a,b,t)$,
\bea
\gamma ^{(i)} _K \, \cJ_\text{DHS} (\xx;\hat a, \hat b, \hat t) & = & \cJ_\text{DHS} (\xx;\hat a, \hat b, \hat t)
\no \\
\gamma ^{(i)} _K \, \cK_\text{E}(\xx; a,b,t) & = & e^{- 2 \pi i b_{iK}} \cK_\text{E}(\xx; a,b,t) \, e^{ 2 \pi i b_{iK}}
\eea
The gauge transformation $\cU (\xx, \yy)$ of (\ref{7.1}) relating the two connections must convert these two monodromies into one another. Therefore, the  monodromy of $\cU(\xx,\yy)$ in any variable $x_i$  must be trivial around $\mA$ cycles, while around a $\mB_K$ cycle we must have,
\bea
\gamma ^{(i)} _K \, \cU(\xx,\yy)  = \cU (\xx, \yy)  \, e^{2\pi i b_{iK}}
 \label{7.2.e}    
\eea
Since the gauge transformation $\cU(\xx,\yy)$ decomposes into the product (\ref{7.2.a}) of mutually commuting factors 
$\cU_i (x_i, y_i)$ that only depend on the variables $x_i,y_i$ and not on $x_j,y_j$ for $j \not= i$, the monodromy conditions of (\ref{7.2.e})  require that the factors $\cU_i(x_i,y_i)$ have trivial $\mA$ monodromy while their $\mB$ monodromy must be given by,
\bea
\cU_i(\mB_K \cdot x_i, y_i) & = & \cU_i(x_i,y_i) \, e^{2 \pi i b_{iK}}
\label{monfactor}
\eea
We manifestly have $\gamma ^{(j)}_K \, \cU_i(x_i,y_i) =\cU_i(x_i,y_i)$ for $j \not= i$ since $\cU_i$ is independent of $x_j$.

\sm

The monodromy conditions for a given value of $i$ in (\ref{monfactor}) are identical to the monodromy conditions for the single-variable case stated in (\ref{mauto.06}) and therefore are solved by the relations (\ref{mauto.09}) of the single-variable case. The expression for $\xi_i^I$ is given in terms of $\hat b_{iK}$ by the triviality of the $\mA$ monodromy,  as follows, 
\bea
\label{7.2.g1}   
\xi^I_i = \pi  \hat b_i^I + \sum_{r=2}^{\infty} {\cal X}^{I J_1 \cdots J_r} \, \hat b_{i J_1} \cdots  \hat b_{i J_r} 
\eea
where $\cX$ obeys the following cyclic property for $r \geq 2$, derived in equation (3.44) of \cite{DHoker:2024desz},
\bea
\label{5.cyclic}
{\cal X}^{I J_1 \cdots J_r}  = {\cal X}^{ J_1 \cdots J_r I} 
\eea
while the $\mB$ monodromies determine $\hat b_{iI}$ in terms of $b_{iK}$ as follows,
\bea
\label{7.2.g}   
\eta _{iI} = \hat b_{iI} & = b_{iI} - \sum_{r=2}^{\infty} {\cal M}_\shuffle^{J_1 \cdots J_r}{}_I(y_i)\, b_{iJ_1} \cdots  b_{iJ_r} 
\eea
The coefficients ${\cal X}^{J I_1 \cdots I_r}$ and  ${\cal M}_\shuffle^{J_1 \cdots J_r}{}_I(y_i)$ are  those that appeared in the single-variable case in  (\ref{mauto.09}).  Both depend on the moduli of the compact Riemann surface $\Sigma$. While ${\cal M}_\shuffle^{J_1 \cdots J_r}{}_I(y_i)$ also depends on the base point $y_i$, 
${\cal X}^{J I_1 \cdots I_r} $ is independent of $y_i$ (see section 3.3.1 of  \cite{DHoker:2024desz}). 

\sm

Finally, each gauge transformation factor $\cU_i$ may be expanded in a Taylor series in $\eta_i=\hat b_i$
\begin{align}
 \label{7.3.a}  
\cU_i^{-1}(x,y) &= 1 + \sum_{r=1}^\infty \cT^{I_1 \cdots I_r}(x, y) \, \eta_{iI_1} \cdots \eta_{i I_r}
 \end{align}
upon eliminating $\xi_i$ via (\ref{7.2.g1}). The components $\cT^{I_1 \cdots I_r}(x,y)$ are  the same combinations of DHS polylogarithms as introduced in \cite{DHoker:2024desz} for the expansion of the inverse of ${\cal U}_{\rm DHS}(x,y)$ in (\ref{mauto.05}).  Recall that its action in the adjoint representation on an arbitrary element $X \in \hat t_{h,n}$ with $H_{iI} X = [\eta_{iI}, X]$ is given~by,
\bea
 \label{7.3.b}  
\cU_i^{-1}(x,y)  X \, \cU_i (x,y) 
= X + \sum_{r=1}^\infty \cT^{I_1 \cdots I_r}(x,y) \, H_{iI_1} \cdots H_{i I_r} X
\eea

\subsection{Dependence on the base point}
\label{sec:aux}

The relation between the generators $(a,b,t)$ and $(\hat a, \hat b, \hat t)$ in (\ref{7.1}) depends on the base point~$\yy$ of the gauge transformation $\cU(\xx,\yy)$. This dependence is implicit in the relation (\ref{7.1}), whose left side has no explicit dependence on $\yy$, and will be made explicit here. It will be convenient to keep the generators $(\hat a, \hat b, \hat t)$ independent of $\yy$ within the arguments of this subsection. Under this assumption, we have the following lemma.

{\lem
\label{5.lem:2}
For fixed generators $(\hat a, \hat b , \hat t)$, the generators $(a,b,t)$ transform to $(a', b', t')$ as follows when the base point of the gauge transformation (\ref{7.1.c}) is moved  from $\yy$ to $\yy'$,
\bea
\label{5.lem.2}
 a_i \to a_i' = \cU^{-1} \, a_i \, \cU 
 \hskip 0.6in 
 b_i \to b_i' = \cU^{-1} \, b_i \, \cU 
 \hskip 0.6in 
 t_{ij}  \to t_{ij}' = \cU^{-1} \, t_{ij}\, \cU 
\eea
where we have used the shorthand $\cU = \cU (\yy,\yy')$ for the gauge transformation. }

\sm

To prove the lemma, we begin by showing that, for fixed $\hat b_i$,  the gauge transformation $\cU(\xx,\yy)$ depends on $\xx$ and $\yy$ only through the endpoints  $\xx$ and $\yy$ of the path-ordered exponential.  This may be established by expressing each factor $\cU_i$ given in (\ref{7.2.c}) in terms of the generators $\hat b_i$ using the first relation of (\ref{7.1.d}), 
\bea
\label{5.UU}
\cU_i  (x, y)  =
{\rm Pexp} \int _{y} ^{x}  \Big( dz\, \bPhi_J(z; \hat B_i) \, \xi_i^J  - \pi \, d \bar z  \, \bar \omega^I(z) \, \hat b_{iI} \Big) 
\eea
where $\hat B_i X = [\hat b_i, X]$ and expressing $\xi_i$ in terms of $\hat b_i$ using (\ref{7.2.g1}). 
The coefficients $\cX$ involved in the relation between $\xi_i$ and $\hat b_i$ are independent of $y_i$. As a result, for fixed $\hat b_i$, the integrand of (\ref{5.UU}) is independent of $x_i$ and $y_i$, so that the only dependence of $\cU_i(x_i, y_i)$ on $x_i$ and $y_i$  is through its endpoints. The remainder of the proof proceeds by writing (\ref{7.1}) for $(a', b', t')$ and base point $\yy'$,
\bea
\label{7.10}
d- \cK_\text{E}  (\xx;a',b',t') = \cU(\xx, \yy') ^{-1} \Big ( d - \cJ_\text{DHS} (\xx; \hat a, \hat b, \hat t) \Big ) \cU(\xx,\yy')
\eea
Having fixed $\hat b_i$, the gauge transformations $\cU(\xx,\yy)$ and $\cU(\xx,\yy')$ that relate the connection $\cJ_\text{DHS}(\xx;\hat a, \hat b, \hat t)$  to either $\cK_\text{E}(\xx;a,b,t)$ or $\cK_\text{E}(\xx;a',b',t')$ differ only in their endpoints. Using the composition rule for path ordered exponentials $\cU(\xx, \yy') = \cU(\xx,\yy) \, \cU(\yy, \yy')$, and eliminating the right sides of (\ref{7.1}) and (\ref{7.10}), we obtain,
\bea
d- \cK_\text{E}  (\xx;a',b',t') = \cU(\yy, \yy')^{-1} \Big ( d- \cK_\text{E}  (\xx;a,b,t)  \Big ) \cU(\yy, \yy')
\label{ktokprime}
\eea
Since the differential $d$ commutes with $\cU(\yy, \yy')$, this equation is  uniquely solved by (\ref{5.lem.2}), which proves the lemma.

\sm

\begin{remark}
\label{5.rmk}
In the sequel, we shall use the result of Lemma \ref{5.lem:2} to choose the particularly convenient base point $\yy_*$ for which  all $y_i$ are set equal to a single value $y_\ast \in \tilde \Sigma$, 
\bea
\label{5.a.1}
y_i = y_* \hskip 0.5in \hbox{ for all } \quad i=1, \cdots , n
\eea
By Lemma \ref{5.lem:1} the residues $\tau_{ij}$ of the simple poles of the connection $\cJ_{\rm DHS}(\zz;\xi, \eta, 0)$ as $z_i \rightarrow z_j$ vanish, and $\cJ_{\rm DHS}(\zz;\xi, \eta, 0)$ is smooth for any $\zz \in \Sigma^n$. Hence, the limit needed to make the choice (\ref{5.a.1}) of integration endpoint in (\ref{7.1.c}) exists and is smooth, since the integrand with $\tau_{ij}=0$ cannot develop any singularity from the collision of the base points $y_i = y_j = y_*$ in (\ref{7.1.c}). Smoothness of the limit $y_i \rightarrow y_j$ can alternatively be seen from
the factorization property (\ref{7.2.a}) of $\cU(\xx, \yy)$ by which the dependence on $y_i$ and $y_j$
occurs in separate series $\cU_i(x_i,y_i)$ and $\cU_j(x_j,y_j)$.
\end{remark}

\subsection{Matching poles and residues}
\label{sec:5.poles}

Henceforth, we shall choose the convenient base point $\yy = \yy_*$ for the gauge transformation $\cU(\xx,\yy)$ of (\ref{7.1.c}), which was defined in (\ref{5.a.1}) by $y_i=y_*$ for all $i = 1, \cdots, n$.  Having matched the $(0,1)$ components on both sides of (\ref{7.1}) with the help of the gauge transformation $\cU(\xx,\yy)$ and the condition $\eta _i = \hat b_i$ from Lemma \ref{5.lem:1}, we may arrange the relation between its $(1,0)$ components  $J^{(1,0)}_i$ in (\ref{defcJ.1}) and $K_i$ in (\ref{7.1.b}) as follows,
\bea
\label{7.2.d.2}
K_i(\xx; a,b,t) = \cU(\xx,\yy_*) ^{-1} J^{(1,0)}_i(\xx; \hat a {-} \xi, \hat b, \hat t) \, \cU(\xx,\yy_*) 
\eea
To establish this equation,  we have used the relation  $\p_i \cU(\xx,\yy) = J_i ^{(1,0)} (\xx; \xi, \eta,0) \, \cU(\xx,\yy) $ with the help of the second relation in (\ref{7.1.d}) of Lemma \ref{5.lem:1}, the relation  $\eta_i = \hat b_i$ of (\ref{7.1.d}),  and the linearity property of the $(1,0)$ components of both connections,
\bea
\label{other72e}
J^{(1,0)}_i(\xx; \hat a, \hat b, \hat t) - J^{(1,0)}_i (\xx; \xi, \eta, 0) = J^{(1,0)}_i(\xx; \hat a {-} \xi, \hat b, \hat t)
\eea
Since $\tau_{ij}=0$,  the gauge transformation $\cU(\xx,\yy_\ast)$ is a smooth function of $\xx \in \tilde \Sigma^n$. The remaining relations needed to  determine the Lie algebra automorphism completely are obtained in the following lemma.

{\lem
\label{5.lem:3}
The map $\phi: (a,b,t) \to (\hat a, \hat b, \hat t)$, for base point $\yy_*$ (see Lemma \ref{5.lem:2} for its relation to arbitrary base point $\yy$), is uniquely determined  by (\ref{7.2.g}), which gives $\hat b_i$ in terms of $b_i$,  together with the following relation giving $\hat t_{ij}$ in terms of  $t_{ij}$,\footnote{Note that, for arbitrary base point $\yy$, the relation becomes $t_{ij} = \cU_i(y_i, y_j) \, \hat t_{ij} \, \cU_i(y_i,  y_j)^{-1}  $ instead.}
\bea
\label{5.lem.3}
\hat t_{ij} = t_{ij} 
\eea
and the relation below giving $\hat a_i$ in terms of $a_i, b_i$ and $\xi_i$  in (\ref{7.2.g1}),
\bea 
\label{mauto.99}
\hat a_i^K{-}\xi_i^K = a_i^K + \sum_{r=1}^{\infty} {\cal M}^{K I_1\cdots I_r}{}_J(y_\ast) 
B_{i I_1} \cdots B_{iI_r} a_i^J \, , \ \ \ \ i=1,\cdots,n
\eea
The coefficients ${\cal M}^{K I_1\cdots I_r}{}_J(y_\ast) $ are those of the single-variable case in (\ref{mauto.13}), evaluated here at the base point $y_*$.}

\sm

To prove (\ref{5.lem.3}), we start by noting that $\hat a_i^K{-}\xi_i^K$ is required to have $a$-degree
one since this is the case for the left side of (\ref{7.2.d.2}) and its right side is linear in $\hat a_i^K{-}\xi_i^K$
(the conjugation by $\cU(\xx,\yy_*) ^{-1}$ merely leads to a series in $B_{iJ}$). We then match the simple pole in $x_i$ and $x_j$ on both sides of (\ref{7.2.d.2}) for all $i \not=j$, and obtain the following relations between their residues,  
\bea
 \label{7.4.a}  
t_{ij} & = & \cU(\xx,\yy_*)^{-1} \hat t _{ij} \, \cU(\xx,\yy_*) \, \Big |_{x_i=x_j} 
\no \\
& = &   \cU_i(x_i, y_*) ^{-1} \, \cU_j(x_i, y_*) ^{-1} \, \hat t_{ij} \, \cU_j(x_i, y_*)  \, \cU_i (x_i, y_*) 
\quad
\eea
where we have used the relations $\big [ \, \cU_k, \hat t_{ij} \big ]=0$  for $k \not= i,j$ to obtain the second equality. Writing out the adjoint action (\ref{7.3.b}) of $\cU_j$ in terms of the generators $\hat b_{iI} = \eta_{iI}$ using (\ref{7.1.d}), as well as the structure relations $\hat B_{jI} \hat t_{ij} = - \hat B_{iI} \hat t_{ij} $ of (\ref{11.lem.2}), we obtain, 
\bea
\label{7.4.b}  
\cU_j(x_i, y_*) ^{-1} \, \hat t_{ij} \, \cU_j (x_i,y_*) 
& = & \hat t_{ij} + \sum_{r=1}^\infty \cT^{I_1 \cdots I_r}(x_i,y_*) \, \hat B_{jI_1} \cdots \hat B_{jI_r} \, \hat t_{ij}  
\no \\ 
& = & \hat t_{ij} + \sum_{r=1}^\infty \cT^{I_1 \cdots I_r}(x_i,y_*) \, (-)^r \hat B_{iI_r} \cdots \hat B_{iI_1} \, \hat t_{ij} 
\eea
The operation carried out on the Lie series in the second line of (\ref{7.4.b}) coincides with  that of the antipode operation $\theta$,
\bea
\theta ( \hat B_{i I_1}  \cdots \hat B_{i I_r}  )  = (-1)^r \hat B_{i I_r} \cdots  \hat B_{i I_1}
\label{defant}
\eea
 which, by Lemma 1.7 of Reutenauer's textbook \cite{Reutenauer}, gives the inverse action of $\cU_i$,  so that we have the following simplification, 
\bea
\label{7.4.c}  
\cU_j(x_i,  y_*) ^{-1} \, \hat t_{ij} \, \cU_j(x_i,y_*)  = \cU_i(x_i, y_*)  \, \hat t_{ij} \, \cU_i(x_i, y_*) ^{-1}
\eea
Combining this result with (\ref{7.4.a}) then completes the proof of (\ref{5.lem.3}).
 
\sm

To prove (\ref{mauto.99}) we make use of the structure relations for both hatted and un-hatted generators (which were imposed as part of the assumptions made to derive the necessary conditions) along with the equality $\hat t_{ij} = t_{ij}$ of (\ref{5.lem.3}), to obtain,
\bea
\label{5.k.99}
[\hat b_{iI}, \hat a_j^J]  = \delta ^J_I \, \hat t_{ij} = \delta ^J_I \, t_{ij} = [b_{iI}, a_j^J] 
\hskip 1in
[\hat b_{iI}, \hat a^I_i ] = [b_{iI} , a^I_i] 
\eea
for all $i,j =1,\cdots, n$ and $j \not= i$. For a fixed value of $i$, the relation on the right in (\ref{5.k.99}) is identical to the corresponding relation in the single-variable case. Accordingly, its solution in (\ref{mauto.99})  inherits the form (\ref{mauto.13}) of the single-variable case and is expressed in terms of the generators $a_i$ and $b_i$ only. 
In the present, multi-variable case, however, it remains to show that the expression in (\ref{mauto.99}) also solves the relations on the left in (\ref{5.k.99}). This result along with the uniqueness of the solution for $\hat a_i$ in (\ref{mauto.99}) is proven in appendix \ref{sec:C1} by means of the generating series,
\bea
\label{C.9}   
{\MM^J_\shuffle \, }{} _I (y) & = & \sum_{r=1}^\infty {\cM_\shuffle ^{J I_1 \cdots I_r}}_I(y) B_{jI_1} \cdots B_{j I_r}
\no \\
\MM^J{}_I(y) & = & \sum_{r=1}^\infty \cM^{J I_1 \cdots I_r}{}_I(y) B_{jI_1} \cdots B_{j I_r} 
\eea
of the coefficients ${\cM_\shuffle ^{J I_1 \cdots I_r}}_I$ and $\cM^{J I_1 \cdots I_r}{}_I $ in (\ref{7.2.g}) and (\ref{mauto.99}), respectively.
This concludes the proof of Lemma \ref{5.lem:3}.

\subsection{Structure relations for $(\xi, \eta, \tau)$ and $(\hat a, \hat b, \hat t)$}
\label{strsec}

In order to complete the proof of the equivalence between the multi-variable connections $\cJ_\text{DHS}$ and $\cK_\text{E}$ by a gauge transformation and an automorphism of $\hat \mt_{h,n}$, it remains to show that the necessary conditions, derived earlier, are in fact sufficient to ensure that the map $\phi: (a,b,t) \to (\hat a, \hat b, \hat t)$ is an automorphism of $\hat \mt_{h,n}$. 

\sm

For this purpose, we will demonstrate in Lemma \ref{5.lem:88a} below that the triplet $(\xi, \eta, \tau)$ defined by Lemma \ref{5.lem:1} and given by (\ref{7.2.g1}) and (\ref{7.2.g}) satisfies the structure relations of~$\hat \mt_{h,n}$. We will further show  in Lemma \ref{5.lem:88b} below that the triplet $(\hat a, \hat b, \hat t)$ defined by (\ref{7.2.g}) and Lemma \ref{5.lem:3} obeys the $\hat \mt_{h,n}$ structure relations of $a$-degree $\leq 1$ in Definition \ref{2.def:1}.  The proof of the $a$-degree two relation in (\ref{11.1}) will be relegated to section \ref{nwsc.57}, where the proof that
$\phi$ is an automorphism of $\hat \mt_{h,n}$ will also be completed.

{\lem
\label{5.lem:88a} The triplet $(\xi, \eta, \tau)$ obeys the structure relations of $\hat \mt_{h,n}$, i.e.
\begin{itemize}
\itemsep=0in
\item[] The map  $(a,b,t) \to (\xi, \eta, \tau)$ is a (non-faithful) homomorphism of the Lie algebra  $\hat \mt_{h,n}$, determined by the relation of (\ref{7.1.d}) together with the explicit expressions for $\xi_i^I$ and $\eta_{iI}=\hat b_{iI}$ in terms of $b_{iK}$ given by (\ref{7.2.g1}) and (\ref{7.2.g}), respectively. 
\end{itemize}}

The triplet $(\xi, \eta, \tau)$ is given in terms of $(a,b,t)$ by Lemma \ref{5.lem:1} and equations (\ref{7.2.g1}) and (\ref{7.2.g}). The structure relations $[b_{iI}, b_{jJ}]=0$ for $i \not= j$ readily imply the following relations,
\bea
\label{5.le.88} 
[\xi _i^I, \xi_j^J]= [\eta_{iI}, \eta _{jJ}]=[\eta_{iI}, \xi ^J_j]=0
\eea
while the cyclic  property of $\cX$ in (\ref{5.cyclic}) implies $[\xi^I_i, \eta_{iI}]=0$. Given that $\tau_{ij}=0$, the triplet $(\xi,\eta,\tau)$ obeys the relations of Definition \ref{2.def:1}  so that the map $(a,b,t) \to (\xi, \eta, \tau)$ is a homomorphism of the Lie algebra $\hat \mt_{h,n}$. This concludes the proof of Lemma \ref{5.lem:88a}.

{\lem
\label{5.lem:88b} The triplet $(\hat a, \hat b, \hat t)$ obeys the following structure relations: the second relation of (\ref{11.1}) as well as the relations of (\ref{11.1a}), (\ref{11.2}), (\ref{11.3}), i.e.\ all the structure relations of $\hat \mt_{h,n}$ in Definition \ref{2.def:1} of $a$-degree 0 and 1 as well as the subset $[\hat a_{i}^I , \hat t_{jk} ]=0$ of $a$-degree 2 relations for $i,j,k$ mutually distinct.
}

\sm

To prove the lemma, we verify that each one of these structure relations is indeed obeyed.
\begin{itemize}
\itemsep=0in
\item The relation $[\hat b_{iI}, \hat b_{jJ}]=0$ for $i \not= j$, i.e.\ the second relation of (\ref{11.1}) for hatted generators,
follows from (\ref{7.2.g}) and the structure relation $[b_{iI}, b_{jJ}]=0$ for the triplet $(a,b,t)$.
\item The relation $[\hat b_{iI}, \hat a_j^J]= \delta ^J_I \, \hat t_{ij}$, i.e.\ (\ref{11.1a}) for the hatted generators, was used in Lemma \ref{5.lem:3} to determine $\hat a_i^I$ in terms of the un-hatted generators and is proven in appendix \ref{sec:C1}.
\item The commutator $[\hat b_{iI}, \hat a^I_i]$ may be computed using the relations $\eta_{iI}=\hat b_{iI}$ and  $[\eta_{iI}, \xi^I_i]=0$ in (\ref{7.1.d}) of Lemma \ref{5.lem:1} in order to equate $[\hat b_{iI}, \hat a^I_i]=[\hat b_{iI}, \hat a^I_i {-}\xi^I_i]$. This relation, in turn, may be evaluated using the expression of (\ref{mauto.99}),
\bea
\label{D.5} 
[\hat b_{iI}, \hat a^I_i] = [\hat b_{iI}, \hat a^I_i -\xi^I_i]
=  [b_{iI}, a^I_i] + \hat B_{iI}  \sum_{s=1}^\infty  \cM^{I J_1 \cdots J_s}{}_K B_{i J_1} \cdots B_{i J_s} a^K_i
\eea
Expressing $ \hat B_{iI}$ as a series in $B_{iL}$ via (\ref{7.2.g}) and using the generating functions (\ref{C.9}),
we may express the commutator as follows,
\bea
\label{D.6} 
[\hat b_{iI}, \hat a^I_i] & = & [b_{iI}, a^I_i] + B_{iI} \Big ( \MM^I{}_K - { \MM_\shuffle ^I}_K - {\MM_\shuffle^I}_J  \, \MM^J{}_K \Big ) a_i^K
\eea
The combination of $\MM$ matrices can be seen to vanish by 
translating equation (B.2) of \cite{DHoker:2024desz} into the language of (\ref{C.9}),
\bea
\label{C.11}   
 {\MM^J_\shuffle }_I  = \MM^J{}_I  - {\MM^J_\shuffle \, }_K  \MM^K {}_I \, 
 \eea 
so that $[\hat b_{iI}, \hat a^I_i] = [ b_{iI}, a^I_i]$. Along with  the relations $t_{ij}= \hat t_{ij}$, this equality proves the structure relation (\ref{11.2}) for the hatted generators.
\item The commutators $[\hat a_{i}^I , \hat t_{jk} ]=[\hat b_{iI} , \hat t_{jk} ]=0$, for $i,j,k$ mutually distinct, follow from taking the commutator of (\ref{mauto.99}) and (\ref{7.2.g}), respectively,  with $\hat t_{jk} = t_{jk}$  and using the structure relations of (\ref{11.3}) for the triplet $(a,b,t)$, which proves the structure relation (\ref{11.3}) for hatted generators and completes the proof of the lemma. 
\end{itemize}

\subsection{Proof of equivalence via the relation (\ref{7.1})}
\label{sec:equiv}

Thus far, we have derived the necessary conditions for the validity of the equivalence relation (\ref{7.1}) between the connections $\cK_\text{E}$ and $\cJ_\text{DHS}$ in Lemmas \ref{5.lem:1} and \ref{5.lem:3}, shown that the map $(a,b,t) \to (\xi, \eta, \tau)$ is a homomorphism of the Lie algebra $\hat \mt_{h,n}$ and demonstrated that the map $(a,b,t) \to (\hat a, \hat b, \hat t)$ preserves the defining structure relations of $\hat \mt_{h,n}$ at $a$ degree 0 and~1.
These properties of the maps $(a,b,t) \to (\xi, \eta, \tau)$ and $(a,b,t) \to (\hat a, \hat b, \hat t)$ suffice to 
prove the existence of the relation (\ref{7.1}), which we shall do 
in this subsection by showing that the combination of these necessary conditions is also sufficient.

{\thm
\label{5.thm:1}
Given the connection $\cK_{\rm E}(\xx;a,b,t)$ for a triplet of generators $(a,b,t)$ of $\mt_{h,n}$, there exists a gauge transformation $\cU$ and a map  $\phi: (a,b,t) \to (\hat a, \hat b, \hat t)$ such that the relation between the connections $\cK_{\rm E}$ and $\cJ_{\rm DHS}$ in (\ref{7.1}) holds. The gauge transformation is given in terms of the triplet $(a,b,t)$ by equations (\ref{7.1.c}), (\ref{7.2.g1}),  (\ref{7.2.g}) and (\ref{7.1.d}), while the map $\phi$
is given by equations (\ref{mauto.99}), (\ref{7.2.g}), and  (\ref{5.lem.3}). Both are unique 
modulo a change in base point $\yy$, as spelled out in Lemma \ref{5.lem:2}.}

\sm

Without loss of generality, we choose the base point $\yy= \yy_*$ in the remainder of this section, as the case of arbitrary $\yy$ may be obtained by applying the automorphism of Lemma~\ref{5.lem:2}.

\sm

To prove the theorem, we begin by recalling that the $(0,1)$ component of $\cK_\text{E}$ vanishes in view of (\ref{7.2.d}), so that the equation for the $(0,1)$ component  of (\ref{7.1}) simplifies as follows, 
\bea
0 = \cU(\xx, \yy_\ast)^{-1} \big ( \bar \p_i - J^{(0,1)}_i (\xx;\hat a, \hat b, \hat t) \big ) \cU(\xx, \yy_\ast)
\eea
In view of the expression for $J_i^{(0,1)}$ given in (\ref{defcJ.2}) and the fact that $\cU(\xx,\yy_*)$ is given by (\ref{7.1.c}) the above equation is solved by $\eta_{iI} = \hat b_{iI}$ of (\ref{7.1.d}).

\sm

Henceforth, we may concentrate on the $(1,0)$ component of (\ref{7.1}), which we express as the vanishing of the difference $\cN$ between the two sides of the equation,
\bea
\label{7.h.1}
\cN = \sum_{i=1}^n dx_i \, \cN_i 
\eea
whose  components $\cN_i$ are given by, 
\bea
\label{7.h.1b}
\cN_i  =   \cU(\xx,\yy_\ast) ^{-1} J_i^{(1,0)}  (\xx; \hat a , \hat b, \hat t) \, \cU(\xx, \yy_\ast) 
- \cU(\xx,\yy_\ast) ^{-1} \p_i \, \cU(\xx, \yy_\ast) - K_i (\xx; a,b,t) 
\eea
Using (\ref{7.1.c}) to evaluate the partial derivatives of $\cU$, we find, 
\bea
\label{7.h.1b}
\cN_i  =   \cU(\xx,\yy_\ast) ^{-1} \Big ( J_i^{(1,0)}  (\xx; \hat a , \hat b, \hat t) -
J_i^{(1,0)}  (\xx; \xi , \eta, \tau) \Big ) \, \cU(\xx, \yy_\ast) 
- K_i (\xx; a,b,t) 
\eea
Finally using $\tau=0$ of (\ref{7.1.d}) and (\ref{other72e}), the expression for $\cN_i$ simplifies as follows, 
\bea
\label{7.h.1a}
\cN_i  =   \cU(\xx,\yy_\ast) ^{-1} J_i^{(1,0)}  (\xx; \hat a {-} \xi, \hat b, \hat t) \, \cU(\xx, \yy_\ast) - K_i (\xx; a,b,t) 
\eea
Expressing $\cN_i$ in terms of the generating functions of (\ref{defcJ.3}) and (\ref{7.1.b}),  
\bea
\label{7.5.a} 
\cN_i & = & \cU^{-1}  \bigg ( {\bf J}_I(x_i,z; H_i) \big ( \hat a_i^I - \xi_i^I \big ) 
+ \sum_{j \neq i} \Big \{  {\bf G}(x_i,x_j; H_i) - {\bf G}(x_i,z; H_i)  \Big \} \hat t_{ij}  \bigg ) \, \cU
\no \\ &&
- {\bf K}_I  (x_i,z;  B_i) a_i^I  - \sum_{j \neq i} \Big \{  {\bf X}(x_i,x_j;B_i) - {\bf X}(x_i,z;B_i)  \Big \}  t_{ij}  
\eea
where the dependence on the point $z$ drops out by the $\hat \mt_{h,n}$ structure relations (\ref{11.2})
of both the hatted and the un-hatted generators.  By construction, the monodromies around $\mA$ and $\mB$ cycles are those of $\bK_J$ while the monodromy when a point $x_i$ is moved around another point $x_j$ is trivial. Our proof will consist in showing that $\cN_i=0$ for any $i=1,\cdots,n$. 

\sm

In addition to the monodromy properties  of $\cN$, the proof of $\cN=0$ will require that the form $\cN$ is closed on $\tilde \Sigma ^n$, which is the subject of Lemma \ref{3.lem:9} below. The proof will also require a special reduced form of $\cN_i$, which is stated in Lemma \ref{7.lem:666} below.

{\lem 
\label{3.lem:9}
The 1-form $\cN$ is closed or, equivalently, its components $\cN_i$ in (\ref{7.h.1a}) satisfy, 
\begin{align}
\bar \p_i \, \cN_i  = 0 \hskip 1in 
\bar \p_j \, \cN_i  =  0 \hskip 1in
\p_j \, \cN_i - \p_i \, \cN_j  =0 
\label{more7.1}
\end{align}
for any $j\neq i$, with $\hat b_{iI}$ and $\hat a_i^I$ in (\ref{7.2.g}) and (\ref{mauto.99}) subject to $\big [ \hat a_i^I, \hat b_{iI} \big ] = \big [ a^I_i, b_{iI}  \big]$}.

\sm

The lemma is proven in appendix \ref{sec:C}. 

{\lem 
\label{7.lem:666}
Each component $\cN_i$ may be reduced to the following sum,
\bea
\cN_i= \sum _{j \not = i} \cN_{ij} (x_i,x_j; B_i) t_{ij}
\label{more7.2}
\eea
where $\cN_{ij} (x_i, x_j ; B_i) $ depends only on $x_i$, $x_j$ and the generators $B_{iI}$ for $I=1,\cdots, h$. It is a $(1,0)$ form in $x_i$, a scalar in $x_j$ and is independent of $x_k$ for all $k \not= i,j$.}

\sm

The lemma is proven in appendix \ref{sec:D}. 

\sm

\subsubsection{Completing the proof of $\cN=0$}

Assuming the validity of Lemmas \ref{3.lem:9} and \ref{7.lem:666}, the proof of the vanishing of $\cN$
in (\ref{7.h.1}) to (\ref{7.5.a}) proceeds by  showing that  $\cN_{ij}=0$ for all $i \not= j$. To do so, we use the fact that $\cN_i$ and thus $\cN_{ij}$ is a holomorphic $(1,0)$ form in $x_i$, admits a formal Taylor series expansion in the generators $b_{iI}$, and has the monodromies of $K_i$ in the variable $x_i$, namely trivial $\mA$ monodromies and $\mB$ monodromies given by,   
\bea
\gamma_K^{(i)} \cN_{ij} = e^{-2\pi i b_{i K}} \cN_{ij} \, e^{2\pi i b_{i K}} 
\label{more7.3}
\eea
The proof proceeds by contradiction and is similar to the proof of Lemma \ref{4.lem:100}, but differs in the use of the $\mA$ periods. Here, we shall assume that $\cN_{ij} \not =0$ so that there exists a non-vanishing term of lowest degree $m$ in the expansion of $\cN_{ij}$ in powers of $b_{iI}$. The monodromy relation (\ref{more7.3})  tells us that this lowest-degree term must be single-valued since the non-trivial effects of the monodromy are of higher order in $b_{iI}$. Combining this with holomorphicity implies that $\cN_{ij}$ must take the following form,
\bea
\cN_{ij} (x_i, x_j; B_i) = \om_K(x_i) \, \hat \cN^{K| I_1 \cdots I_m} (x_j) \, B_{i I_1} \cdots B_{i I_m} + \cO(B_i^{m+1})
\label{more7.4}
\eea 
where $m\geq 0$ and $\hat \cN^{K| I_1 \cdots I_m} (x_j)$ is non-vanishing and depends only on the variable $x_j$. 

\sm

Now evaluate the line integral in the variable $x_i$ around the boundary $\p D_p$ of the preferred fundamental domain $D_p$ of $\Sigma$ (see footnote \ref{foot1} for the definition of $D_p$), keeping the variable $x_j$ fixed in the process. Since $d_{x_i} \, \cN=0$, we have by Stokes's  theorem, 
\bea
0 = \int _{D_p} d_{x_i} \,\cN  & = & \sum_{j \not = i} \oint _{\p D_p} dx_i \, \cN_{ij}(x_i,x_j; B_i) t_{ij}  
\label{more7.5}
\eea
Decomposing the boundary $\p D_p$ into combinations $\bigcup_{K=1}^h \mA^K \mB_K(\mA^K)^{-1} \mB_K^{-1}$ of homotopy curves, the contour integral reduces to $\mA^K$ integrals of $\mB_K$ monodromies and $\mB_K$ integrals of (vanishing) $\mA^K$ monodromies.
Since the families of generators of the Lie algebra obtained by applying strings of $B_i$ to $t_{ij}$ for different values of $j$ are independent of one another, we may drop the $t_{ij}$ factor in the proof and obtain,
\bea  
\sum_{L=1} ^h \oint _{\mA^L} dx_i \Big ( \cN_{ij} (x_i, x_j; B_i) - \gamma _L^{(i)} \cN_{ij} (x_i, x_j; B_i) \Big ) =0 
\label{more7.6}
\eea
for any $j\neq i$. Evaluating the monodromy and its $\mA^L$ period, we obtain, 
\bea
\sum_{L=1}^h \Big ( 1 - e^{-2 \pi i B_{iL} } \Big ) \Big ( \hat \cN^{L| I_1 \cdots I_m} (x_j) B_{iI_1} \cdots B_{iI_m} + \cO(B_i^{m+1}) \Big ) =0
\label{more7.7}
\eea
Expanding to order $m{+}1$ in $B_i$ this gives,
\bea
2 \pi i  \hat \cN^{L| I_1 \cdots I_m} (x_j) B_{iL} B_{iI_1} \cdots B_{iI_m} + \cO(B_i^{m+2})  =0
\label{more7.8}
\eea
Hence we must have $\hat \cN^{L| I_1 \cdots I_m} (x_j) =0$ since the Lie subalgebra of the elements $b_{iI}$ is freely generated.  But this result contradicts our assumption (\ref{more7.4}) in the proof by contradiction, namely that $\cN_{ij} (x_i, x_j; B_i) $ has a first non-zero contribution at some $B$-degree $m\geq 0$. Hence $\cN_{ij}=0$ and thus $\cN_i$ and $\cN$ must vanish, thereby completing the proof of Theorem \ref{5.thm:1}.

\subsection{The map $(a,b,t) \to (\hat a, \hat b, \hat t)$ is a Lie algebra automorphism}
\label{nwsc.57}

In  contrast to the proof of the structure relations for $(\hat a, \hat b, \hat t)$ of $a$-degree zero and one in Lemma \ref{5.lem:88b}, the proof of the structure relation $[\hat a^I_i, \hat a^J_j]=0$ will proceed indirectly.

{\thm
\label{5.thm:35}
The map $\phi: (a,b,t) \to (\hat a, \hat b, \hat t)$, defined by the relations (\ref{mauto.99}), (\ref{7.2.g1}), (\ref{7.2.g}) and (\ref{5.lem.3}) is an automorphism of the  Lie algebra $\hat \mt_{h,n}$.}

\sm

To prove the theorem we need to verify that, if the triplet $(a,b,t)$ satisfies the structure relations of $\mt_{h,n}$ in Definition \ref{2.def:1}, then the triplet $(\hat a, \hat b , \hat t)$ defined by the relations (\ref{mauto.99}), (\ref{7.2.g1}), (\ref{7.2.g}) and (\ref{5.lem.3}) satisfies the structure relations of $\mt_{h,n}$ in Definition \ref{2.def:1} for hatted generators.

\sm

For all structure relations of $a$-degree zero and one, namely the second relation in (\ref{11.1}) and all the relations in (\ref{11.1a}), (\ref{11.2}) and (\ref{11.3}), this property was already proven in Lemma \ref{5.lem:88b}. Therefore, it only remains to prove that the commutators $[\hat a_i^I, \hat a_j^J]$ vanish for any $i\neq j$ in order  to complete the proof of Theorem \ref{5.thm:35}.

\sm

To prove the structure relation $[\hat a_i^I, \hat a_j^J]=0$, we exploit the relation (\ref{7.1}) between the connections $\cK_\text{E}(\xx;a,b,t)$ and $\cJ_\text{DHS}(\xx;\hat a, \hat b, \hat t)$ established in Theorem \ref{5.thm:1}. In particular, flatness of $\cK_\text{E}(\xx;a,b,t)$ for $\xx \in {\rm Cf}_n(\tilde \Sigma)$ (which was proven originally  in \cite{Enriquez:2011} and proven here by different methods in Theorem \ref{7.thm:10})  implies flatness of $\cJ_\text{DHS}(\xx;\hat a, \hat b, \hat t)$ for $\xx \in {\rm Cf}_n( \Sigma)$. Importantly, the proof of Theorem \ref{5.thm:1} made use of structure relations of the triplet $(\hat a, \hat b, \hat t)$ only for $a$-degree zero and one, but did not make use of the relation $[\hat a_i^I, \hat a_j^J]=0$.

\sm

As detailed in section \ref{sec:2.4prf}, the flatness conditions on $\cJ_\text{DHS}$ imply the vanishing of the commutators $[J^{(1,0)} _i (\xx;\hat a, \hat b, \hat t), J^{(1,0)} _j (\xx;\hat a, \hat b, \hat t)]$.  Integrating these commutators against arbitrary anti-holomorphic $(0,1)$ forms in $x_i$ and $x_j$ gives the combination defined in (\ref{B.2}) but this time for the triplet $(\hat a, \hat b, \hat t)$,
\bea
\label{D.7} 
 \hat \cS_{ij}^{KL} = 
 \int _\Sigma d^2 x_i \, \bar \om^K(x_i) \int _\Sigma d^2 x_j \, \bar \om^L(x_j) \, \big [ J^{(1,0)}_i(\xx; \hat a, \hat b, \hat t) , J^{(1,0)}_j (\xx; \hat a , \hat b, \hat t) \big ] 
\eea
The vanishing of the commutator requires this quantity to vanish. Its explicit form was given in (\ref{B.2}) for the triplet $(a,b,t)$ and is easily transcribed for the triplet $(\hat a, \hat b, \hat t)$ as follows,
\bea
\label{D.8}
\hat \cS_{ij}^{KL} & = &  
\int _\Sigma d^2 x_i \, \bar \om^K(x_i)  \int _\Sigma d^2 x_j \, \bar \om^L(x_j)  \Big [ \bPhi _I (x_i; \hat B_i) \, \hat a^I_i, \bPhi _J (x_j; \hat B_j) \, \hat a^J_j \Big ]
 \\ &&
+ \int _\Sigma d^2 x_i \, \bar \om^K(x_i)  \int _\Sigma d^2 x_j \, \bar \om^L(x_j) 
\bigg \{ \Big [ \bPhi _I (x_i; \hat B_i) \, \hat a^I_i,  \sum_{k \not = j} \bG (x_j,x_k; \hat B_j) \, \hat t_{jk} \Big ]  - ( i \leftrightarrow j) \bigg \}
\no \\ &&
+ \int _\Sigma d^2 x_i \, \bar \om^K(x_i)  \int _\Sigma d^2 x_j \, \bar \om^L(x_j) 
\bigg [  \sum_{k \not = i} \bG (x_i,x_k; \hat B_i) \, \hat t_{ik} , \sum_{\ell \not = j} \bG (x_j,x_\ell; \hat B_j) \, \hat t_{j\ell} \bigg ] 
\no
\eea
and integrated with the help of the formulas of (\ref{B.4}). As in the treatment of (\ref{B.2}), all the integrals can be carried out right away, except for the contribution to the last line from $k=j$ and $\ell=i$, and we obtain the following simplified expression, 
\bea
\label{D.9}
\hat \cS_{ij} ^{KL}
=  [\hat a^I_i , \hat a^J_j] + \int _\Sigma d^2 x_i \, \bar \om^K(x_i)  \, \int _\Sigma d^2 x_j \, \bar \om^L(x_j)
\Big [  \bG (x_i,x_j; \hat B_i) \, \hat t_{ij} ,  \bG (x_j,x_i; \hat B_j) \, \hat t_{ij} \Big ] 
\eea
It was shown in appendix \ref{sec:B} that the double integral in $\cS_{ij}^{KL}$ vanishes with the help of the structure relations that involve only $t_{ij}$, $b_i $ and $b_j$. Transcribing the generators from $t_{ij}$, $b_i $ and $b_j$ to $\hat t_{ij}$, $\hat b_i$ and $\hat b_j$ and using the fact that the hatted generators satisfy the same structure relations of $a$-degree $\leq 1$ as the un-hatted ones by Lemma \ref{5.lem:88b}, it follows that the double integral in $\hat \cS_{ij}^{KL}$ of (\ref{D.9}) vanishes as well. Therefore, the vanishing of the commutator  $[J^{(1,0)} _i (\xx;\hat a, \hat b, \hat t), J^{(1,0)} _j (\xx;\hat a, \hat b, \hat t)]$ implies the vanishing of $\hat \cS_{ij}^{KL}$ which in turn implies the structure relation $[\hat a^I_i , \hat a^J_j]=0$. Together with the remaining $\hat \mt_{h,n}$ structure relations proven in Lemma \ref{5.lem:88b}, this completes our proof of Theorem \ref{5.thm:35}.

\sm

With the results of Theorems \ref{5.thm:1} and \ref{5.thm:35} in place, we arrive at the following 
corollary which is one of the main results of this work:
\begin{cor}
\label{exuniq}
There exists a gauge transformation $\cU$ and an automorphism $\phi: (a,b,t) \to (\hat a, \hat b, \hat t)$
of the Lie algebra $\hat \mt_{h,n}$ relating the connections $\cK_{\rm E}$ and $\cJ_{\rm DHS}$ as in (\ref{7.1}). The pair $(\cU,\phi)$ is unique up to a change in base point $\yy$ of ${\cal U}$, as spelled out in Lemma \ref{5.lem:2}.
The explicit form of ${\cal U}$ and $\phi$ can be obtained from (\ref{7.1.c}), (\ref{7.1.d}),
 (\ref{7.2.g1}),  (\ref{7.2.g}) and (\ref{7.2.g}), (\ref{5.lem.3}), (\ref{mauto.99}), respectively.
\end{cor}
The proof of the corollary is mostly based on Theorem \ref{5.thm:1} where the statement of the corollary 
is made while referring to $\phi$ as a map, without claiming it to be an isomorphism of~$\hat \mt_{h,n}$. 
The isomorphism property of $\phi$ is established in Theorem \ref{5.thm:35} which completes the proof of the corollary.

\subsection{Relations amongst $\cM$ tensors induced by the automorphism}

In view of Theorem \ref{5.thm:35}, the map $\phi: (a,b,t) \to (\hat a , \hat b, \hat t)$ is a Lie algebra automorphism of~$\hat \mt_{h,n}$, namely the triplet $(\hat a, \hat b, \hat t)$ satisfies the structure relations of $\mt_{h,n}$. Some of these structure relations imply new and non-trivial relations between the coefficients $\cM$ that occur in the Lie series expansion of $\hat a$  in terms of $b$ and $a$ given in (\ref{mauto.99}). The relations fall into two categories, linear and quadratic in $\cM$, which follow from the expansion in $B_{iI}$ of the $a$-degree two terms in the $\hat \mt_{h,n}$ relations $[\hat t_{ij}, \hat a^K_i + \hat a^K_j ] = 0$ and  $[\hat a^I_i , \hat a^J_j] = 0$, respectively. Since the goal of this is section is merely to give a flavor of the relations among $\cM$, we omit the combinatorially more involved quadratic relations and only discuss linear relations in detail.\footnote{Given that the quadratic relations among ${\cal M}$ are generated by the only fundamental $\hat \mt_{h,n}$ relation $[\hat a^I_i , \hat a^J_j] = 0$ of Definition \ref{2.def:1} which is not yet settled by Lemma \ref{5.lem:88b}, we do not expect any independent cubic or higher-order relations among ${\cal M}$. Their linear relations in turn are generated by the corollary $[\hat t_{ij}, \hat a^K_i + \hat a^K_j ] = 0$ of Definition \ref{2.def:1} and thus follow from the quadratic relations. An investigation of whether the linear relations conversely imply all quadratic relations is beyond the scope of this work.}
The result for the linear relations may be summarized in the following proposition.

{\prop
\label{5.prop:77}
The combination of the automorphism $\phi$ and the structure relation $[\hat t_{ij}, \hat a^K_i + \hat a^K_j]=0$ implies the following relations between the expansion coefficients $\cM$ of $\hat a_i^K {-} \xi^K_i$ in terms of $B_{iL}$ given in (\ref{mauto.99}), 
\bea
\label{eqprop:77}
\sum_{s=0}^r (-)^s \Big ( \cM ^{K( I_1 \cdots I_s \, \shuffle \, I_r \cdots I_{s+1} L)}{}_L 
+ \cM ^{K( I_1 \cdots I_s L \,  \shuffle \, I_r \cdots I_{s+1} )}{}_L \Big ) =0
\eea 
where the index $L$  is to be summed following the standard Einstein convention.}

\sm

We shall use vector notation for the indices and write  $\vI = (I_1 \cdots I_r)$ and $B_{i \vI} = B_{iI_1} \cdots B_{iI_r}$ in order  to shorten the notations in this subsection. To prove the proposition, we evaluate the commutator $[\hat t_{ij}, \hat a^K_i + \hat a^K_j]$ by means of the expansion of $\hat a_i^K$ in (\ref{mauto.99}) along with $\hat t_{ij} = t_{ij}$ and $[ \hat t_{ij}, \xi^K_i + \xi_j^K]=0$, 
\bea
\big [\hat t_{ij}, \hat a^K_i + \hat a^K_j \big ] & = & 
\sum _{\vI} \cM^{K \vI }{}_L \big [ t_{ij}, B_{i \vI} \, a^L_i + B_{j \vI} \, a^L_j  \big ]
\label{hatthata}
\eea
Throughout, the notation $\sum_{\vI}$ refers to a sum over words $I_1\cdots I_r$ of arbitrary length $r\geq 0$.
We shall now make use of a combinatorial Lie algebra relation proven in appendix E of \cite{DS-Fay},
\bea
\big [ t_{ij}, B_{i \vI} \, a^L_i + B_{j \vI} \, a^L_j  \big ] 
= \sum _{\vI = \vP M \vQ, \, \vQ \not= \emptyset} \delta ^L_M \sum_{\vP = \vX \, \shuffle \, \vY} 
\Big [ B_{i \vX} \big ( B_{i \vQ} + B_{i \theta(\vQ)} \big ) B_{i\theta( \vY)}  t_{ij} , t_{ij} \Big ]
\eea
where $\theta$ is the antipode whose action was defined in (\ref{defant}) and $\sum_{\vI = \vP M \vQ}$ refers to a sum over all deconcatenations of $\vec{I}$ into words $\vP,\vQ$ with a single letter $M$ in between. This identity
allows us to recast the commutator (\ref{hatthata}) as follows, 
\bea
\big [\hat t_{ij}, \hat a^K_i + \hat a^K_j \big ] & = &
\sum_{\vQ, \vX, \vY, \, \vQ\not= \emptyset} \cM^{K (\vX \, \shuffle \, \vY)  L \vQ}{}_L
\Big [ B_{i \vX} \big ( B_{i \vQ} + B_{i \theta (\vQ)} \big ) B_{i \theta ( \vY)} \, t_{ij} , t_{ij} \Big ]
\eea
Using the summation formula (see for instance appendix A.2 of \cite{DS-Fay}),
\bea
\sum_{\vI = \vX \vQ, \, \vQ \not = \emptyset} (\vX \shuffle \vY) L \vQ = \vY L \shuffle \vI - ( \vY \shuffle \vI) L
\eea
we obtain,
\bea
\big [\hat t_{ij}, \hat a^K_i + \hat a^K_j \big ] & = &
\sum_{\vI, \vY} \cM^{K ( \vY L \, \shuffle \, \vI - (\vY \, \shuffle \, \vI) L)}{}_L 
\Big [ B_{i \vI} B_{i \theta (\vY)} \, t_{ij} + B_{i \vY} B_{i \theta (\vI)} \, t_{ij}, t_{ij} \Big ]
\no \\ & = &
\sum_{\vI, \vY} \cM^{K ( \vY L \, \shuffle \, \vI - (\vY \, \shuffle \, \vI) L)}{}_L 
\Big [ B_{i \vI} B_{j \vY} \, t_{ij} + B_{i \vY} B_{j \vI} \,t_{ij}, t_{ij} \Big ]
\eea
From the second line above, it is clear that the contribution from $(\vY \shuffle \vI)$ cancels since $(B_{iI} + B_{jI}) t_{ij}=0$. Combining the remaining  contributions in the form of the first line above by letting $\vI \to \vX$ in the first term and $\vY \to \vX$ and $\vI \to \vY$ in the second term, we obtain, 
 \bea
\big [\hat t_{ij}, \hat a^K_i + \hat a^K_j \big ] & = &
\sum_{\vX, \vY} \Big ( \cM^{K ( \vX \, \shuffle \, \theta (\vY) L) }{}_L 
+ \cM^{K (  \vX L \, \shuffle \, \theta (\vY) )}{}_L  \Big ) 
\Big [ B_{i \vX \vY} t_{ij}, t_{ij} \Big ]
\eea
Since the elements $[B_{i\vI} \, t_{ij}, t_{ij}]$ for different words $\vI$ are linearly  independent, we may identify the coefficients which, in components, gives (\ref{eqprop:77}) and proves Proposition \ref{5.prop:77}. 

\sm

Note that the simplest non-trivial instances of (\ref{eqprop:77}) occur for $r=2,3$ where we find,
\begin{align}
0 &= \cM^{K L I_1 I_2}{}_L + \cM^{K L I_2 I_1}{}_L 
\\
0 &= \cM^{K L I_1 I_2 I_3}{}_L - \cM^{K L I_3 I_2 I_1}{}_L  
- \cM^{K I_3 L ( I_1 \shuffle I_2 ) }{}_L
+ \cM^{K I_1 L ( I_2 \shuffle I_3 ) }{}_L
\notag
\end{align}

\newpage

\appendix

\section{Proof of Lemma \ref{2.lem:3}}
\label{sec:A}

To prove equation (\ref{3.2}) of Lemma \ref{2.lem:3}, we evaluate the difference between its left and right sides and use the Jacobi identity to simplify the outcome as follows,
\bea
\label{A.1}
\bar \p_i \big [ J^{(1,0)}_i, J^{(1,0)}_j \big ]  - \Big [ J^{(0,1)} _i, \big [ J^{(1,0)}_i, J^{(1,0)}_j \big ]  \Big ]
& = &
\Big [ \bar \p_i J_i^{(1,0)} - [ J^{(0,1)} _i, J^{(1,0)}_i], J_j ^{(1,0)} \Big ]
\no \\ &&
- \Big [ \bar \p_i J_j^{(1,0)} - [ J^{(0,1)} _i, J^{(1,0)}_j], J_i ^{(1,0)} \Big ]
\qquad
\eea
The covariant derivatives $\bar \partial_i X - [ J^{(0,1)} _i, X]$ inside the commutators on the right side may be evaluated for the first term by setting $X =J^{(1,0)}_i$ and using (\ref{2.derJi}) and for the second term by setting $X=J^{(1,0)}_j$ and using (\ref{2.derJj}) to obtain,
\bea
\label{A.1a}
\bar \p_i \big [ J^{(1,0)}_i, J^{(1,0)}_j \big ]  - \Big [ J^{(0,1)} _i, \big [ J^{(1,0)}_i, J^{(1,0)}_j \big ]  \Big ]
& = &
- \pi \sum_{k \not= i} \delta (x_i,x_k) \big [ t_{ik}, J_j^{(1,0)} \big ]
\no \\ &&
+ \pi \delta (x_i,x_j) \big [ J_i^{(1,0)} , t_{ij} \big ]
\eea
Next, we use the following relations for $i \not = j$,
\bea
\label{A.1b}
\sum_{k \not= i,j} \delta (x_i,x_k) \big [ t_{ik}, J_j^{(1,0)} \big ]
& = & \sum_{k \not= i,j} \delta(x_i,x_k) \Big [ t_{ik}, \bPhi_K (x_j;B_j) a_j^K + \sum _{\ell \not=j} \bG (x_j, x_\ell;B_j) t_{j \ell} \Big ]
\no \\ & = &
\sum_{k \not= i,j} \delta(x_i,x_k)  \bG (x_j, x_i;B_j) \big [ t_{ik},  t_{ij}  + t_{jk} \big ]
\eea
where in going from the first line on the right side to the second line, we have used the relations $[t_{ik}, a_j^K]= [t_{ik}, b_{jK}]=0$ to cancel the first term inside the commutator, and $[t_{ik}, t_{j\ell}]=0$ for $\ell \not= i,k$ to cancel the corresponding contributions from the second term. In the remaining terms, $x_i=x_k$ thanks to the $\delta$-function, and the Green function may be pulled out from the commutator using again $[t_{ik}, b_{jK}]=0$. The last line of (\ref{A.1b}) cancels in view of the relation $[t_{ik}, t_{ij}+ t_{jk}]=0$. Using this result in (\ref{A.1a}), we obtain,
\bea
\label{A.2}
\bar \p_i \big [ J^{(1,0)}_i, J^{(1,0)}_j \big ]  - \Big [ J^{(0,1)} _i, \big [ J^{(1,0)}_i, J^{(1,0)}_j \big ]  \Big ]
= 
- \pi \delta (x_i, x_j) \, J^{(1,0)}_{ij} 
\eea
where $J^{(1,0)}_{ij}$ is defined by the following limit,
\bea
\label{A.3}
J^{(1,0)} _{ij} =  \lim_{x_j \to x_i} \Big [ t_{ij} , \, J^{(1,0)}_i +  J^{(1,0)}_j \Big ]
\eea 
The definition may be rendered more explicit by expressing $J^{(1,0)} _{ij} $ in terms of generating functions using (\ref{2.gen.1}) to obtain, 
\bea
\label{A.4}
J^{(1,0)} _{ij} & = & \lim _{x_j \to x_i} \Big [ t_{ij}, \, \bPhi _J(x_i;B_i) a^J_i + \bPhi _J(x_j;B_j) a^J_j
\no \\ && \qquad ~
+ \sum _{k \not= i} \bG(x_i,x_k;B_i) t_{ik} + \sum _{k \not= j} \bG(x_j,x_k;B_j) t_{jk}  \Big ]
\eea
The contributions in the second line of (\ref{A.4}) for which $k \not= i,j$ admit regular limits as $x_j \to x_i$ while keeping the other points fixed,  and  cancel one another. This may be established by converting the generators $B_i$ and $B_j$ into $B_k$ using (\ref{11.lem.2}) and then pulling the common factor $\theta ( \bG(x_i,x_k;B_k) )$ out of the commutator, where $\theta ( \hat B_{i I_1}  \cdots \hat B_{i I_r}  )  = (-1)^r \hat B_{i I_r} \cdots  \hat B_{i I_1}$ is the antipode, see (\ref{defant}). The remaining commutator $[t_{ij}, t_{ik}{+}t_{jk}]$ vanishes by (\ref{11.lem.3}). Thus, we are left with,
\bea
\label{A.5}
J^{(1,0)} _{ij} & = & \lim _{x_j \to x_i} \Big [ t_{ij}, \, \bPhi _J(x_i;B_i) a^J_i + \bPhi _J(x_j;B_j) a^J_j
\no \\ && \qquad ~
+ \bG(x_i,x_j;B_i)  t_{ij}  +  \bG(x_j,x_i;B_j)  t_{ij}   \Big ]
\qquad
\eea  
To take the limit carefully and to obtain a more explicit formula, we expand in powers of $B_i$ and $B_j$. 
Using the fact that the terms of degree zero in $B$ cancel in the limit, thanks to the relation $[t_{ij}, a^J_i {+} a^J_j]=0$ of (\ref{11.lem.1}), we find,
\bea
\label{A.6}
J^{(1,0)} _{ij} & = & \sum_{r=1}^\infty \bigg \{ \p_i \Phi ^{I_1 \cdots I_r}{}_J (x_i) 
\Big [ t_{ij}, \, B_{iI_1} \cdots B_{iI_r}  a^J_i + B_{jI_1} \cdots B_{jI_r}  a^J_j \Big ] 
\no \\ && \qquad
+ \lim _{x_j \to x_i}   \p_i \cG^{I_1 \cdots I_r} (x_i, x_j)  \Big [ t_{ij}, \, B_{iI_1} \cdots B_{iI_r} t_{ij} \Big ]
\no \\ && \qquad
+ \lim _{x_j \to x_i}  \p_j \cG^{I_1 \cdots I_r} (x_j, x_i) \Big [ t_{ij}, \, B_{jI_1} \cdots B_{jI_r} t_{ij} \Big ] \bigg \}
\eea
Using (\ref{11.lem.2}) and the second relation of (\ref{2.a.8}) in the following forms,
\bea
\label{A.7}
\p_j \cG^{I_1 \cdots I_r} (x_j, x_i) & = & (-)^r
 \p_j \cG^{I_r \cdots I_1} (x_i, x_j)
\no \\
B_{jI_1} \cdots B_{jI_r} t_{ij} & = & (-)^r B_{iI_r} \cdots B_{iI_1} t_{ij}
\eea
and suitably relabeling  the summation indices $I$, we obtain,
\bea
\label{A.8}
J^{(1,0)} _{ij} & = & \sum_{r=1}^\infty \bigg \{ \p_i \Phi ^{I_1 \cdots I_r}{}_J (x_i) 
\Big [ t_{ij}, \, B_{iI_1} \cdots B_{iI_r}  a^J_i + B_{jI_1} \cdots B_{jI_r}  a^J_j \Big ] 
 \\ && \qquad
+ \lim _{x_j \to x_i} \Big (  \p_i \cG^{I_1 \cdots I_r} (x_i, x_j) +  \p_j \cG^{I_1 \cdots I_r} (x_i, x_j) \Big )
\Big [ t_{ij}, \, B_{iI_1} \cdots B_{iI_r} t_{ij} \Big ] \bigg \}
\no
\eea
For $r \geq 2$, the limit on the second line is continuous and combines using the relation,
\bea
\label{A.9}
 \lim _{x_j \to x_i} \Big (  \p_i \cG^{I_1 \cdots I_r} (x_i, x_j) +   \p_j \cG^{I_1 \cdots I_r} (x_i, x_j) \Big ) 
 = \p_i \cG^{I_1 \cdots I_r} (x_i, x_i)
 \eea
 For $r=1$, the limit fails to exist for each individual term, due to an angular singularity, but the limit of the sum is continuous and vanishes thanks to the identity $\cG^I(x_i,x_i) =0$ which follows from (\ref{2.a.8}). Collecting all ingredients,  we obtain our final formula for $J^{(1,0)} _{ij}$,
 \bea
 \label{A.10}
 J^{(1,0)} _{ij} & = & \sum_{r=1}^\infty \p_i \Phi ^{I_1 \cdots I_r} {}_J(x_i) 
\Big [ t_{ij}, \, B_{iI_1} \cdots B_{iI_r}  a^J_i + B_{jI_1} \cdots B_{jI_r}  a^J_j \Big ] 
\no \\ && 
+ \sum_{r=2}^\infty \p_i \cG^{I_1 \cdots I_r} (x_i, x_i) \Big [ t_{ij}, \, B_{iI_1} \cdots B_{iI_r} t_{ij} \Big ]
\eea
We observe that $ J^{(1,0)} _{ij} $ is a single-valued $(1,0)$ form in $x_i$ which is in the range of $\p_i$. It follows that its integral over $\Sigma$ against an arbitrary anti-holomorphic form   $\bar \om^K(x_i)$ vanishes, 
\bea
\label{A.12}
\int _\Sigma d^2 x_i \, \bar \om^K(x_i) J^{(1,0)} _{ij} =0
\eea
It remains to show that $ J^{(1,0)} _{ij} $ is holomorphic in $x_i$. To compute its $\bar \p_i$ derivative it will be convenient to use formula (\ref{A.3}) prior to taking the limit. To obtain the full derivative, we then need to differentiate in both $x_i$ and $x_j$, 
\bea
\label{A.13}
\bar \p_i J^{(1,0)} _{ij}  = \lim _{x_j \to x_i} \Big ( \bar \p_i + \bar \p_j \Big ) 
\Big [ t_{ij}, J^{(1,0)} _i + J^{(1,0)}_j \Big ]
\eea
These derivatives can be worked out using (\ref{2.derJi}) and (\ref{2.derJj}) and we find, 
\bea
\label{A.14}
\bar \p_i J^{(1,0)} _{ij}  & = &
\lim _{x_j \to x_i} \bigg \{ \Big [ t_{ij}, \big [ J^{(0,1)}_i + J^{(0,1)} _j , J^{(1,0)}_i + J^{(1,0)} _j \big ] \Big ]
\no \\ && \qquad
- \pi \sum_{k \not=i} \delta (x_i, x_k) [t_{ij}, t_{ik} ] - \pi \sum_{k \not=j} \delta (x_j, x_k) [t_{ij}, t_{j k} ] \bigg \}
\eea
The sum of the $\delta$ functions vanishes in the limit $x_j \to x_i$  in view of the relation 
(\ref{11.lem.3}), and the first line may be rearranged using the relation, 
\bea
\label{A.15}
\lim _{x_j \to x_i} \big [ t_{ij}, J^{(0,1)}_i + J^{(0,1)} _j \big] = - \pi \bar \om^I (x_i) [ t_{ij} , b_{iI} + b_{jI} ]=0
\eea
which follows from (\ref{11.lem.1}), as follows,
\bea
\label{A.16}
\bar \p_i J^{(1,0)} _{ij}  & = &
\Big [ \lim_{x_j\to x_i} ( J^{(0,1)}_i + J^{(0,1)} _j ), J^{(1,0)} _{ij} \Big ]
\eea
The proof of the vanishing of $J^{(1,0)} _{ij}$ may be completed by the use of Lemma \ref{4.lem:300} for the single-valued $(1,0)$ form $\nu(z)= J_{ij} ^{(1,0)}$ in $z=x_i$ and $\beta_I$ running over the $2h$ components of $b_{iI}$ and $b_{jI} $. The conditions for Lemma \ref{4.lem:300} are met by (\ref{A.12}) and (\ref{A.16}) from which we conclude that $J^{(1,0)} _{ij}=0$, which completes the proof of Lemma \ref{2.lem:3}.

\newpage

\section{Proof of Lemma \ref{2.lem:4}}
\label{sec:B}

In this appendix, we prove Lemma \ref{2.lem:4}, which states the vanishing of the integrals, 
\bea
\label{B.1}
   \newL_{ij}^K = \int _\Sigma d^2 x_i \, \bar \om^K(x_i) \big [ J^{(1,0)}_i, J^{(1,0)}_j \big ] 
\eea
for all $K=1,\cdots, h$ and $i,j=1,\cdots, n$ where $J_i^{(1,0)} $ and $J_j^{(1,0)}$ are given by (\ref{defcJ.3}). Since $J^{(1,0)}_i$ is a single-valued scalar in $x_j$ for $j \not= i$ and $J^{(1,0)}_j$ is a single-valued $(1,0)$ form in $x_j$, it follows that $\newL_{ij}^K$ is a single-valued $(1,0)$ form in $x_j$. 

\sm

To prove the vanishing of $\newL_{ij}^K$ we shall make use of two intermediate results, which will be proven below. First, that $\newL_{ij}^K$ satisfies the following differential equation for any $x_k \in \Sigma$, 
\begin{align}
\bar \p_j \, \newL^K_{ij} &= \Big [ J^{(0,1)} _j , \, \newL^K_{ij} \Big ]\label{BB.1}
\end{align}
Secondly, that its integral  in $x_j$ against any anti-holomorphic Abelian differential $\bar \om^L(x_j)$ must vanish, namely $\newL_{ij}^{KL} =0$ with,
\bea
\label{BB.2}
\newL^{KL}_{ij} = \int _\Sigma d^2 x_j \, \bar \om^L(x_j) \, \newL_{ij} ^K 
= \int _\Sigma d^2 x_i \, \bar \om^K(x_i) \int _\Sigma d^2 x_j \, \bar \om^L(x_j) \, \big [ J^{(1,0)}_i, J^{(1,0)}_j \big ] 
\eea
The relations (\ref{BB.1}) and (\ref{BB.2}), together with the facts that $\cS_{ij}^K$ is a single-valued $(1,0)$ form in $x_j$ and that it admits a formal Taylor series in $b_{jI}$, ensure that the conditions of Lemma \ref{4.lem:300} are satisfied with $\nu (z) = \cS_{ij}^K$ for $z=x_j$ and $\beta _I = b_{jI}$. As a result, we have $\cS_{ij}^K=0$ which completes the proof of Lemma \ref{2.lem:4}.

\sm

It remains to establish (\ref{BB.1}) and to show that $\newL_{ij}^{KL}=0$ in (\ref{BB.2}). Equation (\ref{BB.1}) follows from integrating the differential equation (\ref{3.2}) of Lemma \ref{2.lem:3}  (with $i$ and $j$ swapped), 
\begin{align}
\bar \p_j \big [ J^{(1,0)}_i, J^{(1,0)}_j \big ]  &=
\Big [ J^{(0,1)} _j , [  J^{(1,0)}_i , J^{(1,0)}_j] \Big ] + \pi \delta(x_i,x_j) \big[t_{ij} , J^{(1,0)}_i+J^{(1,0)}_j\big]
\label{detb2}\\
&\quad  + \pi \sum_{k\neq i,j} \delta(x_j,x_k) \big[ t_{jk} , J^{(1,0)}_i\big] 
\notag
\end{align}
in $x_i$ over $\Sigma$  against $\bar \om^K(x_i)$. The delta distribution in the first line gives rise to the 
limit $J^{(1,0)}_{ij}$ which was defined in (\ref{A.3}) and shown to cancel in the remainder of appendix~\ref{sec:A}.
The coefficients of the delta distributions in the second line of (\ref{detb2}) 
vanish by the arguments below (\ref{A.1b}).
To establish $\newL_{ij}^{KL}=0$ in (\ref{BB.2}), we begin by expressing $J^{(1,0)}_i$ and $J^{(1,0)}_j$ as sums over $\bPhi$, $\bG$ contributions. Three different types of Lie algebra generators~arise, 
\bea
\label{B.2}
 \newL_{ij}^{KL} & = &  
\int _\Sigma d^2 x_i \, \bar \om^K(x_i)  \int _\Sigma d^2 x_j \, \bar \om^L(x_j)  \Big [ \bPhi _I (x_i; B_i) \, a^I_i, \bPhi _J (x_j; B_j) \, a^J_j \Big ]
\no \\ &&
+ \int _\Sigma d^2 x_i \, \bar \om^K(x_i)  \int _\Sigma d^2 x_j \, \bar \om^L(x_j) 
\bigg \{ \Big [ \bPhi _I (x_i; B_i) \, a^I_i,  \sum_{k \not = j} \bG (x_j,x_k;B_j) \, t_{jk} \Big ]  - ( i \leftrightarrow j) \bigg \}
\no \\ &&
+ \int _\Sigma d^2 x_i \, \bar \om^K(x_i)  \int _\Sigma d^2 x_j \, \bar \om^L(x_j) 
\bigg [  \sum_{k \not = i} \bG (x_i,x_k;B_i) \, t_{ik} , \sum_{\ell \not = j} \bG (x_j,x_\ell;B_j) \, t_{j\ell} \bigg ] 
\eea
Some of the  integrals may be evaluated with the help of the following formulas,
\bea
\label{B.4}
\int _\Sigma d^2 x_i \, \bar \om^K(x_i) \, \bPhi _I (x_i; B_i) & = & \delta ^K_I 
\no \\
\int _\Sigma d^2 x_i \, \bar \om^K(x_i) \, \bG(x_i, x_j;B_i) & = & 0
\eea
which easily follows from integrating (\ref{2.gen.1}).
As a result, the contributions to $\newL_{ij}^{KL}$ arising from the first line in (\ref{B.2}) vanish  with the further help of $[a_i^I, a_j^J]=0$ of the first relation in (\ref{11.1}); the contribution from the first term inside the braces on the second line of (\ref{B.2}) vanishes by first integrating over $x_j$ and then over $x_i$  and reversing the order of these integrations  for the second term under the braces; the only non-vanishing contribution arising  from the last line is for $k=j$ and $\ell=i$, so that we are left with, 
\bea
\newL_{ij} ^{KL}
=  \int _\Sigma d^2 x_i \, \bar \om^K(x_i)  \, \int _\Sigma d^2 x_j \, \bar \om^L(x_j)
\Big [  \bG (x_i,x_j;B_i) \, t_{ij} ,  \bG (x_j,x_i;B_j) \, t_{ij} \Big ] 
\eea
To simplify this expression, we expand it in powers of $B_i$ and $B_j$ as follows, 
\bea
\newL_{ij}^{KL}  & = &
\sum _{{r,s=0 \atop r+s\geq 1}}^\infty \int _\Sigma d^2 x_i \, \bar \om^K(x_i) \, \big ( \anotherL _{ij}^L  \big )_{r,s}
\no \\
\big ( \anotherL _{ij}^L  \big )_{r,s} & = & 
 \int _\Sigma d^2 x_j \, 
\p_i \cG ^{I_1 \cdots I_r} (x_i, x_j) \, \bar \om^L(x_j)  \, \p_j \cG^{J_1 \cdots J_s} (x_j,x_i) 
\no \\ && \qquad \times 
\Big [ B_{iI_1} \cdots B_{iI_r} t_{ij}, \, B_{jJ_1} \cdots B_{jJ_s} t_{ij} \Big ] 
\eea
where the term $r=s=0$ cancels in view of $[t_{ij}, t_{ij}]=0$. The integral over $x_j$ may be performed 
 using the following integral relation based on the $\delta_{I_r}^J$ trace of (\ref{2.a.5}),
\bea
\int _\Sigma d^2 x_j \, 
\p_i \cG ^{I_1 \cdots I_r} (x_i, x_j) \, \bar \om^L(x_j)  \, \p_j \cG^{J_1 \cdots J_s} (x_j,x_i) =
\lim_{z \to x_i} \p_i \cG^{I_1 \cdots I_r L J_1 \cdots J_s} (x_i,z)
\eea
where the limit on the right side was proven to exist for $r{+}s \geq 1$ in subsections 8.2 and 8.3 of \cite{DHoker:2024ozn}, so that we have, 
\bea
\big ( \anotherL _{ij}^L  \big )_{r,s}  = 
\lim_{z \to x_i} \p_i \cG^{I_1 \cdots I_r L J_1 \cdots J_s} (x_i,z)
\Big [ B_{iI_1} \cdots B_{iI_r} t_{ij}, \, B_{jJ_1} \cdots B_{jJ_s} t_{ij} \Big ] 
\eea 
Now consider the following sum, and conveniently relabel the indices $I \leftrightarrow J$ as follows,
\bea
\big ( \anotherL _{ij}^L  \big )_{r,s} + \big ( \anotherL _{ij}^L  \big )_{s,r}
& = &
\lim_{z \to x_i} \bigg \{  \p_i \cG^{I_1 \cdots I_r L J_1 \cdots J_s} (x_i,z) 
\Big [ B_{iI_1} \cdots B_{iI_r} t_{ij}, \, B_{jJ_1} \cdots B_{jJ_s} t_{ij} \Big ] 
\no \\ && \qquad
+  \p_i \cG^{J_s \cdots J_1 L I_r \cdots I_1} (x_i,z) 
\Big [ B_{iJ_s} \cdots B_{iJ_1} t_{ij}, \, B_{jI_r} \cdots B_{jI_1} t_{ij} \Big ]  \bigg \}
\qquad
\eea
Using the following properties in the second line,
\bea
\cG ^{J_s \cdots J_1 L I_r \cdots I_1} (x_i, z) & = & (-)^{r+s+1} \cG ^{I_1 \cdots I_r L J_1 \cdots J_s} (z,x_i)
\no \\
B_{iJ_s} \cdots B_{iJ_1} t_{ij} & = & (-)^s B_{jJ_1} \cdots B_{jJ_s} t_{ij}
\no \\
B_{jI_r} \cdots B_{jI_1} t_{ij} & = & (-)^r B_{iI_1} \cdots B_{iI_r} t_{ij}
\eea
we obtain, 
\bea
\big ( \anotherL _{ij}^L  \big )_{r,s} + \big ( \anotherL _{ij}^L  \big )_{s,r}
& = &
\lim_{z \to x_i} \bigg \{  \p_i \cG^{I_1 \cdots I_r L J_1 \cdots J_s} (x_i,z)  
+ \p_i \cG ^{I_1 \cdots I_r L J_1 \cdots J_s} (z,x_i) \bigg \} 
\no \\ && \qquad \times 
\Big [ B_{iI_1} \cdots B_{iI_r} t_{ij}, \, B_{jJ_1} \cdots B_{jJ_s} t_{ij} \Big ] 
\eea
The limit is continuous for $r{+}s\geq 1$ since the limit of each term is. 
The sum of the limits is a total $x_i$ derivative of the coincident limit, 
\bea
\big ( \anotherL _{ij}^L  \big )_{r,s} + \big ( \anotherL _{ij}^L  \big )_{s,r} & = &
 \p_i  \cG^{I_1 \cdots I_r L J_1 \cdots J_s} (x_i,x_i)    
\Big [ B_{iI_1} \cdots B_{iI_r} t_{ij}, \, B_{jJ_1} \cdots B_{jJ_s} t_{ij} \Big ] 
\eea
This shows that the sum of the contributions from each pair $(r,s)$ and $(s,r)$ is in the range of $\p_i$, which immediately implies $\newL_{ij}^{KL}=0$ and, as argued previously using Lemma \ref{4.lem:300}, implies that $\newL^K_{ij}=0$. This completes the proof of Lemma \ref{2.lem:4}.

\newpage

\section{Proof of the solution for $\hat a_i^K - \xi_i^K$ in Lemma \ref{5.lem:3}}
\label{sec:C1}

In this appendix, we complete the proof of the existence and uniqueness of the solution given in equation (\ref{mauto.99}) of Lemma \ref{5.lem:3} for $\hat a_i^K {-} \xi_i^K$.

\subsection{Existence of the solution}
 
To prove that the expression for $\hat a_i^K{-} \xi_i^K$ given in (\ref{mauto.99}) solves the relation on the left side of (\ref{5.k.99}),  we begin by translating (\ref{7.2.g}) into its adjoint action, given in terms of $B_{iI} X = [b_{iI}, X]$ and $\hat B_{iI} X = [\hat b_{iI}, X]$ for $X \in \hat \mt_{h,n}$,
\bea
\label{C.3}   
\hat B_{iI} = B_{iI} - \sum_{r=2}^{\infty} {\cal M}_\shuffle^{I_1 \cdots I_r}{}_I(y_\ast)\, B _{iI_1} \cdots  B_{iI_r} 
\eea
Applying the expression for $\hat B_{iI}$ given in (\ref{C.3}) to the expression for $\hat a_j^J {-} \xi^J_j$ given in (\ref{mauto.99}) for $j \not=i$ and using $\eta_{iI}= \hat b_{iI}$ in $[\eta_{iI}, \xi_j^J]=0$ to get $[\hat b_{iI}, \xi^J_j]=0$,  we obtain,
\bea
\label{C.4}   
{} \big [ \hat b_{iI}, \hat a^J_j - \xi ^J_j \big ] & = & [ b_{iI}, a_j^J ] + Z_1+Z_2+Z_3
\eea
where the terms $Z_1, Z_2, Z_3$ are given by (suppressing the dependence of $\cM$ and $\cM_\shuffle$ on $y_\ast$),
\bea
\label{C.5}   
Z_1 & = & - \sum_{r=2}^\infty {\cM_\shuffle ^{J_1 \cdots J_r}}_I \, B_{iJ_1} \cdots B_{i J_r} a_j^J
= - \sum_{r=1}^\infty {\cM_\shuffle ^{J_1 \cdots J_r J }}_I \, B_{iJ_1} \cdots B_{i J_r} t_{ij} 
\no \\
Z_2 & = & \sum_{s=1}^\infty  \cM^{J J_1 \cdots J_s}{}_K B_{iI} B_{jJ_1} \cdots B_{jJ_s} a^K_j
= \sum_{s=1}^\infty  \cM^{J J_1 \cdots J_s}{}_I B_{jJ_1} \cdots B_{jJ_s} t_{ij}
\no \\
Z_3 & = & - \sum _{r=2} ^\infty \sum_{s=1}^\infty {\cM_\shuffle ^{I_1 \cdots I_r}}_I \, 
\cM^{J J_1 \cdots J_s}{}_K \, B_{iI_1} \cdots B_{iI_r} B_{j J_1} \cdots B_{jJ_s} a^K_j
\eea
In passing from the first expression for $Z_1$ to the second, we have used the relation $B_{i J_r} a^J_j = \delta ^J_{J_r} t_{ij}$ of (\ref{11.1a}) and then relabeled $ r \to r{+}1$. In passing from the first to the second sum in $Z_2$ we have simply used $[B_{iI}, B_{jJ}]=0$ and $B_{iI} a^K_j = \delta ^K_I t_{ij}$.   The shuffle properties of ${\cM_\shuffle ^{I_1 \cdots I_r}} _I$ in section 3.3 of \cite{DHoker:2024desz} imply the following antipode relation (see \cite{Reutenauer})
\bea
\label{C.6}   
{\cM_\shuffle ^{I_r \cdots I_1}} _I = (-)^{r+1} {\cM_\shuffle ^{I_1 \cdots I_r}} _I
\eea
Together with the relation $B_{iJ_1} \cdots B_{i J_r} t_{ij}  = (-)^r B_{jJ_r} \cdots B_{j J_1} t_{ij} $ in (\ref{11.lem.2}), and relabeling the summation indices gives the following expression for $Z_1$,
\bea
\label{C.7}   
Z_1  = - \sum_{r=1}^\infty {\cM_\shuffle ^{J J_1 \cdots J_r }}_I \, B_{jJ_1} \cdots B_{j J_r} t_{ij} 
\eea
which is minus the expression (\ref{C.5}) for $Z_2$ with $\cM_\shuffle$ in the place of $\cM$.
Moving $B_{i I_r}$ past the $B_j$ factors in $Z_3$, using $B_{iI_r}  a^K_j = \delta _{I_r}^K t_{ij}$, relabeling $r\to r{+}1$, and converting the remaining $B_i$ factors into $B_j$ factors gives,  
\bea
\label{C.8}   
Z_3 = - \sum_{r,s=1}^\infty \cM^{J J_1 \cdots J_s}{}_K {\cM_\shuffle ^{K I_1 \cdots I_r}}_I
B_{j J_1} \cdots B_{j J_s} B_{jI_1} \cdots B_{jI_r} t_{ij}
\eea
In terms of the generating functions of (\ref{C.9}), 
we may write the sum $Z_1 + Z_2+Z_3$ as follows,
\bea
\label{C.10}   
Z_1 + Z_2+Z_3 = \Big ( {-} {\MM^J_\shuffle }_I  + \MM^J{}_I  - \MM^J {}_K \, {\MM^K_\shuffle \, }_I \Big ) t_{ij}
\eea
which vanishes by (\ref{C.11}). Therefore, we have $[ \hat b_{iI}, \hat a^J_j {-} \xi ^J_j] =  [ \hat b_{iI}, \hat a^J_j ]  = [ b_{iI}, a_j^J ] $ which, combined with $t_{ij} = \hat t_{ij}$, proves that (\ref{mauto.99}) solves the equations of (\ref{5.k.99}).

\subsection{Uniqueness of the solution}
 
Next, we prove that (\ref{mauto.99}) is the unique solution for $\hat a_j^J{-} \xi_j^J$ to (\ref{5.k.99}) that has $a$-degree one. We start by noting that the relations of (\ref{5.k.99}), where $a_j^J, b_{iI}$ and $\hat b_{iI}$ are viewed as given, are linear inhomogeneous equations for $\hat a_j^J$ for which we have obtained a particular solution in (\ref{mauto.99}). Therefore, to show uniqueness,  it remains to show that, for a given value of $i$, the intersection of the kernels of $\hat B_{iI}$ for all $i \not=j$ is trivial. To do so, we write the most general expression for a Lie algebra element of $a$-degree one, 
\bea
\label{C.12}   
C_j = \sum_k F_{j kL}(B) a^L_k + \sum _{k \not= j} G_{jk} (B) t_{jk} + \sum _{k \not= \ell \not= j} H_{jk\ell}(B) t_{k\ell}
\eea 
where $F_{jkL}, G_{jk}$ and $H_{jk\ell}$ are Lie series which may depend on all possible generators $B$ as well as points on $\tilde \Sigma$ and moduli of $\Sigma$. Expressing (\ref{C.3}) in terms of the generating function ${\MM^J _\shuffle }_I (B)= {\MM^J _\shuffle }_I $ introduced in (\ref{C.9}),
\bea
\label{C.13a}
\hat B_{iI} = B_{iI} - B_{iJ} \, {\MM^J _\shuffle }_I (B_i)
\eea
and using the fact that the matrix series $\delta^J_I- {\MM^J_\shuffle}_I$ is invertible, the kernel of $\hat B_{iI}$ equals the kernels of $B_{iI}$. To determine the intersection of the kernels of $B_{iI}$ for $i \not=j$,  we need to solve the following system of equations for given value of $j$ and all $i \not= j$,
\bea
\label{C.13}   
B_{iI} \, C_j =0
\eea
In the first term, all dependence of $F_{jkL}$ on $B_\ell$ with $\ell \not=k$ can be converted into contributions of the form of the second and third terms using $[b_{iI}, a^L_k]=\delta ^L_I \, t_{ik}$ and the fact that $[B_{iI}, B_{\ell L}]=0$ for all $\ell \not= i$. Therefore, we may assume without loss of generality that $F_{jkL}$ depends only on $B_k$. Furthermore, the trace relation allows us to recast $[b_{kL}, a^L_k]$ in terms of the $G$ and $H$ terms, so we can assume that the trace part in $F$ has been removed. 
Since $[B_{\ell L}, t_{jk}]=0$ for all $\ell \not= k,j$, the function $G_{jk}$ can only depend on $B_j$ and $B_k$, but the latter may be converted to $B_j$ via (\ref{11.lem.2}), so effectively we have $G_{j k}(B_j)$. Applying the same reasoning to the function $H_{jk\ell}$, we find that it can depend only on $B_k$, or equivalently on $B_\ell$.  Thus, we end up with the following reduced dependence of $C_j$,
\bea
\label{C.14}   
C_j = \sum_k F_{j kL}(B_k) a^L_k + \sum _{k \not= j} G_{jk} (B_j) t_{jk} + \sum _{k \not= \ell \not= j \not= k} H_{jk\ell}(B_k) t_{k\ell}
\eea 
Applying $B_{iI}$ with $i \not= j$ gives,
\bea
\label{C.15}   
B_{iI} C_j & = & B_{iI} F_{j i L}(B_i) a^L_i +  \sum_{k \not= i}  F_{j kI }(B_k) t_{ik} + B_{iI} G_{ji} (B_j) t_{ji} 
\no \\ &&
+ \sum _{k \not= i, j} H_{jki}(B_k) B_{iI} t_{ik} + \sum_{\ell \not= i,j} B_{iI} H_{ji\ell} (B_i) t_{i \ell}
\eea 
Since only the first term involves $a_i$ and the trace part of $F$ has been removed, we find that $F_{jiL}=0$ for all $i \not=j$. Therefore the second term must also vanish, and we are left with only the $G$ and $H$ terms. The $G$ term is the only one that contains $t_{ij}$ as opposed to the $t_{im}$ with $m\neq i,j$ in the
second line of (\ref{C.15}), so application of $B_{jL}$ which annihilates the second line shows that 
the function $G$ must vanish as well. 
The contribution from the $H$ terms may be rewritten as follows,
\bea
\label{C.16}   
B_{iI} C_j & = & 
B_{iI}  \sum _{k \not= i, j} \Big (  \theta H_{jki}(B_i) + H_{jik} (B_i) \Big ) t_{ik}
\eea 
in terms of the antipode $\theta$ of (\ref{defant}). Since the subalgebra generated by $b_i$ and $t_{ik}$ is free we must have,
\bea
\label{C.17}   
\theta H_{jki}(B_i) + H_{jik} (B_i) =0
\eea
for all $k \not=i,j$ and $i \not=j$. 
Inspection of (\ref{C.14}) show that this relation cancels the contribution from the $H$ terms, so that $C_j=0$ and the intersection of the kernels of all $B_{iI}$ for $i \not=0$ is trivial.  

\newpage

\section{Proof of Lemma \ref{3.lem:9}}
\label{sec:C}

In this appendix, we provide a proof of Lemma \ref{3.lem:9}, which states that the 1-form $\cN$, defined in (\ref{7.h.1}) in terms of its components given by (\ref{7.5.a}), is closed,
\begin{align}
\bar \p_i \, \cN_i  = 0 \hskip 1in 
\bar \p_j \, \cN_i  =  0 \hskip 1in
\p_j \, \cN_i - \p_i \, \cN_j  =0 
\label{ceq.1}
\end{align}
for any $j\neq i$, with $\hat b_{iI}$ and $\hat a_i^I$ in (\ref{7.2.g}) and (\ref{mauto.99}) subject to $\big [ \hat a_i^I, \hat b_{iI} \big ] = \big [ a^I_i, b_{iI}  \big]$.

\subsection{Proof of $\bar \p_i \, \cN_i  = 0$}

The claim may be proven by direct calculation and  applying $\bar \p_i$ to $\cN_i$,
\bea
\bar \p_i \cN_i & = & 
- \pi \cU^{-1} \bigg \{  \bar \om^I(x_i) H_{iI} \Big ( {\bf J}_K(x_i,z; H_i) \big ( \hat a_i^K - \xi_i^K \big ) 
+ \sum_{j \neq i} \big \{  {\bf G}(x_i,x_j; H_i) - {\bf G}(x_i,z; H_i)  \big \} \hat t_{ij}  \Big ) 
\no \\ && \hskip 0.5in
-  \, \bar \om^I(x_i) H_{iI} \bJ_K(x_i,z;H_i) (\hat a^K_i - \xi^K_i) 
+ \sum_{j \not= i} \Big (  \delta (x_i,x_j) - \bar \om^I(x_i) H_{iI} \bG(x_i,x_j;H_i)
\no \\ && \hskip 1.5in
+ \, \bar \om^I(x_i) H_{iI} \bG(x_i,z;H_i) \Big ) \hat t_{ij} \bigg \} \, \cU
+ \pi  \sum_{j \not= i} \delta (x_i,x_j) t_{ij} 
\label{ceq.2}
\eea
where the first line on the right side arises from differentiating $\cU$ and $\cU^{-1}$ in (\ref{7.5.a}) while the last two lines arise from differentiating the terms inside the outer parentheses in (\ref{7.5.a}) and its last line. More specifically, the generating functions of DHS and Enriquez kernels entering ${\cal N}_i$ in (\ref{7.5.a}) are differentiated via (\ref{2.gen.3}) and (\ref{3.Kder}), respectively, where contributions proportional to $ \kappa(x_i)$ due to term-by-term application of (\ref{2.gen.3}) cancel from $\bar \p_i( {\bf G}(x_i,x_j; H_i) - {\bf G}(x_i,z; H_i))$. The Dirac $\delta$-functions also cancel one another thanks to  (\ref{7.4.a}). The remaining contributions cancel term by term inside the curly bracket.

\subsection{Proof of $\bar \p_j \, \cN_i  =  0$}

The claim may be proven by applying $\bar \p_j$ with $j\neq i$ to both sides of (\ref{7.5.a}),
\bea
\bar \p_j \cN_i 
& = &
\pi \cU^{-1} \bigg \{ \Big ( \delta (x_i,x_j) - \bar \om^K(x_j) \bJ_K(x_i,x_j;H_i) \Big ) \hat t_{ij} 
\no \\ && \qquad ~
+ \,\bar \om^I(x_j) H_{jI} \bJ_K(x_i,z;H_i)  (\hat a^K_i - \xi^K_i) 
\no \\ && \qquad ~
+\Big ( \bG(x_i,x_j;H_i) - \bG(x_i,z;H_i) \Big ) H_{jI} \hat t_{ij} \bigg \} \, \cU
- \pi \delta(x_i,x_j) t_{ij} 
\label{ceq.3}
\eea
where the first line arises from differentiating the terms inside the outer parentheses in (\ref{7.5.a}) and the second and third  lines arise from differentiating $\cU$ and $\cU^{-1}$ and the last line in (\ref{7.5.a}). The Dirac $\delta$-functions cancel one another thanks to (\ref{7.4.a}). On the second line, we may move the left factor of $H_{jI}$ past $\bJ_K$ using the fact that $[H_{jI}, H_{iJ}]=0$ for $j \not=i$, and then use $H_{jI} (\hat a^J_i {-} \xi^J_i) = [\hat b_{jI}, \hat a^J_i] =  \delta ^J_I \hat t_{ij}$ (as proven in Lemma \ref{5.lem:88b}) to express the second line as $ \bar \om^K(x_j) \bJ_K(x_i,z;H_i) \hat t_{ij}$. Combining this expression with the $\bJ_K$ term on the first line cancels the last line using (\ref{2.gen.2}) and $ H_{jI} \hat t_{ij}= - H_{iI} \hat t_{ij}$ and concludes the proof of $\bar \p_j \, \cN_i  =  0$.

\subsection{Proof of $\p_j \, \cN_i - \p_i \, \cN_j  =0$}

Since we have already shown $\p_i K_j - \p_j K_i=0$ in (\ref{E.3.2}), the remaining contribution is,
\bea
\p_j \cN_i-\p_i \cN_j & = & 
\p_j \bigg \{ \cU^{-1}  \Big ( \bPhi_J(x_i; H_i) \big ( \hat a_i^J - \xi_i^J \big ) 
+ \sum_{k \neq i}  {\bf G}(x_i,x_k; H_i) \, \hat t_{ik}  \Big ) \, \cU \bigg \}  - (i \leftrightarrow j) 
\qquad
\label{ceq.4}
\eea
Using the relation,
\bea
\p_j \bG(x_i,x_j; H_i) \hat t_{ij} = \p_i \bG(x_j,x_i;H_j) \hat t_{ij}
\label{ceq.5}
\eea
the contribution from the derivative of the terms inside the large parentheses is seen to vanish by the antisymmetrization in $i,j$. The remaining contribution arises solely from the derivative of the gauge transformation and may be evaluated with the help of the following consequence of (\ref{7.2.c}),
\bea
\p_j \, \cU = \Big ( \bPhi _K (x_j; H_j) \,\xi_j^K \Big ) \, \cU
\hskip 0.8in
\p_j \, \cU^{-1} = - \, \cU^{-1} \,  \Big ( \bPhi _K (x_j; H_j) \, \xi_j^K \Big ) 
\label{ceq.6}
\eea  
After multiplying (\ref{ceq.4}) to the left by $\cU$ and to the right by $\cU^{-1}$,  we are left with the condition of vanishing of the following combination, 
\bea
V_{ij}(x_i,x_j) =  \bigg [ \bPhi_J(x_i; H_i) \big ( \hat a_i^J - \xi_i^J \big ) 
+ \sum_{k \neq i}  {\bf G}(x_i,x_k; H_i)\,  \hat t_{ik}, \bPhi_K(x_j;H_j) \,\xi^K_j \bigg ] - (i \leftrightarrow j)
\quad
\label{ceq.7}
\eea
Since $\xi_j$ is a Lie series in $\eta_j = \hat b_j$ and independent of $\eta_k$ for all $k \not= j$, all terms above with $k\not=j$ vanish, and we are left with, 
\bea
V _{ij} (x_i,x_j) =  \Big [ \bPhi_J(x_i; H_i) \big ( \hat a_i^J - \xi_i^J \big ) 
+  {\bf G}(x_i,x_j; H_i)\,  \hat t_{ij}, \bPhi_K(x_j;H_j) \,\xi^K_j \Big ] - (i \leftrightarrow j)
\label{ceq.8}
\eea
This expression is a single-valued  $(1,0)$ form in both $x_i$ and $x_j$. Its $\bar \p_i$ derivative is given by,
\bea
\bar \p_i V_{ij} (x_i,x_j) & = & 
- \pi \delta (x_i, x_j) \Big  [ \hat t_{ij} , \bPhi_K(x_j;H_j) \, \xi^K _j  + \bPhi_K(  x_j ;H_i) \, \xi^K_i   \Big ]
\no \\ &&
- \pi \bar \om^K(x_i) H_{iK} V_{ij}(x_i,x_j)
\label{ceq.9}
\eea
Since $\xi_i$ is a Lie series in $\eta_i$, so is $\bPhi_K(x_j;H_i) \xi^K_i $ and therefore, in view of (\ref{7.1.d}), we have $\bPhi_K(x_j;H_j) \xi^K _j  + \bPhi_K(x_j;H_i) \xi^K_i = \bPhi_K(x_j;H_i+H_j) (\xi_i^K+ \xi_j^K)$, whose lowest order term in $H$ is of degree one in the generators $\xi, \eta$. As a result, the coefficient of the $\delta$-function cancels,
\bea
\Big  [ \hat t_{ij} , \bPhi_K(x_j;H_j) \, \xi^K _j  + \bPhi_K(x_j;H_i) \, \xi^K_i   \Big ]
= \big[ \hat t_{ij} , \bPhi_K(x_j;H_i+H_j) (\xi_i^K+ \xi_j^K) \big] =0
\label{ceq.10}
\eea 
using the relations $\eta_i = \hat b_i$,  $[\hat b_{iI}, \hat b_{jJ}]=0$ and $[ \hat b_{iI} {+} \hat b_{jI} , \hat t_{ij}] = 0$.
The remaining equation, 
\bea
\bar \p_i  V_{ij}(x_i,x_j) = - \pi \bar \om^K(x_i) H_{iK}  V_{ij} (x_i,x_j)
\label{ceq.11}
\eea
implies the following relation for the single-valued $(1,0)$ form $V_{ij}^J(x_i)$, 
\bea
\bar \p_i V_{ij} ^J (x_i) = - \pi \bar \om^K(x_i) H_{iK}  V_{ij} ^J (x_i)
\hskip 0.6in
V^J_{ij} (x_i) = \int _\Sigma d^2x_j \, \bar \om^J(x_j)  V_{ij}(x_i, x_j)
\label{ceq.12}
\eea
To proceed, we shall also need the following result, 
\bea
\int _\Sigma d^2 x_i \, \bar \om^I(x_i)  V_{ij}^J (x_i) = 
\int _\Sigma d^2 x_i \, \bar \om^I(x_i) \int _\Sigma d^2 x_j \, \bar \om^J(x_j) V_{ij}(x_i,x_j) =0
\label{ceq.13}
\eea
To prove it, we evaluate the double integral using (\ref{ceq.8}) by performing first the integral over $x_i$ in the first term and the integral over $x_j$ in the $( i \leftrightarrow j)$ term, and subsequently performing the remaining integral. The result is as follows, 
\bea
\int _\Sigma d^2 x_i \, \bar \om^I(x_i) V_{ij} ^J (x_i) =
\big [ \hat a ^I_i - \xi_i^I, \xi_j^J \big ] - \big [ \hat a ^J_j  - \xi_j^J, \xi_i^I \big ]
=  \big [ \xi_i^I , \hat a ^J_j \big ] - \big [ \xi_j^J ,  \hat a ^I_i \big ] 
\eea
where we have used the structure relations $[\xi_i^I, \xi_j^J ]=0$ for $i \not= j$ to get the last equality. Since $\xi_i^I$ is given by a Lie series in $\hat b_i$ in (\ref{7.2.g1}), the adjoint action of $\xi_i^I$ is given as follows,
\bea
\big [ \xi_i^I , \hat a ^J_j \big ] =
\sum_{r=1}^\infty \cX ^{I K_1 \cdots K_r} \hat B_{i K_1 } \cdots \hat B_{i K_r} \hat a^J_j
= \sum^\infty_{r=0} \cX^{I K_1 \cdots K_r J} \hat B_{i K_1 } \cdots \hat B_{i K_r} \hat t_{ij}
\eea
where we set $\cX^{JK} = \pi Y^{JK}$ for $r=1$,  $\hat B_i X = [\hat b_i, X]$ for all $X \in \hat \mt_{h,n}$, and used the structure relation $\hat B_{i K_r} \hat a_j^J = \delta^J_{K_r} \hat t_{ij}$. To evaluate the commutator with $(i, I) \leftrightarrow (j,J)$, we use the cyclic and antipode relations of the Lie series (\ref{7.2.g1}),
\bea
\label{DDD}
\cX^{I K_1 \cdots K_r} = \cX^{K_1 \cdots K_r I}
\hskip 1in
\cX^{I K_r \cdots K_1} = (-)^{r-1} \cX^{I K_1 \cdots K_r}
\eea
to recast the commutator as follows,
\bea
\big [ \xi_j^J , \hat a_i^I \big ] 
= \sum^\infty_{r=0} \cX^{J K_1 \cdots K_r I} \hat B_{j K_1 } \cdots \hat B_{j K_r} \hat t_{ij}
= \sum^\infty_{r=0} (-)^r \cX^{J K_r \cdots K_1 I} \hat B_{i K_1 } \cdots \hat B_{i K_r} \hat t_{ij}
\eea
Combining the two relations in (\ref{DDD}) implies the relation $\big [ \xi_j^J , \hat a_i^I \big ] = \big [ \xi_i^I , \hat a ^J_j \big ] $ which completes the proof of the relation (\ref{ceq.13}). 

\sm

To prove the vanishing of $V_{ij} ^J(x_i)$, we employ Lemma \ref{4.lem:300} with $\nu(z) = V_{ij} ^J(x_i)$  for $z=x_i$ and $\beta_I=\eta_{iI}$ whose conditions are met by (\ref{ceq.12}) and (\ref{ceq.13}).  The vanishing of $V_{ij}^J(x_i)$ then provides the vanishing of the integral of $V_{ij}(x_i,x_j)$ which, in combination with (\ref{ceq.11}), allows us to apply Lemma \ref{4.lem:300} afresh to $\nu(z) = V_{ij}(x_i,x_j)$ for $z=x_i$ and $\beta_I=\eta_{iI}$. On these grounds, we conclude that also $V_{ij}(x_i,x_j)=0$.

\newpage

\section{Proof of Lemma \ref{7.lem:666}}
\label{sec:D}

In this appendix we provide a proof of Lemma \ref{7.lem:666}, which states the following property of the components $\cN_i$ of $\cN$ given by (\ref{7.5.a}), 
\bea
\cN_i= \sum _{j \not = i} \cN_{ij} (B_i) t_{ij}
\label{deq.1}
\eea
where $\cN_{ij} (B_i) $ depends only on the variables $x_i, x_j$ and the generators $B_{iI}$; it is a $(1,0)$ form in $x_i$ and a scalar in $x_j$ and is independent of $x_k$ for all $k \not= i,j$. 

\sm
 
 The starting point for the proof  is the expression for $\cN_i$ given in (\ref{7.5.a}). Since $\cU_i$ depends only on $\eta_i = \hat b_i$ and is independent of both $\hat a_i$ and $\hat a_k$ with $k\neq i$, we have $[ H_i, \cU_j]=0$ for $j \not=i$, so that (\ref{7.5.a}) may be rewritten as follows,
\bea
\label{7.5.b} 
\cN_i & = & 
\hat \cU_i^{-1} \Big ( \cU_i^{-1} \,  {\bf J}_I(x_i,z; H_i) \big ( \hat a_i^I - \xi_i^I \big ) \, \cU_i   \Big ) \, \hat \cU_i 
\no \\ && 
+ \sum_{j \neq i} \Big ( \cU_i^{-1} \big \{  {\bf G}(x_i,x_j; H_i) - {\bf G}(x_i,z; H_i)  \big \} \cU_i \Big ) 
\Big (  \cU_i^{-1} \, \cU_j^{-1} \, \hat t_{ij} \, \cU_j \, \cU_i \Big ) 
\no \\ &&
-  {\bf K}_I  (x_i,z;  B_i) a_i^I  - \sum_{j \neq i} \Big \{  {\bf X}(x_i,x_j;B_i) - {\bf X}(x_i,z;B_i)  \Big \}  t_{ij}  
\eea
where $\hat a_i{ -} \xi_i$ is given by (\ref{mauto.99}),  $\cU_i$ is given by (\ref{7.3.a}),  and we have defined, 
\bea
\hat \cU_i = \cU \, \cU_i^{-1} = \prod _{j \not = i} \cU_j
\label{deq.2}
\eea
By construction of the single-variable case, we have,
\bea
{\bf K}_I(x_i,y_*;  B_i) a_i^I  = \cU_i^{-1} \Big (  {\bf J}_I(x_i,y_*; H_i) \big ( \hat a_i^I - \xi_i^I \big ) \Big ) \, \cU_i 
 \label{deq.3}
 \eea
after setting the arbitrary point $z$ in (\ref{7.5.b}) to the common base point $y_\ast$ of the gauge transformation
 (\ref{7.1.c}) with $\yy_*=(y_\ast,\cdots,y_\ast)$. Eliminating $\bJ_I$ in favor of $\bK_I$ gives, 
\bea
\label{7.66}
\cN_i & = &
 \sum_{j \neq i} \Big ( \cU_i^{-1} \big \{  {\bf G}(x_i,x_j; H_i) - {\bf G}(x_i,y_\ast; H_i)  \big \} \cU_i \Big ) 
\Big (  \cU_i^{-1} \, \cU_j^{-1} \,\hat  t_{ij} \, \cU_j \, \cU_i \Big ) 
\no \\ &&
- {\bf K}_I  (x_i,y_\ast;  B_i) \Big ( a_i^I - \hat \cU_i^{-1} \, a^I_i \, \hat \cU_i \Big )  
- \sum_{j \neq i} \Big \{  {\bf X}(x_i,x_j;B_i) - {\bf X}(x_i,y_\ast;B_i)  \Big \}  t_{ij}  
\eea
In the first line, the terms inside the parentheses depend only on $H_i$ which may be converted into $B_i$. But the factor inside the second parentheses depends on $H_i$ and $H_j$. To evaluate the action of $\cU_j$ we use (\ref{7.3.a}), 
\bea
\cU_j^{-1} \, \hat t_{ij} \, \cU_j  & = &
\hat t_{ij} + \sum _{r=1} ^\infty \cT^{I_1 \cdots I_r} (x_j,y_\ast) H_{jI_1} \cdots H_{jI_r} \hat t_{ij}
\no \\
& = &
\hat t_{ij} + \sum _{r=1} ^\infty \cT^{I_1 \cdots I_r} (x_j,y_\ast) (-)^r H_{iI_r} \cdots H_{iI_1} \hat t_{ij}
\label{deq.4}
\eea
This operation coincides with the action of the antipode (\ref{defant}) and, together with $\cU_i$,  defines a gauge transformation from $x_i$ to $x_j$, 
\bea
\cU_{ij} = {\rm Pexp} \int ^{x_i} _{x_j} \Big ( dy \, \cJ_i(y; \xi_i, \eta_i, 0) - \pi d\bar y \, \bar \om^I(y) \, \eta_{iI} \Big )
\label{deq.5}
\eea
which is independent of $y_\ast$ and we have,
\bea
\cU_i^{-1} \, \cU_j^{-1} \, \hat t_{ij} \, \cU_j \, \cU_i = \cU_{ij}^{-1} \, \hat t_{ij} \, \cU_{ij}
\label{deq.6}
\eea
Thus, the combination $\cU_i^{-1} \, \cU_j^{-1} \, \hat t_{ij} \, \cU_j \, \cU_i $ may be expressed in terms of an action of generators $H_i$ on $\hat t_{ij}$ or, equivalently as the action of generators $B_i$ on $t_{ij}$. This guarantees that the first line in (\ref{7.66}) is of the form of the lemma. 

\sm

For the second line of (\ref{7.66}), only the first term is not yet manifestly in the form of the lemma. To show that it actually is, we compute $\hat \cU_i^{-1} \, a^J_i \, \hat \cU_i$ as follows, first for $i=1$,
\bea
\hat \cU_1^{-1} \, a^J_1 \, \hat \cU_1 & = & \cU_n^{-1} \cdots \cU_2^{-1} a^J_1 \, \cU_2 \cdots \cU_n
\no \\ & = &
\cU_n^{-1} \cdots \cU_3^{-1} \Big ( a_1^J + \big \{ \cU_2^{-1} a^J_1 \, \cU_2 - a_1 ^J \big \} \Big ) \, \cU_3 \cdots \cU_n
\label{deq.7}
\eea
The term inside the braces is proportional to $t_{12}$, acted upon repeatedly by factors of $H_2$ and therefore commutes with $\cU_3, \cdots, \cU_n$,
\bea
\hat \cU_1^{-1} \, a^J_1 \, \hat \cU_1 & = &  
\cU_n^{-1} \cdots \cU_3^{-1} \, a_1^J \,  \cU_3 \cdots \cU_n
+ \big \{ \cU_2^{-1} a^J_1 \, \cU_2 - a_1 ^J \big \} 
\label{deq.8}
\eea
Carrying out the recursion, and generalizing to arbitrary values of $i$, we obtain, 
\bea
\hat \cU_i^{-1} \, a^J_i \, \hat \cU_i & = & a_i^J + \sum _{j\not =i} \big ( \cU_j^{-1} a^J_i \, \cU_j - a_i ^J \big ) 
\label{deq.9}
\eea
Expanding $\cU_j$ in powers of $B_j$, we denote the corresponding coefficients by $\tilde \cT$ (as opposed to its expansion in powers of $H_j$ whose coefficients were denoted by $\cT$) and we have,
\bea
\cU_j^{-1} \, a^J_i \, \cU_j - a^J_i  = \sum _{r=1} ^\infty \tilde \cT^{I_1 \cdots I_r} B_{jI_1} \cdots B_{jI_r} a^J_i
= \sum _{r=1} ^\infty \tilde \cT^{I_1 \cdots I_{r-1} J} B_{jI_1} \cdots B_{jI_{r-1} } t_{ij} 
\label{deq.10}
\eea
Hence this term is also of the form of the lemma and concludes its proof.

\newpage

\section{Proof of Lemma \ref{E.lem:1}}
\label{sec:55}

In this last appendix, we provide a proof of Lemma \ref{E.lem:1}, which is at the heart of Theorem~\ref{7.thm:10}. 
Item (a) follows by construction, while  item (b) holds since $\bK_J$ and $\XX$ have vanishing $\mA$ monodromy and the $\mB$ monodromy follows from decomposing (\ref{3.Kmon3}) into $K_i$ and $K_j$  for the monodromy operator $\gamma ^{(k)}_L$.

\subsection{Proof of item (c)}

To prove item (c), i.e.\ holomorphicity of $[K_i,K_j]$ in  all its variables $x_k$ with $k=1,\cdots,n$, we first demonstrate the vanishing of the following commutator,
\bea
{} [t_{ik}, K_j] \Big |_{x_i=x_k} = 0 \qquad  {\rm for } \qquad  i,k\neq j 
 \label{vanikj}
\eea
This can be shown by expressing $K_i$ in terms  of the generating functions $\OO_J$ and $\XX$ using the first line of (\ref{7.1.b}) and then substituting the result into the above commutator with $t_{ik}$ for $i,j,k$ mutually distinct, 
\bea
{} \big  [t_{ik}, K_j(\xx;a,b,t) \big ] = \Big [ t_{ik}, {\bf W}_J(x_j;B_j)a_j^J + \sum_{\ell \neq j} {\bf X}(x_j,x_\ell;B_j) t_{j\ell} \Big ]
\label{kjwithw}
\eea
The contributions from $\OO_J$ and the terms with $\ell \not= i,k$ vanish in view of (\ref{11.3}) and (\ref{11.lem.4}), respectively, leaving only the contributions from $\ell=i,k$, 
\bea
\label{5.d.2}
\big [ t_{ik}, K_j \big ]  = 
{} \big [t_{ik}, \XX(x_j,x_i ;B_j) t_{ij} + \XX(x_j,x_k;B_j) t_{jk} \big ] 
\eea
Setting $x_i=x_k$ allows us to move a common factor $\XX(x_j,x_k;B_j)$ out of the commutator, which then 
vanishes in view of $[t_{ik} , b_{jI}] = 0$ and (\ref{11.lem.3}), thereby establishing (\ref{vanikj}). 

\sm

We then use (\ref{vanikj}) and (\ref{3.Kder}) to evaluate the derivative $\bar \p_k$ of $[K_i, K_j]$ for the case where  $k \not= i,j$, 
\bea
\bar \p_k [K_i, K_j] & = & \pi \delta (x_k, x_i) [t_{ik}, K_j] - \pi \delta (x_k,x_j) [t_{jk}, K_i] = 0
\eea
so that $[K_i, K_j]$ is holomorphic in $x_k$ for $k \not= i,j$. Evaluating also its $\bar \p_i$ derivative gives,
\bea
\label{5.d.1}
\bar \p_i [K_i, K_j] =
- \pi \delta (x_i,x_j) [t_{ij}, K_i+K_j ] - \pi \sum _{k \not= i,j} \delta (x_i,x_k) [t_{ik}, K_j] 
\eea
Since the summands $\delta (x_i,x_k) [t_{ik}, K_j] $ have $j\neq i,k$ and vanish by (\ref{vanikj}),
we are left with the following contribution from (\ref{5.d.1}),
\bea
\label{5.d.4}
\bar \p_i \big  [K_i, K_j \big ] = - \pi \delta (x_i,x_j) K_{ij}
\eea
where we have defined,
\bea
\label{5.d.9}
K_{ij} = \lim _{x_j \to x_i} \Big [ t_{ij}, K_i + K_j \big ]
\eea
To complete the proof of holomorphicity of $[K_i, K_j]$ we need to show that $K_{ij}=0$. 
We note that this quantity was proven to vanish in \cite{Enriquez:2011} using its simplicial structure.

\subsubsection{The vanishing of $K_{ij}$}

The vanishing of $K_{ij}$ is established in the following lemma.

{\lem
\label{E.lem:5}
The combination $K_{ij}$ defined in (\ref{5.d.9}) satisfies the following properties,
\begin{description}
\itemsep=-0.05in
\item (i) it is a holomorphic $(1,0)$ form in $x_i$ which is independent of $x_k$ for all $k \not= i$;
\item (ii) its $\mA$ monodromy vanishes while its $\mB$ monodromy is given by, 
\bea
\label{E.4.1}
\gamma ^{(i)} _L K_{ij} = e^{-2 \pi i (B_{iL} + B_{jL})} K_{ij}
\eea
\item (iii) its $\mA$ periods vanish,
\bea
\label{E.4.2}
\oint _{\mA^L} dx_i \,  K_{ij} =0
\eea
\item (iv) $K_{ij}=0$.
\end{description}}

Item \textit{(i)} is proven by substituting the expressions for $K_i$ and $K_j$ into (\ref{5.d.9}), using (\ref{kjwithw}) and rearranging the result as follows, 
\bea
\label{E.4.3}
K_{ij} = K^\OO _{ij} + K^\XX_{ij} 
+ \sum_{k \not = i,j}  \Big [ t_{ij} ,  \XX (x_i,x_k;B_i)  t_{ik}  +  \XX(x_i,x_k;B_j)  t_{jk}  \Big ]
\eea
where,
\bea
\label{E.4.4}
K^\OO_{ij} & = & \Big [ t_{ij}, \OO_J(x_i;B_i) a^J_i + \OO_J(x_i; B_j) a^J_j \Big ]
\no \\
K^\XX_{ij} & = & \lim_{x_j \to x_i} \Big [ t_{ij}, \XX(x_i;x_j; B_i) t_{ij}  + \XX (x_j, x_i; B_j) t_{ij} \Big ]
\eea
The contributions from the sum over $k \not = i,j$  in (\ref{E.4.3}) vanish. This may be seen using the relation $\XX (x_i,x_k;B_i)  t_{ik} = \theta( \XX (x_i,x_k;B_k) ) t_{ik}$ in terms of the antipode $\theta$ defined in (\ref{defant}), to pull the common factor $\theta( \XX (x_i,x_k;B_k))$ out of both commutators, and then using the identity $[t_{ij}, t_{ik} + t_{jk}]=0$ to show the vanishing of all the $k \not= i,j$ contributions. The remaining contributions $K^\OO _{ij}$ and $ K^\XX_{ij}$ depend only on $x_i$ and therefore, so does $K_{ij}$. Furthermore, $K_{ij}$ is holomorphic in $x_i$ because $K^\OO_{ij}$ is holomorphic, and the pole in $x_j$ at $x_i$ in the commutator giving  $K^\XX_{ij}$ cancels out in view of $[t_{ij} , t_{ij}]=0$.  These results together prove item \textit{(i)}. 

\sm

Item \textit{(ii)} follows from combining the monodromies of $K_i$ and $K_j$ given by (\ref{3.Kmon3}) via (\ref{5.d.9}).

\sm

Item \textit{(iii)} may be shown by calculating the integrals involved with the help of (\ref{3.Kper}) and the corresponding integrals for $\OO_J$ and $\XX$, given by, 
\begin{align}
\label{E.4.5}
\oint _{\mA^L} dz \, \OO_J(z;B) & =  \sfZ(B_L) \delta ^L_J - { 1 \over h} \sfY(B_L)  B_J & 
\no \\
\oint _{\mA^L} dz \, \XX(z,y;B) & =   - { 1 \over h} \sfY(B_L)  & 
\end{align}  
where the function $\sfZ(B)$ was defined in (\ref{3.Kper1})  and $\sfY(B)$ is given as follows,
\bea
\label{E.ZY}
\sfY(B) & = & { \sfZ(B) -1 \over B}  = - 2 \pi i \sum _{r=1}^\infty {{\rm Ber}_r \over r!} \, (- 2 \pi i B)^{r-1}
\eea
We note that the functions $\sfY(B)$ and $\sfZ(B)$ satisfy the following relations, 
\bea
\label{E.4.8}
\sfZ(B) - \sfZ(-B) = 2 \pi i B \hskip 1in \sfY(B) + \sfY(-B) = 2 \pi i
\eea

\sm

$\bullet$ \textit{Evaluating the $\mA$ periods of $K^\OO_{ij}$ in (\ref{E.4.4})} we find, 
\bea
\label{E.4.6}
\oint _{\mA^L} dx_i \,  K^\OO_{ij} 
& = &
\Big [ t_{ij} , \sfZ(B_{iL}) a_i^L  - { 1\over h} \sfY(B_{iL}) B_{iJ} a_i^J 
+ ( i \leftrightarrow j) \Big ]
\eea
The relation $B_{iJ} a^J_i = - \sum _{k\not= i} t_{ik}$ makes the contribution in $1/h$ proportional to, 
\bea
\label{E.4.7}
\Big [ t_{ij} ,  \sfY(B_{iL}) t_{ij} + \sfY(B_{jL})  t_{ij} \Big ] 
+ \sum_{k \not =i,j } \Big [ t_{ij} , \sfY(B_{iL}) t_{ik} + \sfY(B_{jL}) t_{jk} \Big ] 
\quad
\eea
The sum over $k$ vanishes as may be seen by using $B_{iL} t_{ik} = - B_{kL} t_{ik}$ and $B_{jL} t_{jk} = - B_{kL} t_{jk}$ to convert all $B_i$ and $B_j$ to $B_k$, pulling the common factor out of the commutator, and using the relation  $[t_{ij}, t_{ik} + t_{jk}]=0$. The first commutator of (\ref{E.4.7}) may be seen to vanish by converting the prefactor of the second term to $\sfY(-B_{iL})$ and then using the second relation of (\ref{E.4.8}).
As a result, (\ref{E.4.6}) simplifies further and may be expanded in a Taylor series in $B$  as follows,
\bea
\label{E.4.9}
\oint _{\mA^L} dx_i \,  K^\OO_{ij} 
=
\Big [ t_{ij} , \sfZ(B_{iL}) a_i^L  + \sfZ(B_{jL}) a_j^L \Big ]
=
 \sum _{r=0}^\infty \sfR_{r,0} \Big [ t_{ij}, (B_{iL})^r a_i^L + (B_{jL})^r a_j^L \Big ]
\eea
where we use the standard expansion for $\sfZ(B)$ in terms of Bernoulli numbers $\Ber_m$,
\bea
\label{E.ZR}
\sfZ(B) = \sum _{r=0}^\infty \sfR_{r,0} B^r
\hskip 1in
\sfR_{m,n} = ( -2 \pi i )^{m+n} (-)^n  {\Ber _{m+n} \over m! \, n !}
\eea
Using the notation $B_{ijL} = B_{iL} + B_{jL}$ and $B_{ijL} t_{ij}=0$, one readily shows, 
\bea
\big [ t_{ij}, (B_{iL})^r a_i^L + (B_{jL})^r a_j^L \big ]
& = & B_{ijL} \big [ t_{ij}, (B_{iL})^{r-1} a_i^L + (B_{jL})^{r-1} a_j^L \big ]
\no \\ &&
- \big [ t_{ij}, (B_{iL})^{r-1} t_{ij} + (B_{jL})^{r-1} t_{ij} \big ]
\eea
The Bernoulli numbers $\Ber_r$ and thus $\sfR_{r,0}$ are non-vanishing only for either $r=1$ or even $r \geq 0$. For $r=1$, the above combination manifestly vanishes, while for $r$ even, the second line vanishes. Iterating this equation a second time will produce a further term $B_{ijL} \big [ t_{ij}, (B_{iL})^{r-2} t_{ij} + (B_{jL})^{r-2} t_{ij} \big ]$ which vanishes in view of $B_{ijL} t_{ij}=0$. Setting $r=2s$ and solving the recursion relation gives, 
\bea
\sum _{r=0}^\infty \sfR_{r,0} \big [ t_{ij}, (B_{iL})^{2s} a_i^L + (B_{jL})^{2s} a_j^L \big ]
= \sum _{r=0}^\infty \sfR_{r,0} (B_{ijL})^{2s}  \big [ t_{ij}, a_i^L + a_j^L \big ] =0
\eea 
This completes our proof of,
\bea
\label{FF.100}
\oint _{\mA^L} dx_i \,  K^\OO_{ij} =0
\eea

\sm

$\bullet$ \textit{Evaluating the $\mA$ periods of $K^\XX_{ij}$ in (\ref{E.4.4})} is done by expanding in powers of $B$,
\bea
\oint _{\mA^L} dx_i \, K_{ij}^\XX = \sum_{r=2}^\infty \oint _{\mA^L} dx \, \chi^{I_1 \cdots I_r} (x,x) 
\Big [  t_{ij}, B_{iI_1} \cdots B_{iI_r} t_{ij} + B_{jI_1} \cdots B_{jI_r} t_{ij} \Big ]
\eea
where we have omitted the contributions for $r=0$ and $r=1$ since the corresponding Lie algebra elements vanish. 
To evaluate the integrals for $r \geq 2$ we use  a trick employed in \cite{Enriquez:2011}, 
\bea
\chi ^{I_1 \cdots I_r} (x,x) = \chi^{I_1 \cdots I_r} (x,p) + \int _p ^x d z \, \p_z \chi^{I_1 \cdots I_r}(x,z)
\label{chiprim}
\eea
for an arbitrary point $p$.  The contribution from the first term is given by,
\bea
- {1 \over h} \sum^\infty_{r=2} \sfR_{r+1,0}  \Big [ t_{ij}, (B_{iL})^r t_{ij} + (B_{jL})^r t_{ij} \Big ]
\eea
Since the Bernoulli numbers are non-vanishing only for  odd $r \geq 2$ and the commutator vanishes for odd $r$, each  summand is zero. The remaining terms are as follows,
\bea
\oint _{\mA^L} dx_i \, K_{ij}^\XX = 
\sum_{r=2}^\infty \int _p ^{\mA^L \cdot p} \!\! dx \, \int ^x _p \!\! dz \, 
\p_z \chi^{I_1 \cdots I_r} (x,z)  \Big [  t_{ij}, B_{iI_1} \cdots B_{iI_r} t_{ij} + B_{jI_1} \cdots B_{jI_r} t_{ij} \Big ]
\qquad
\eea
Recasting the second term in the commutator in terms of $B_i$ using $B_{jI} t_{ij} = - B_{iI} t_{ij}$, reversing the order of $x$ and $z$,  and relabeling the indices using both equations in (\ref{E.3.3}), we obtain,  
\bea
\oint _{\mA^L} dx_i \, K_{ij}^\XX & = & \sum_{r=2}^\infty \big [  t_{ij}, B_{iI_1} \cdots B_{iI_r} t_{ij}  \big ] 
\no \\ && \quad \times 
\bigg \{  \int _p ^{\mA^L \cdot p}  \!\! dx \, \int ^x _p dy + \int _p ^{\mA^L \cdot p}  \!\! dy \, \int ^y _p dx \bigg \} 
 \p_y \chi^{I_1 \cdots I_r} (x,y) 
\eea
The sum of the two integrals is the integral over the square,
\bea
\oint _{\mA^L} dx_i \, K_{ij}^\XX & = & \sum_{r=2}^\infty  \big [  t_{ij}, B_{iI_1} \cdots B_{iI_r} t_{ij}  \big ] 
 \int _p ^{\mA^L \cdot p}  dx  \int _p ^{\mA^L \cdot p}  dy \,  \p_y \chi^{I_1 \cdots I_r} (x,y) =0
\eea
This shows that,
\bea
\label{FF.200}
\oint _{\mA^L} dx_i \, K_{ij}^\XX=0
\eea
Combining the results of (\ref{FF.100}) and (\ref{FF.200}) with the fact that the sum over $k$ in (\ref{E.4.3}) is identically vanishing proves (\ref{E.4.2}) and thus item \textit{(iii)}. 

\sm

Item \textit{(iv)} follows from using items \textit{(i)}, \textit{(ii)} and \textit{(iii)} to satisfy the assumptions of Lemma~\ref{4.lem:100} with $\nu(z) = K_{ij}$ for $z=x_i$ and $\beta_I = b_{iI}+ b_{jI}$ which then implies $K_{ij}=0$. This completes the proof of Lemma \ref{E.lem:5} and thus also of item (c) in Lemma \ref{E.lem:1}.

\subsection{Proof of  item (d)}
\label{sec:46}

To establish item (d) of  Lemma \ref{E.lem:1}, we need to show the vanishing of the $\mA$ period $\cV^K_{ij}$ of the commutator $[K_i, K_j]$, given in (\ref{E.7.q}).  In view of item (a) of Lemma \ref{E.lem:1}, the combination $\cV_{ij}^K$ is a holomorphic $(1,0)$ form in $x_j$, namely $\bar \p_j \cV_{ij}^K = 0$. Furthermore, its monodromy is known since the monodromy of the commutator is known from (\ref{E.7.p}). Thus, it remains to prove that the $\mA$-period of $\cV_{ij}^K$ in the variable $x_j$,\footnote{Since $[K_i,K_j]$ is a holomorphic $(1,0)$ form in both $x_i$ and $x_j$, the order in which the contour integrals are being carried out, including for $L=K$, is immaterial. As a result, the double integral satisfies $ \cV_{ji}^{LK} = - \cV_{ij}^{KL}$.  In particular, for $L=K$ we must have $\cV_{ji} ^{KK} = - \cV_{ij}^{KK}$. We now proceed to evaluate $\cV_{ij}^{KL}$. }
 \bea
 \label{4.Vij}
 \cV^{KL}_{ij}  =  \oint _{\mA^L} dx_j \, \oint _{\mA^K} dx_i \, [K_i, K_j]
 \eea
 vanishes. Indeed, with those three properties of $\cV^K_{ij}$, namely holomorphicity in $x_j$, monodromy in $x_j$ inherited from (\ref{E.7.p}), and vanishing $\mA$ periods in $x_j$, the assumptions of Lemma \ref{4.lem:100} are satisfied so that $\cV^K_{ij}=0$. 
 
 \sm
 
 The remainder of this appendix is dedicated to calculating the double integral (\ref{4.Vij}) and showing that it vanishes.

\subsubsection{Decomposing the double integral $\cV^{KL}_{ij}$} 
 
To compute $\cV^{KL}_{ij}$, we isolate the single term in $K_i$ that depends on $x_j$,
 \bea
 K_i = S_{ij} + \XX(x_i,x_j;B_i) t_{ij}
 \eea
 where $S_{ij}$ is independent of $x_j$ and given by,
 \bea
S_{ij} =  \OO_J(x_i;B_i) a^J_i +  \sum_{k \neq i,j}\XX (x_i,x_k;B_i)  t_{ik}  
\eea 
 The integral (\ref{4.Vij}) may be decomposed into a sum of integrals, as follows, 
 \bea
 \label{E.x.2}
 \cV_{ij}^{KL} & = & \cW^{KL}_{ij}  + 
\lim _{\ep \to 0}  \oint _{\mA^L_\ep } dx_j \oint _{\mA^K} dx_i \, \bigg \{ 
\big [S_{ij}, S_{ji} \big ] + \big [ \XX(x_i,x_j;B_j) t_{ij} , S_{ji} \big ]
\no \\ && \hskip 2.5in
+ \big [ S_{ij}, \XX(x_j,x_i;B_j) t_{ij} \big ]
 \bigg \}
\eea
where we have defined,
\bea
\label{4.Wij}
 \cW_{ij}^{KL} = \lim _{\ep \to 0}  \oint _{\mA^L_\ep} dx_j \oint _{\mA^K} dx_i \, \Big [ \XX(x_i,x_j;B_i) t_{ij} , \XX(x_j, x_i;B_j) t_{ij} \Big ]
 \eea
 The integration contour $\mA^L_\ep$ homotopic to the homology cycle $\mA^L$ is obtained by 
 infinitesimally displacing the latter into the interior of the preferred fundamental domain $D_p$ as drawn in
 figure 3 of \cite{DHoker:2025dhv}. Next, we proceed to evaluating these integrals using (\ref{E.4.5}), with the further help of,
 \bea
 \label{E.x.1}
  \oint _{\mA^K} dx_i \, S_{ij} = \sfZ(B _{iK}) \, a_i^K + { 1 \over h} \sfY(B_{iK})  t_{ij}
 \eea 
 The integrand $[S_{ij}, S_{ji} ]$ contains no poles at $x_i=x_j$, so that the limit $ \ep \to 0$ is trivial and the limit instruction can be omitted. It may be evaluated using (\ref{E.x.1}),
\bea
\oint _{\mA^L } dx_j \oint _{\mA^K} dx_i \,  \big [S_{ij}, S_{ji} \big ]
= \left [ \sfZ(B _{iK}) \, a_i^K + { 1 \over h} \sfY(B_{iK})  t_{ij}, \sfZ(B _{jL}) \, a_j^L + { 1 \over h} \sfY(B_{jL})  t_{ij} \right ]
\qquad
\eea
To evaluate the integral of the second term in the  integrand of (\ref{E.x.2}), we first  carry out the integral over $x_i$ of the first argument of the commutator since the second argument is independent of $x_i$. The result of the integral over $x_i$ is independent of $x_j$ which allows us to evaluate the integral over $x_j$, and we obtain, 
\bea
\oint _{\mA^L } dx_j \oint _{\mA^K} dx_i \, \big [ \XX(x_i,x_j;B_j) t_{ij} , S_{ji} \big ]
= - { 1 \over h} \left [ \sfY(B_{iK}) t_{ij}, \sfZ(B _{jL}) \, a_j^L + { 1 \over h} \sfY(B_{jL}) t_{ij} \right ]
\qquad
\eea
Finally, to compute the integral of the third term in the integrand, we need to carry out the integral over $x_j$ first, but that is not the order in which the integrations appear. Thus, we must interchange the order, at the cost of picking up the residue of the pole at $x_i=x_j$ when $L=K$, and we obtain,
\begin{align}
\lim_{\ep \to 0} \oint _{\mA^L _\ep} dx_j & \oint _{\mA^K} dx_i \, \big [ S_{ij}, \XX(x_j,x_i;B_j) t_{ij} \big ]
\no \\ & = 
 \left [ \sfZ(B _{iK}) \, a_i^K + { 1 \over h} \sfY(B_{iK})  t_{ij}, - { 1 \over h} \sfY(B_{jL})  t_{ij} + 2 \pi i  \delta^K_L t_{ij} \right ]
\label{swapij}
\end{align}
where a careful calculation of the sign of the residue for the contour prescription for $\mA^L _\ep$ \cite{DHoker:2025dhv} determines the sign of the last term. The sum of these contributions gives, 
 \bea
 \label{E.x.3}
 \cV_{ij} ^{KL} & = & \cW^{KL}_{ij} + \big [ \sfZ(B _{iK} ) \, a^K_i, \, \sfZ(B _{jL}) \, a^L_j \big ]
 - { 1 \over h^2} \left [ \sfY(B _{iK})  t_{ij}, \sfY(B_{jL})  t_{ij}  \right ]
 \no \\ && \hskip 0.4in
 + 2 \pi i  \delta^{KL} \left [ \sfZ(B _{iK}) \, a_i^K + { 1 \over h} \sfY(B_{iK})  t_{ij}, t_{ij} \right ]
 \eea 
 where $\cW_{ij}^{KL}$ is given in (\ref{4.Wij}).

 \subsubsection{Convolution of $\chi$-forms}
 
 It remains to evaluate the contribution $\cW_{ij}^{KL}$ for which both factors in the integrand depend non-trivially on both $x_i$ and $x_j$. To this end, we begin by proving the following corollary of Theorem 3 of \cite{DHoker:2025dhv},
 {\cor
 \label{4.cor:1}
 The following convolution integral holds (no Einstein summation convention for $K$ such that $ \varpi ^{J_1 \cdots J_{\ell-1} K}{}_K(y)$ in the second line does not vanish), 
  \bea
 \oint _{\mA^K}  dx_i \, \chi^{J_1 \cdots J_s }  (y,x_i)  \, \chi^{I_1 \cdots I_r }(x_i,z) 
& = &
 + { 1 \over h}   \sum _{\ell=0}^s  \sfR_{r+1,s+1-\ell} \, \delta _K ^{J_s \cdots J_{\ell+1} I_1 \cdots I_r} \, 
 \varpi ^{J_1 \cdots J_\ell}{}_K(y) \label{chilemm}
  \\ &&
 - { 1 \over h} \sum_{\ell=1}^{s+1} \sfR_{r+1,s+1-\ell} \, \delta ^{J_s \cdots J_{\ell} I_1 \cdots I_r} _K \, 
 \varpi ^{J_1 \cdots J_{\ell-1} K}{}_K(y)
  \no \\ &&
 +   \sum_{\ell=0}^s \sum_{k=0}^r \sfR_{k, s-\ell} \, \delta _K^{J_s \cdots J_{\ell+1} I_1 \cdots I_k} \,  
 \chi^{J_1 \cdots J_\ell K I_{k+1} \cdots I_r } (y,z)
 \no \quad
 \eea 
 where $ \sfR_{m,n}$ was defined in (\ref{E.ZR}).}
 
 \sm

 To prove the corollary, we make use of the following convolution established in equation (17) of Theorem 3 in \cite{DHoker:2025dhv}, here rendered symmetrical in the factors of the integrand by singling out the index $C$ in the second factor,
 \begin{align}
 \label{4.cor.1}
 \oint _{\mA^K}  dx_i \, g^{J_1 \cdots J_s A}{}_B  (y,x_i) & \, g^{I_1 \cdots I_r C}{}_D(x_i,z)
 = \sfR_{r+1,0} \, \delta _D^{I_1 \cdots I_r CK} \, g^{J_1 \cdots J_s A}{}_B (y,z)
 \no \\ &
 + \delta ^A_B \sum _{\ell=0}^s  \sfR_{r+1,s+1-\ell} \, \delta _D ^{J_s \cdots J_{\ell+1} I_1 \cdots I_r CK} \, g^{J_1 \cdots J_\ell}{}_K(y,z)
 \no \\ &
 - \delta ^A_B  \sum_{\ell=1}^{s+1} \sfR_{r+1,s+1-\ell} \, \delta ^{J_s \cdots J_{\ell} I_1 \cdots I_rC} _K \, 
 g^{J_1 \cdots J_{\ell-1} K}{}_D(y,z)
  \no \\ &
 - \delta ^A_B \sum_{\ell=0}^s \sum_{k=0}^r \sfR_{k, s-\ell} \, \delta _K^{J_s \cdots J_{\ell+1} I_1 \cdots I_k} \,  
 g^{J_1 \cdots J_\ell K I_{k+1} \cdots I_r C}{}_D (y,z)
 \end{align}
Contracting with $\delta ^B_A \, \delta ^D_C$, and writing $g$ in terms of $\varpi$ and $\chi$ in the second and third lines, we see that the $\chi$ part cancels between the two for all $1 \leq \ell \leq s$, leaving only the $\ell=s{+}1$ contribution, which in turn cancels against the first line and gives  (\ref{chilemm}).

\subsubsection{Calculation of $\cW_{ij}^{KL}$}
 
Expanding the integral $\cW_{ij}^{KL}$ in powers of $B$, we obtain,
\bea
\cW_{ij}^{KL} & = & \sum_{r,s=0}^\infty \oint _{\mA^L} dx_j \oint _{\mA^K} dx_i \, 
\chi ^{J_1 \cdots J_s} (x_j, x_i) \, \chi^{I_1 \cdots I_r}(x_i,x_j)
\no \\ && \hskip 1in \times
{} \Big [ B_{iI_1} \cdots B_{iI_r} t_{ij}, B_{jJ_1} \cdots B_{jJ_s} t_{ij} \Big ]
\eea
and using the convolution above to carry out the integral over $x_i$, we arrange the result according to the
three terms in the convolution (\ref{chilemm}), 
\bea
\cW_{ij}^{KL} & = & W_1+W_2
\eea
where, 
\bea
W_1 & = & { 1 \over h}  \sum_{r,s=0}^\infty \sum _{\ell=0}^s  \sfR_{r+1,s+1-\ell} \, 
\oint _{\mA^L} dx_j  \,  \varpi^{J_1 \cdots J_\ell}{}_K(x_j)
\Big [ (B_{iK})^r  t_{ij}, B_{jJ_1} \cdots B_{jJ_\ell } (B_{jK})^{s-\ell} t_{ij} \Big ]
\no \\ &&
 - { 1 \over h}  \sum_{r,s=0}^\infty   \, \sum_{\ell=1}^{s+1} \sfR_{r+1,s+1-\ell} \, 
 \oint _{\mA^L} dx_j  \, \varpi^{J_1 \cdots J_{\ell-1} K}{}_K(x_j)
 \no \\ && \hskip 1in \times
 \Big [ (B_{iK})^r  t_{ij}, B_{jJ_1} \cdots B_{jJ_{\ell-1}} (B_{jK})^{s-\ell+1} t_{ij} \Big ]
\eea
and 
\bea
W_2 & = & \sum_{r,s=0}^\infty \sum_{k=0}^r \sum _{\ell=0}^s \sfR_{k, s-\ell} \, 
\oint _{\mA^L} dx_j  \chi^{J_1 \cdots J_\ell K I_{k+1} \cdots I_r } (x_j, x_j)
\no \\ && \hskip 1in \times 
\Big [ (B_{iK})^k B_{iI_{k+1}} \cdots B_{iI_r} t_{ij}, B_{jJ_1} \cdots B_{jJ_\ell} (B_{jK})^{s-\ell} t_{ij} \Big ]
\eea

\subsubsection*{$\bullet$ The vanishing of $W_2$}

In $W_2$, we change summation variables $ \ell \to s-\ell$, and redefine the subscripts of the $I$ labels $I_{k+1} \to I_1, \cdots ,I_r \to I_{r-k}$, to obtain,  
\bea
W_2 & = & \sum_{r,s=0}^\infty \sum_{k=0}^r \sum _{\ell=0}^s \sfR_{k, \ell} \, 
\oint _{\mA^L} dx_j \, \chi^{J_1 \cdots J_{s-\ell}K I_{1} \cdots I_{r-k} } (x_j, x_j)
\no \\ && \hskip 1in \times 
\Big [ (B_{iK})^k B_{iI_{1}} \cdots B_{iI_{r-k}} t_{ij}, B_{jJ_1} \cdots B_{jJ_{s-\ell}} (B_{jK})^\ell t_{ij} \Big ]
\eea
Next, we change summation variables from $(r,s ) \to (\rho, \sigma)$ with $r= k + \rho$ and $s = \ell + \sigma$, let $x_j \to x$, and recast the integral as a double integral as in (\ref{chiprim}),
\bea
W_2 & = & \sum_{k,\ell, \rho, \sigma =0} ^\infty  \sfR_{k, \ell} \, 
\oint _{\mA^L} dx  \Big \{ \chi ^{J_1 \cdots J_\sigma K I_1 \cdots I_\rho } (x,p) +  \int _p ^{x} d y \, \p_{y} \chi^{J_1 \cdots J_\sigma K I_1 \cdots I_\rho } (x,y) \Big \} 
\no \\ && \hskip 1in \times 
\Big [ (B_{iK})^k B_{iI_1} \cdots B_{iI_\rho } t_{ij}, B_{jJ_1} \cdots B_{jJ_\sigma} (B_{jK})^\ell t_{ij} \Big ]
\label{w3expr}
\eea
First,  consider the sum of the double integral  terms paired as follows, $(k, \rho) \leftrightarrow (\ell, \sigma)$, 
\bea 
\sfR_{k,\ell} \oint _{\mA^L} dx  \int _p ^{x} d y \, \p_{y} \chi^{J_1 \cdots J_\sigma K I_1 \cdots I_\rho } (x,y)
\Big [ (B_{iK})^k B_{iI_1} \cdots B_{iI_\rho } t_{ij}, B_{jJ_1} \cdots B_{jJ_\sigma} (B_{jK})^\ell t_{ij} \Big ]
\no \\
+ \sfR_{\ell,k} \oint _{\mA^L} dx  \int _p ^{x} d y \, \p_{y} \chi^{I_1 \cdots I_\rho K J_1 \cdots J_\sigma } (x,y)
\Big [ (B_{iK})^\ell B_{iJ_1} \cdots B_{iJ_\sigma } t_{ij}, B_{jI_1} \cdots B_{jI_\rho} (B_{jK})^k t_{ij} \Big ]
\quad
\eea 
where we have relabeled $I \leftrightarrow J$. In the second line, we use,
 \bea
&&
\Big [ (B_{iK})^\ell B_{iJ_1} \cdots B_{iJ_\sigma } t_{ij}, B_{jI_1} \cdots B_{jI_\rho} (B_{jK})^k t_{ij} \Big ]
\no \\ && \qquad =  
(-)^{k+\ell+\rho+\sigma+1} \Big [ (B_{iK})^k B_{iI_\rho} \cdots B_{iI_1 } t_{ij}, B_{jJ_\sigma} \cdots B_{jJ_1} (B_{jK})^\ell t_{ij} \Big ]
\quad
\eea 
and
\bea
\p_{y} \chi^{I_1 \cdots I_\rho K J_1 \cdots J_\sigma } (x,y)
= (-)^{\rho+\sigma+1} \p_{x} \chi^{J_\sigma  \cdots J_1  K I_\rho \cdots I_1 } (y,x)
\eea 
as well as $\sfR_{\ell,k} = (-)^{k+\ell} \sfR_{k,\ell}$. Upon suitably relabeling $I$ and $J$ the commutator factors become identical, as are the factor of $\sfR_{k,\ell}$ while the integrals add up to,
\bea
\oint _{\mA^L} dx  \int _p ^{x} d y \, \p_{y} \chi^{J_1 \cdots J_\sigma K I_1 \cdots I_\rho } (x,y) 
+ 
\oint _{\mA^L} dx  \int _p ^{x} d y \, \p_{x} \chi^{J_1  \cdots J_\sigma  K I_1 \cdots I_\rho } (y,x)
\eea 
swapping integration variables, the sum is equal to,
\bea
\Big ( \oint _{\mA^L} dx  \int _p ^{x} d y \, + \oint _{\mA^L} dy  \int _p ^{y} d x \Big ) 
\p_{y} \chi^{J_1 \cdots J_\sigma K I_1 \cdots I_\rho } (x,y) =0
\eea
It remains to evaluate the contribution to (\ref{w3expr}) from the terms involving only a single integral
which can be straightforwardly performed via components of (\ref{E.4.5}), 
\bea
W_2 & = & - { 1 \over h} \delta^K_L \sum_{k,\ell, \rho, \sigma =0} ^\infty  \sfR_{k, \ell} \,  \sfR_{\rho+\sigma+2,0} \, 
\delta_L^{J_1 \cdots J_\sigma  I_1 \cdots I_\rho }  
\Big [ (B_{iK})^{k+\rho}   t_{ij}, (B_{jK})^{\ell+\sigma}  t_{ij} \Big ]
\eea
Taking again the pair $(k, \rho) \leftrightarrow (\ell, \sigma)$ in the summand, gives,
\bea
(-)^\ell \Big [ (B_{iK})^{k+\rho}   t_{ij}, (B_{jK})^{\ell+\sigma}  t_{ij} \Big ]
+ (-)^k \Big [ (B_{iK})^{\ell+\sigma }   t_{ij}, (B_{jK})^{k+\rho }  t_{ij} \Big ]
\no \\ =
(-)^\ell \Big ( 1 - (-)^{\rho+\sigma} \Big ) \Big [ (B_{iK})^{k+\rho}   t_{ij}, (B_{jK})^{\ell+\sigma}  t_{ij} \Big ]
\eea
Since the prefactor $\sfR_{\rho+\sigma+2,0}$ is proportional to ${\rm Ber}_{\rho+\sigma+2}$ it must vanish for $\rho+\sigma$ odd and hence this contribution vanishes. We conclude that $W_2=0$.

\subsubsection*{$\bullet$ Calculating $W_1$}

Recalling the integrals (no Einstein summation conventions for $K$ in the second line), 
\bea
\oint _{\mA^L} dx \, \varpi ^{J_1 \cdots J_\ell} {}_K(x) & = &
\sfR_{\ell,0} \Big ( \delta ^{J_1 \cdots J_\ell L}_K - { 1 \over h} \delta ^{J_\ell}_K \, \delta ^{J_1 \cdots J_{\ell-1}} _L \Big )
\no \\
\oint _{\mA^L} dx \, \varpi ^{J_1 \cdots J_{\ell -1} K } {}_K(x) & = &
\sfR_{\ell ,0} \Big ( \delta ^{J_1 \cdots J_{\ell -1} L}_K - { 1 \over h}  \delta ^{J_1 \cdots J_{\ell-1} } _L \Big )
\eea
we can carry out the integrals to obtain, 
\bea
W_1 & = & { 1 \over h}  \sum_{r,s=0}^\infty \bigg \{ \sum _{\ell=0}^s  R_{r+1,s+1-\ell} \, 
\sfR_{\ell,0} \Big ( \delta ^{J_1 \cdots J_{\ell-1} L}_K - { 1 \over h} \delta ^{J_1 \cdots J_{\ell-1}} _L \Big )
\no \\ && \hskip 0.5in
- \sum_{\ell=1}^{s+1} R_{r+1,s+1-\ell} \, 
\sfR_{\ell ,0} \Big ( \delta ^{J_1 \cdots J_{\ell-1}  L}_K - { 1 \over h}  \delta ^{J_1 \cdots J_{\ell-1} } _L \Big ) \bigg \} 
 \no \\ && \hskip 1in \times
 \Big [ (B_{iK})^r  t_{ij}, B_{jJ_1} \cdots B_{jJ_{\ell-1}} (B_{jK})^{s-\ell+1} t_{ij} \Big ]
 \label{exprw1}
\eea
The $1/h$ contribution may be rearranged as follows, 
\bea
\label{FF.16}
{ 1 \over h} \delta^{KL}  \sum_{r,s=0}^\infty \Big ( \sfR_{r+1,s+1} - \sfR_{r+1,0} \sfR_{s+1,0} \Big )  \Big  [ (B_{iK})^r t_{ij}, \, (B_{jK})^s t_{ij} \Big  ]
\eea
The sum of the first term inside the parentheses may be simplified  by symmetrizing the summand in $r,s$ and using the relation $\sfR_{s+1,r+1} = (-)^{r+s} \sfR_{r+1,s+1}$ as follows, 
\begin{align}
\Big  [ (B_{iK})^r t_{ij}, & \, (B_{jK})^s t_{ij} \Big  ] 
+ (-)^{r+s} \Big  [ (B_{iK})^s t_{ij}, \, (B_{jK})^r t_{ij} \Big  ] 
\no \\ & = 
\Big  [ (B_{iK})^r t_{ij},  \, (B_{jK})^s t_{ij} \Big  ] 
+  \Big  [ (B_{jK})^s t_{ij}, \, (B_{iK})^r t_{ij} \Big  ] =0
\end{align}
so that the contribution from the first term vanishes identically. The sum of the second term inside the parentheses may be simplified by symmetrizing the commutators in $r,s$,
\begin{align}
\Big  [ (B_{iK})^r t_{ij}, & \, (B_{jK})^s t_{ij} \Big  ] 
+ \Big  [ (B_{iK})^s t_{ij}, \, (B_{jK})^r t_{ij} \Big  ] 
\no \\ & = 
\big ( 1 + (-)^{r+s+1} \big )  \Big  [ (B_{iK})^r t_{ij}, \, (B_{jK})^s t_{ij} \Big  ]
\end{align}
In view of the prefactor on the right side, the only non-vanishing contributions can come from $r+s$ odd so that either $r$ is even and $s$ is odd or the reverse.   It follows that $\sfR_{r+1,0} \sfR_{s+1,0}$ vanishes unless either $r$ or $s$ vanishes. Since $\sfR_{1,0} = \pi i $, (\ref{FF.16}) reduces to,
\bea
- {  \pi i \over h} \delta^{KL}  \sum_{r=0}^\infty (1-(-)^r) \sfR_{r+1,0} \Big  [ (B_{iK})^r t_{ij}, \, t_{ij} \Big  ]
= - {2 \pi i \over h} \delta^{KL} [ \sfY(B_{iK}) t_{ij}, t_{ij} ]
\eea

The $1/h^2$ contributions in the expression (\ref{exprw1}) for $W_1$ cancel one another for all $\ell=1, \cdots, s$. For $\ell=0$, the contribution from the first term in $W_1$ actually vanishes. Therefore, the only remaining contribution is from the $\ell=s+1$ term in the second line of $W_1$ which~gives, 
\bea
 { 1 \over h^2} \sum_{r,s=0}^\infty \sfR_{r+1,0} \sfR_{s+1,0}
\Big [ (B_{iK})^r t_{ij}, \, (B_{jL})^s  t_{ij} \Big ]
= { 1 \over h^2} \Big [ \sfY(B_{iK} ) t_{ij} , \, \sfY(B_{jL} ) t_{ij} \Big ]
\eea
leading to
 \bea
 \cW_{ij}^{KL} = - {2 \pi i \over h} \delta^{KL} [ \sfY(B_{iK}) t_{ij}, t_{ij} ]
 +  { 1 \over h^2} \Big [ \sfY(B_{iK} ) t_{ij} , \, \sfY(B_{jL} ) t_{ij} \Big ]
 \label{E.new}
 \eea

 \subsubsection{Summary}
 \label{summsec}
 
 Putting together all contributions in (\ref{E.x.3}) and (\ref{E.new}), the terms in $1/h$ and $1/h^2$ both cancel against $ \cW_{ij}^{KL} $ and we~have, 
  \bea
 \cV_{ij} ^{KL}  =  
 \Big [ \sfZ(B _{iK} ) \, a^K_i, \, \sfZ(B _{jL}) \, a^L_j + 2 \pi i \, \delta ^{KL}  t_{ij} \Big ] 
\label{vklij}
 \eea 
 For $L \not=K$, this formula simplifies by dropping the second term in the commutator, while the remainder may be expanded as follows, 
 \bea
\cV^{KL}_{ij}   \, \Big|_{K\neq L} 
 = \sum_{r,s=0}^\infty \sfR_{r,0} \sfR_{s,0}  \Big [ (B_{iK})^r a_i^K, (B_{jL})^s a_j^L \Big ] 
 \eea
Since $L \not=K$ and $j \not = i$, the elements $B_{iK}$ and $B_{jL}$ commute with one another and we also have $B_{iK} a^L_j=0$ so that,
 \bea
 \big [ (B_{iK})^r a_i^K, (B_{jL})^s a_j^L \big ] = (B_{iK})^r (B_{jL})^s \big [  a_i^K,  a_j^L \big ] =0
 \eea
 which proves the vanishing of $ \cV_{ij} ^{KL} $ in (\ref{vklij}) for $L \not= K$. We address the $L=K$ case next.

 \subsubsection{Calculation of $\cV_{ij}^{KK}$}
 
It remains to show $\cV_{ij}^{KL}=0$ in the case $K=L$, for which we have the following lemma.  
{\lem  
\label{4.lem:7}
We have the following algebraic identity, 
\bea
 \cV_{ij} ^{KK} =   \Big [ \sfZ(B _{iK} ) \, a^K_i, \, \sfZ(B _{jK}) \, a^K_j + 2 \pi i \,  t_{ij} \Big ] =0
 \eea
 where the function $\sfZ(B)$ was defined in (\ref{3.Kper1}). }
 
 \sm
   
We shall provide a constructive proof by calculating the two contributions,
\begin{align} 
\cV_{ij}^{KK} & = 2 \pi i \cV^{(1)} _{ij}  + \cV^{(2)}_{ij} &
\cV^{(1)} _{ij} & =   \Big [ \sfZ(B _{iK} ) \, a^K_i,  t_{ij} \Big ] 
\label{defv1v2} \\ 
&& \cV^{(2)}_{ij} & =   \Big [ \sfZ(B _{iK} ) \, a^K_i, \, \sfZ(B _{jK}) \, a^K_j \Big ] \notag
\end{align}
and eventually showing that $\cV^{(2)} _{ij}  = -2\pi i \cV^{(1)} _{ij} $.
In various instances, we shall make use of the following two versions of Leibniz's rule, 
\bea
B^r \big [ X,Y \big ] & = & \sum _{k=0}^r  \binom{r}{k} \, \left [  B^{r-k} X, B^k Y \right ]
\no \\
\big [ B^r X, Y \big ] & = & \sum _{k=0}^r (-)^k \binom{r}{k} \, B^{r-k} \left [ X, B^k Y \right ]
\label{leibrule}
\eea

\subsubsection{Calculation of $\cV^{(1)} _{ij}$}

Expanding $\sfZ(B_{iK})$ in powers of $B_{iK}$, and using the second version of Leibniz's rule in order to move $B_{iK}$ to either outside of the commutator, or onto $t_{ij}$ where we can convert it to $B_{jK}$ using $B_{iK}^k t_{ij} = (-)^k B_{jK}^k t_{ij}$, we obtain, 
\bea
\cV^{(1)} _{ij} & = & 
\sum _{r=0}^\infty \sfR_{r,0} \sum _{k=0}^r  \binom{r}{k} B_{iK}^{r-k} \Big [ a_i^K, \, B_{jK}^k  t_{ij} \Big ]
\eea
Now use the second version of Leibniz's rule again to move $B_{jK}$ to either outside of the commutator, or onto $a_i^K$, 
 \bea
\cV^{(1)} _{ij} & = & 
\sum _{r=0}^\infty \sfR_{r,0} \sum _{k=0}^r  \binom{r}{k}  
\sum_{p=0}^k (-)^p \binom{k}{p} B_{iK}^{r-k}B_{jK}^{k-p} \big [ B_{jK}^p a_i^K, \,  t_{ij} \big ]
\eea
We separate the $p=0$ contribution and, for $p \geq 1$, use $B_{jK}^p a_i^K = B_{jK}^{p-1} t_{ij}$,
\bea
\label{4.ZZ}
\cV^{(1)} _{ij}  & = & 
\sum _{r=0}^\infty \sfR_{r,0} \sum _{k=0}^r  \binom{r}{k} B_{iK}^{r-k} B_{jK}^{k} \big [ a_i^K, \,  t_{ij} \big ]
\no \\ &&
+ \sum _{r=0}^\infty \sfR_{r,0} \sum _{k=1}^r  \binom{r}{k} 
\sum_{p=1}^k (-)^p \binom{k}{p} B_{iK}^{r-k}  B_{jK}^{k-p} \big [ B_{jK}^{p-1} t_{ij}, \,  t_{ij} \big ]
\eea
In the second line, we convert $B_{iK} ^{r-k} \to (-)^{r-k} B_{jK}^{r-k}$, since the operators effectively always act on $t_{ij}$ where they may be converted into one another, and interchange the orders of summation over $p$ and $k$,
\bea 
\sum _{r=0}^\infty \sfR_{r,0} \sum _{p=1}^r   (-)^{r+p} \binom{r}{p} B_{jK}^{r-p} \big [ B_{jK}^{p-1} t_{ij}, \,  t_{ij} \big ]
\sum_{k=p}^r (-)^{k}   \binom{r-p}{r-k} 
\label{cavbinom}
\eea
The last sum evaluates to $ (-1)^r \delta_{r,p}$ which, together with the first line of (\ref{4.ZZ}), gives
\bea
\cV^{(1)} _{ij}  & =  
\sum_{r=0}^\infty \sfR_{r,0} 
B_{ijK}  ^r  \big [ a_i^K, t_{ij} \big ] 
 + \sum_{r=2}^\infty \sfR_{r,0}  \big [ t_{ij},  B_{iK}^{r-1}  t_{ij} \big ] 
 \label{cv1exp}
 \eea
 where we continue to use the abbreviation $B_{ijK}= B_{iK} + B_{jK}$ and dropped the $r=1$ term of the
 second sum since $[ t_{ij}, (B_{iK})^{r-1} t_{ij}  ] $ vanishes in that case.

\subsubsection{Calculation of $\cV^{(2)} _{ij}$}

Expanding both $\sfZ(B_{iK})$ and $\sfZ(B_{jK})$ in powers of their arguments, and  
using the second version of Leibniz's rule in (\ref{leibrule}) on both $B_{iK}$ and $B_{jK}$, we bring
$\cV^{(2)} _{ij}$ in (\ref{defv1v2}) into the~form, 
\bea
\cV^{(2)} _{ij} & = & 
\sum _{r,s=0}^\infty \sfR_{r,0} \, \sfR_{s,0} \sum_{k=0}^r \sum_{\ell=0}^s (-)^{k+\ell} \binom{r}{k} \binom{s}{\ell} 
B_{iK} ^{r-k} B_{jK}^{s-\ell} \big [ B_{jK}^\ell a_i^K, \,B_{iK}^k a_j^K \big ]
\eea
Changing variables from $r$ to $r=k+\rho$ and $s$ to $s=\ell+\sigma$, 
\bea
\cV^{(2)} _{ij} & = & 
\sum_{k,\ell, \rho, \sigma=0} ^\infty \sfR_{k+\rho,0} \, \sfR_{\ell+\sigma,0} \, (-)^{k+\ell} 
\binom{k+\rho}{k} \binom{\ell+\sigma}{\ell} B_{iK} ^\rho B_{jK}^\sigma \big [ B_{jK}^\ell a_i^K, \,B_{iK}^k a_j^K \big ]
\eea
The contribution with $(k,\ell)=(0,0)$ manifestly vanishes. The remaining contributions are split according to whether $k$ and/or $\ell$ is non-vanishing, 
\bea
\cV^{(2)} _{ij} & = & U_1+U_2+U_3
\eea
where,
\bea
U_1 & = & \sum_{k=1}^ \infty \sum_{\rho, \sigma=0} ^\infty \sfR_{k+\rho,0} \, \sfR_{\sigma,0} \, (-)^{k} 
\binom{k+\rho}{k} B_{iK} ^\rho B_{jK}^\sigma \big [ a_i^K, \,B_{iK}^{k-1} t_{ij} \big ]
\label{defUUs} \\
U_2 & = &  \sum_{\ell=1}^ \infty \sum_{\rho, \sigma=0} ^\infty \sfR_{\rho, 0} \, \sfR_{\ell+\sigma,0} \, (-)^{\ell} 
\binom{\ell+\sigma}{\ell} B_{iK} ^\rho B_{jK}^\sigma \big [ B_{jK}^{\ell-1} t_{ij},  a_j^K \big ]
\no \\ 
U_3 & = & \sum_{k,\ell=1}^ \infty \sum_{\rho, \sigma=0} ^\infty \sfR_{k+\rho,0} \, \sfR_{\ell+\sigma,0} \, (-)^{k+\ell} 
\binom{k+\rho}{k} \binom{\ell+\sigma}{\ell} B_{iK} ^\rho B_{jK}^\sigma \big [ B_{jK}^{\ell-1} t_{ij}, \,B_{iK}^{k-1} t_{ij}  \big ]
\no
\eea
We proceed to evaluate each one in turn.

\subsubsection*{ $\bullet$ Calculation  of $U_3$}

We apply Leibniz's rule to each one of the two factors outside the commutator, 
\bea
\label{FF.17}
U_3 & = & \sum_{k,\ell=1}^ \infty \sum_{\rho, \sigma=0} ^\infty \sum _{\a =0}^\rho \sum_{\b =0}^\sigma 
\sfR_{k+\rho,0} \, \sfR_{\ell+\sigma,0} \, (-)^{\a+\b} 
\binom{k+\rho}{k} \binom{\ell+\sigma}{\ell} \binom{\rho}{\a} \binom{\sigma}{\b} 
\no \\ && \hskip 1in \times
\big [ B_{iK}^{\rho-\a +\b + \ell-1} t_{ij}, \,B_{jK}^{\sigma - \b + \a +k-1} t_{ij}  \big ]
\eea
Next, we pair up the contributions obtained by swapping $(k,\rho,\a) \leftrightarrow (\ell, \sigma, \b)$, 
\bea
&&\big [ B_{iK}^{\rho-\a +\b + \ell-1} t_{ij}, \,B_{jK}^{\sigma - \b + \a +k-1} t_{ij}  \big ]
+ 
\big [ B_{iK}^{\sigma - \b +\a + k-1} t_{ij}, \,B_{jK}^{\rho - \a + \b +\ell-1} t_{ij}  \big ]
\no \\ && \hskip 0.7in
=
\Big ( 1 - (-)^{\rho+\sigma+k+\ell} \Big ) \big [ B_{iK}^{\rho-\a +\b + \ell-1} t_{ij}, \,B_{jK}^{\sigma - \b + \a +k-1} t_{ij}  \big ]
\eea
Since only terms with $\rho+\sigma+k+\ell$ odd contribute, we must have either $k+\rho$ odd while $\ell+\sigma$ is even, or the reverse. These two contributions are identical, so that we shall evaluate the contributions from $\ell+\sigma$ even and $k+\rho$ odd and include an overall factor of 2. Since the factor $\sfR_{k+\rho,0}$  in (\ref{FF.17}) vanishes for all odd $k+ \rho$ except $k+\rho=1$, the only contribution to the sum arises from $k + \rho=1$. Since the sum over $k$ in (\ref{FF.17}) starts at $k=1$, the sum restricts to $k=1$ and $\rho=0$, and we obtain, 
\bea
U_3 = 2 \pi i \sum_{\ell=1}^ \infty \sum_{ \sigma=0} ^\infty  \sum_{\b =0}^\sigma 
 \, \sfR_{\ell+\sigma,0} \, (-)^{\b} 
\binom{\ell+\sigma}{\ell}  \binom{\sigma}{\b} 
\big [ B_{iK}^{\b + \ell-1} t_{ij}, \,B_{jK}^{\sigma - \b } t_{ij}  \big ]
\eea
Next let $\b \to \sigma -\beta$, change variables $\sigma \to \mu = \ell+\sigma$, and swap the summations over $\beta, \ell$, 
\bea
U_3 = 2 \pi i \sum_{\mu=1}^ \infty (-)^\mu \sfR_{\mu,0}   
 \sum_{\b =0}^{\mu-1}  \big [ B_{iK}^{\mu-1-\b} t_{ij}, \,B_{jK}^\b t_{ij}  \big ] 
 \sum_{\ell=1} ^{ \mu-\b}  (-)^{\b+\ell}  \binom{\mu}{\ell}  \binom{\mu-\ell}{\b} 
\label{sumlim}
\eea
The sum over $\ell$ evaluates as follows, 
\bea
\sum_{\ell=1} ^{\mu-\b}  (-)^{\b+\ell}  \binom{\mu}{\ell}  \binom{\mu-\ell}{\b} 
& = &
(-)^\mu \delta_{\beta, \mu}- (-)^{\beta}  \binom{\mu}{\b} 
\label{sumlim2}
\eea
The first term on the right does not contribute as $\beta=\mu$ lies outside the summation range. Thus we are left with,
\bea
U_3 = - 2 \pi i \sum_{\mu=1}^ \infty (-)^\mu \sfR_{\mu,0}  
 \sum_{\b =0}^{\mu-1} (-)^\beta   \binom{\mu}{\b}  \big [ B_{iK}^{\mu-1-\b} t_{ij}, \,B_{jK}^\b t_{ij}  \big ] 
 \label{sumlim3}
\eea

\subsubsection*{$\bullet$ Calculation of $U_1+U_2$}

To compute the sum of $U_1$, $U_2$ in (\ref{defUUs}), we change variables in $U_2$ from $(\ell, \sigma, \b) $ to $(k, \rho, \a)$, regroup the sums as follows, and use $B_{iK}^{k-1} t_{ij} = (-)^{k-1} B_{jK}^{k-1} t_{ij}$ in both commutators,
\bea
U_1+U_2 & = & 
 - \sum_{k=1}^ \infty \sum_{\rho, \sigma=0} ^\infty \sfR_{k+\rho,0} \, \sfR_{\sigma,0} \, 
\binom{k+\rho}{k} 
\no \\ && \qquad \times 
\Big \{ B_{iK} ^\rho B_{jK}^\sigma \big [ a_i^K, \,B_{jK}^{k-1} t_{ij} \big ] 
+ B_{iK} ^\sigma B_{jK}^\rho \big [ B_{iK}^{k-1} t_{ij} , a_j^K \big ]  \Big \}
\eea
Next, we use the second version of Leibniz's rule on each term,
\bea
U_1+U_2 & = & 
 - \sum_{k=1}^ \infty \sum_{\rho, \sigma=0} ^\infty \sfR_{k+\rho,0} \, \sfR_{\sigma,0} \, 
\binom{k+\rho}{k} \sum_{\a=0}^{k-1} (-)^\a \binom{k-1}{\a} 
\no \\ && \quad \times 
\Big \{ B_{iK} ^\rho B_{jK}^{\sigma+k-1-\a}  \big [ B_{jK}^\a a_i^K, \, t_{ij} \big ] 
+ B_{iK} ^{\sigma +k-1-\a}  B_{jK}^\rho \big [  t_{ij} , B_{iK}^\a a_j^K \big ]  \Big \}
\eea
Denoting the sum of the terms with $\a =0$ by $U_0$ and the sum of terms with $\a \geq 1$ and thus $k \geq 2$ by $U'$, we have,
\bea
U_1+U_2= U_0+U'
\eea
where we have defined,  
\bea
U_0 & = &  
- \sum_{k=1}^ \infty \sum_{\rho, \sigma=0} ^\infty \sfR_{k+\rho,0} \, \sfR_{\sigma,0} \,  \binom{k+\rho}{k}  
\Big \{ B_{iK} ^\rho B_{jK}^{\sigma+k-1}  \big [ a_i^K, \, t_{ij} \big ] 
+ B_{iK} ^{\sigma +k-1}  B_{jK}^\rho \big [  t_{ij} , a_j^K \big ]  \Big \}
\no \\
U' & = & 
 - \sum_{k=2}^ \infty \sum_{\rho, \sigma=0} ^\infty \sum_{\a=1}^{k-1} (-)^{\a+\rho} \sfR_{k+\rho,0} \, \sfR_{\sigma,0}  
\binom{k+\rho}{k}  \binom{k-1}{\a} 
\no \\ && \qquad \times 
\Big \{ B_{jK}^{\rho+\sigma+k-1-\a}  \big [ B_{jK}^{\a-1}  t_{ij} , \, t_{ij} \big ] 
+ B_{iK} ^{\rho+ \sigma +k-1-\a}   \big [  t_{ij} , B_{iK}^{\a-1}  t_{ij}  \big ]  \Big \}
\eea

\sm

$\star$ We begin by evaluating $U'$ by converting all $B_{jK}$ to $B_{iK}$ in the first term inside the braces so that the sum factors as follows,
\bea
\Big \{ \cdots \Big \} = \Big ( 1 - (-)^{\rho+\sigma+k}) \Big ) B_{iK} ^{\rho+\sigma+k-1-\a}   \big [  t_{ij} , B_{iK}^{\a-1}  t_{ij}  \big ] 
\eea
Therefore, only terms with $\rho+\sigma+k$ odd contribute, which implies that either $k + \rho$ is even and $\sigma$ is odd, or $k+\rho$ is odd and $\sigma $ is even. Since $k \geq 2$, $\rho \geq 0$ and $\sfR_{k+\rho,0}$ vanishes for $k+ \rho$ odd we must have $\sigma $ odd and $k + \rho$ even. But since $\sfR_{\sigma,0}$ vanishes for all odd $\sigma$, except $\sigma =1$, the only contribution can come from $\sigma=1$ and we obtain,
\bea
U' & = & 
 - 2 \pi i \sum_{k=2}^ \infty \sum_{\rho=0} ^\infty \sum_{\a=1}^{k-1} (-)^{\a+\rho} \sfR_{k+\rho,0} 
\binom{k+\rho}{k}  \binom{k-1}{\a} 
 B_{iK} ^{\rho +k-\a}   \big [  t_{ij} , B_{iK}^{\a-1}  t_{ij}  \big ]  
\eea
Changing variable from $k$ to $\mu=\rho+k$, 
\bea
U' & = & 
 - 2 \pi i \sum_{\mu=2}^ \infty \sum_{\rho=0} ^{\mu-2} \sum_{\a=1}^{\mu-\rho-1} (-)^{\a+\rho} \sfR_{\mu,0} 
\binom{\mu}{\rho}  \binom{\mu-\rho-1}{\a} 
 B_{iK} ^{\mu-\a}   \big [  t_{ij} , B_{iK}^{\a-1}  t_{ij}  \big ]  
\eea
Swapping the summations over $\rho$ and $\alpha$,  and carrying out the sum over $\rho$ using,
\bea
\sum_{\rho=0}^{\mu-1-\a} (-)^{\a+\rho} \binom{\mu}{\rho} \binom{\mu-\rho-1}{\a} = (-)^{\mu+1}
\eea
we obtain, 
\bea
U'=  2 \pi i \sum_{\mu=2}^ \infty \sum_{\a=1}^{\mu-1} (-)^{\mu} \sfR_{\mu,0} 
 B_{iK} ^{\mu-\a}   \big [  t_{ij} , B_{iK}^{\a-1}  t_{ij}  \big ]  
\eea
Applying Leibniz's rule to the $B$ factors outside the commutator, 
\bea
U'=  2 \pi i \sum_{\mu=2}^ \infty (-)^{\mu} \sfR_{\mu,0}  \sum_{\a=1}^{\mu-1} \sum_{\beta=0}^{\mu-\a} \binom{\mu-\a}{\b} 
  \big [  B_{iK} ^{\b}  t_{ij} , B_{iK}^{\mu -\b -1}  t_{ij}  \big ]  
\eea
Swapping the summations of $\a$ and $\b$ with some care,
\bea
\label{4.swap}
\sum_{\a=1} ^{\mu-1} \sum_{\b =0}^{\mu -\a} = \sum_{\b=0}^{\mu-1} \sum_{\a=1} ^{{\rm min}(\mu-1,\mu-\b)} 
= \sum_{\b=0}^{\mu-1} \sum_{\a=1} ^{\mu-\b } \Big ( 1 - \delta _{\b,0} \delta _{\a,\mu} \Big)
\eea
we obtain,
\bea
U'=  2 \pi i \sum_{\mu=2}^ \infty (-)^{\mu} \sfR_{\mu,0} \left (  - [t_{ij}, B_{iK}^{\mu-1} t_{ij} ]  + \sum_{\b=0}^{\mu-1}  \sum_{\a=1}^{\mu-\b} \binom{\mu-\a}{\b} 
  \big [  B_{iK} ^{\b}  t_{ij} , B_{iK}^{\mu -\b -1}  t_{ij}  \big ] \right )
\eea
where the first term inside the big parentheses is produced by the term $- \delta _{\b,0} \delta _{\a,\mu}$ in (\ref{4.swap}).   The sum over $\a$ may be performed using, 
\bea
\sum_{\a=1}^{\mu-\b} \binom{\mu-\a}{\b}  = \binom{\mu}{\b+1}
\eea
so that,
\bea
U'=  2 \pi i \sum_{\mu=2}^ \infty (-)^{\mu} \sfR_{\mu,0} \left (  - [t_{ij}, B_{iK}^{\mu-1} t_{ij} ]  
+ \sum_{\b=0}^{\mu-1}  \binom{\mu}{\b+1} 
  \big [  B_{iK} ^{\b}  t_{ij} , B_{iK}^{\mu -\b -1}  t_{ij}  \big ]  \right )
\eea
Finally, let $\beta \to \mu-1-\beta$, 
\bea
U'=  2 \pi i \sum_{\mu=2}^ \infty (-)^{\mu} \sfR_{\mu,0} \left (  - [t_{ij}, B_{iK}^{\mu-1} t_{ij} ]   +
\sum_{\b=0}^{\mu-1}  \binom{\mu}{\b} 
  \big [  B_{iK} ^{\mu-1-\b}  t_{ij} , B_{iK}^{\b }  t_{ij}  \big ]  \right )
\eea
Recall the expression (\ref{sumlim3}) for $U_3$ with $B_{jK}$ converted to $B_{iK}$,
\bea
U_3 = - 2 \pi i \sum_{\mu=1}^ \infty (-)^\mu \sfR_{\mu,0}  
 \sum_{\b =0}^{\mu-1}  \binom{\mu}{\b}  \big [ B_{iK}^{\mu-1-\b} t_{ij}, \,B_{iK}^\b t_{ij}  \big ] 
\eea
Their sum is as follows, 
\bea
U_3+U'= - 2 \pi i \sum_{\mu=2}^ \infty (-)^{\mu} \sfR_{\mu,0}  [t_{ij}, B_{iK}^{\mu-1} t_{ij} ]  
\eea
Since $\mu \geq 2$ only even $\mu$ contribute so that the sign factor equals 1 and we recover
$-2\pi i$ times the second term in the expression (\ref{cv1exp}) for $\cV^{(1)}_{ij}$. Hence, it
remains to show that the leftover term $U_0$ in $\cV^{(2)}_{ij}= U_0{+}U'{+}U_3$
equals $-2\pi i $ times the first term of (\ref{cv1exp}).

\sm

$\star$ To evaluate $U_0$, change variables from $k $ to $\nu=k+\sigma-1$, 
\bea
U_0 & = &  
- \sum_{\nu, \rho=0}^ \infty \Big \{ B_{iK} ^\rho B_{jK}^{\nu}  + B_{iK} ^\nu  B_{jK}^\rho  \Big \} \big [ a_i^K, \, t_{ij} \big ] 
\sum_{\sigma =0}^{\nu} \sfR_{\nu + \rho -\sigma+1,0} \, \sfR_{\sigma,0} \,  \tbinom{\nu + \rho - \sigma +1}{\rho} 
\eea
or
\bea
U_0 & = &  
- \sum_{\nu, \rho=0}^ \infty  f(\nu, \rho) B_{iK} ^\rho B_{jK}^\nu \big [ a_i^K, \, t_{ij} \big ] 
\eea
where the function $f(\nu, \rho)$ is defined by, 
\bea
\label{4.ff}
f(\nu, \rho) & = & 
\sum_{\sigma =0}^\nu \sfR_{\nu+\rho -\sigma+1,0} \, \sfR_{\sigma,0} \,  \tbinom{\nu + \rho -\sigma+1}{\rho}  
+ \sum_{\sigma =0}^{\rho} \sfR_{\nu+\rho -\sigma+1,0} \, \sfR_{\sigma,0} \,  \tbinom{\nu+ \rho -\sigma+1}{\nu}
\eea

{\prop
\label{4.prop:5}
We have the following equivalent identities,
\bea
\label{4.lem:ff}
f(\nu, \rho) =  2 \pi i \sfR_{\nu + \rho,0} \, \binom{\nu + \rho}{\rho}
\eea
or explicitly in terms of Bernoulli numbers given by equation (\ref{6.bernoulli}). 
}

\sm

With the help of the proposition, $U_0$ simplifies as follows, upon setting $\mu=r+1$,
\bea
U_0 & = &
- 2 \pi i \sum_{r=0}^ \infty \sfR_{r,0} \Big ( B_{iK} + B_{jK} \Big )^r   \big [ a_i^K, \, t_{ij} \big ] 
\eea
This result indeed reproduces $-2\pi i$ times the first term of $\cV^{(1)}_{ij}$ in (\ref{cv1exp}),
implies $\cV^{(2)}_{ij} = -2\pi i \cV^{(1)}_{ij}$ in (\ref{defv1v2}) and therefore concludes the proof of Lemma \ref{E.lem:1}.

\subsubsection{Proof of Proposition \ref{4.prop:5}}

The formula (\ref{4.lem:ff}) is equivalent to the following relation, 
\bea
\label{6.bernoulli}
\sum _{\sigma=0}^{\nu} { \Ber _{\sigma} \Ber_{\nu+ \rho + 1  -\sigma} \over \rho ! \, \sigma ! \, (\nu +1 - \sigma)! }
+ \sum _{\sigma=0}^{\rho} { \Ber _{\sigma} \Ber_{\nu +\rho +1 -\sigma} \over \nu ! \, \sigma ! \, (\rho +1  - \sigma)!}
+ { \Ber _{\nu + \rho} \over \nu ! \, \rho ! } =0
\eea
as may be seen by  expressing $\sfR_{m,n}$ in (\ref{4.lem:ff}) in terms of Bernoulli numbers using (\ref{E.ZR}). In turn, the product of (\ref{6.bernoulli}) with $\nu ! \, \rho!$ gives the formula quoted in section \ref{sec:quote}.   Thus,  proving Proposition \ref{4.prop:5} is equivalent to proving (\ref{6.bernoulli}). To prove (\ref{6.bernoulli}), we  consider the generating function $h(x)$  for the Bernoulli numbers and its translate $h(x{+}y)$
 for complex~$x,y$,
\bea
\label{F.4.2}
h(x) = { x \over e^x -1} = \sum _{n=0}^\infty {\Ber _n \over n!} x^n
\hskip 1in
h(x+y) =  \sum _{m,n=0}^\infty {\Ber _{m+n} \over m! \, n!} \, x^m \, y^n
\eea
where the series for $h(x)$ and $h(x+y)$ converge absolutely inside the discs $|x| < 2 \pi$ and $|x+y|< 2 \pi$, respectively. Using the Taylor expansions of (\ref{F.4.2}) for $|x|, |y|, |x+y| < 2 \pi$, we expand the following combination in a double Taylor series in $x$ and $y$, 
\bea
\label{F.4.4}
{ 1 \over y}  h(y) \Big (h(x+y) - h(x) \Big )  
=
\sum _{\sigma=0} ^\infty { \Ber _\sigma \over \sigma !} y^{\sigma-1} 
\sum _{\nu ,\lambda =0}^\infty  { \Ber _{\nu+\lambda+1} \over \nu ! \, (\lambda+1)!} \, x^\nu y^{\lambda+1}
\eea
Changing summation variable from $\lambda$ to $\rho = \lambda + \sigma$, while retaining the variables $\sigma$ and $\nu$, this expression becomes, 
\bea
\label{F.4.5}
{ 1 \over y}  h(y) \Big (h(x+y) - h(x) \Big )  
=
\sum_{\nu, \rho=0}^\infty \sum _{\sigma=0} ^\rho 
{ \Ber _\sigma \, \Ber _{\nu+\rho+1-\sigma} \over \nu ! \, \sigma ! \,  (\rho+1-\sigma)! } \, x^\nu y^\rho
\eea
Comparing with the identity (\ref{6.bernoulli}) that is to be proven, we observe that the coefficient of $x^\nu y^\rho$ in the sum of (\ref{F.4.5}) is precisely the second term in (\ref{6.bernoulli}). Therefore, using the symmetry of (\ref{F.4.5}) under swapping $(x,\nu)$ with $(y, \rho)$, the generating function for the left side of the identity (\ref{6.bernoulli}) for all $\nu, \rho \geq 0$ is given by,
\bea
&&
 { 1 \over x}  h(x) \Big (h(x+y) - h(y) \Big )  + { 1 \over y}  h(y) \Big (h(x+y) - h(x) \Big )  + h(x+y) 
\no \\ && \quad =
{ 1 \over e^x -1} \Big ( { x+y \over e^{x+y}-1} - { y \over e^y-1} \Big ) 
+ { 1 \over e^y -1} \Big ( { x+y \over e^{x+y}-1} - { x \over e^x-1} \Big ) 
+{ x+y \over e^{x+y}-1}
\eea
The last line is readily seen to vanish, thereby proving (\ref{6.bernoulli}) and Proposition \ref{4.prop:5}.

\end{document}